\documentclass[11pt]{article}

\usepackage[margin=1in]{geometry}
\usepackage{amsmath,amsthm,amsfonts,amssymb}
\usepackage{bm}
\usepackage{graphicx}
\usepackage{float}
\usepackage{booktabs}
\usepackage{array}
\usepackage{enumitem}
\usepackage{natbib}
\IfFileExists{newtxtext.sty}{\usepackage{newtxtext}}{}
\IfFileExists{newtxmath.sty}{\usepackage[subscriptcorrection]{newtxmath}}{}
\usepackage[colorlinks=true,linkcolor=black,citecolor=black,urlcolor=black]{hyperref}
\graphicspath{{./art/}}

\newenvironment{keywords}{\par\smallskip\noindent\textbf{Keywords:}\ }{\par\medskip}

\newcommand{\R}{\mathbb R}
\newcommand{\Ycal}{\mathcal Y}
\newcommand{\Acal}{\mathcal A}
\newcommand{\Lcal}{\mathcal L}

\newcommand{\Ical}{\mathcal I}
\newcommand{\Hcal}{\mathcal H}

\newcommand{\LR}{\mathrm{LR}}
\newcommand{\T}{\mathrm{\scriptscriptstyle T}}
\newcommand{\F}{\mathrm F}
\newcommand{\op}{\mathrm{op}}
\newcommand{\indep}{\perp\!\!\!\perp}

\DeclareMathOperator{\tr}{tr}

\DeclareMathOperator*{\argmax}{arg\,max}
\DeclareMathOperator*{\argmin}{arg\,min}

\newtheorem{theorem}{Theorem}
\newtheorem{lemma}{Lemma}
\theoremstyle{definition}
\newtheorem{assumption}{Assumption}
\theoremstyle{remark}

\begin{document}

\title{Multimodal domain adaptation under label shift and blockwise missing modalities}

\author{%
Zebin Wang$^{1}$, Ziang Dou$^{2}$, Molei Liu$^{2}$, and Tianxi Cai$^{1}$\\[0.75em]
\small $^{1}$Department of Biostatistics, Harvard University, Boston, MA 02115, USA\\
\small $^{2}$Department of Biostatistics, Peking University Health Science Center, Beijing 100083, China%
}
\date{}

\maketitle

\begin{abstract}
\noindent Multimodal domain adaptation uses labeled source datasets to predict outcomes in an unlabeled target population. Here, different sources may observe different subsets of modalities and have distribution shifts from the target. In this scenario, the cross-modal patterns used for alignment depend on the outcome distribution, so aligning unadjusted source data can misrepresent the target. Nevertheless, the distribution shift of the outcome cannot be measured directly when labels are unavailable in the target sample and the missing modality blocks prevent simple pooling. We propose a reference-anchored domain adaptation method. A reference modality observed in every source and the target is used to estimate the target outcome distribution and reweight source observations before alignment. The auxiliary modalities are then mapped to a common, target-defined representation obtained by canonical correlation analysis (CCA) in the target and reproduced in each source by ridge-regression maps. Outcome information is transferred through density-ratio models on the aligned representation. When gold-standard source labels are sparse, a surrogate-label-assisted approach is developed to enable robust domain adaptation. We establish a unified-rotation match-up result for the target CCA and source ridge maps and consistency of the target conditional outcome distribution. Simulations and a renal cell carcinoma (RCC) application show improved calibration and stable prediction under distributional shift and blockwise missing modalities.
\end{abstract}


\section{Introduction}
\label{sec:intro}

\subsection{Background}
\label{sec:background}

Multimodal biomedical studies increasingly combine structured electronic health records (EHRs), clinical text, radiology reports, medical images, and molecular measurements \citep{lahat2015multimodal,baltrusaitis2019multimodal}. In practice, however, these data are often assembled across hospitals, cohorts, or calendar periods rather than collected under a common protocol. One hospital may collect imaging but not molecular data, while another may provide only structured records and clinical text. Within one hospital, new modalities may also become available only in later periods. Missingness therefore often occurs by entire modality blocks and differs systematically across data sources. At the same time, gold-standard outcomes, such as disease recurrence, may require a specialist review and prolonged follow-up. Such outcomes may be observed in some external or historical source cohorts, but they are not available in the target population where prediction is needed. This naturally leads to a multimodal domain-adaptation problem: using labeled, partially observed source data to predict outcomes in an unlabeled target population.

A further complication is the distributional shift. Disease prevalence, patient composition, and referral patterns may differ across institutions or over time. We focus on label shift, under which the outcome distribution differs between the source and target populations, while the class-conditional feature distribution is stable \citep{lipton2018labelshift,garg2020unified}. For multivariate outcomes, this may involve changes in both individual outcome frequencies and their dependence. Even when the distribution of the measurements conditional on the outcome is stable, differences in the outcome mix change the marginal feature distributions and the associations between modalities. Consequently, direct pooling of source data may produce poorly calibrated target predictions.

The central challenge is that label shift and blockwise missingness cannot be addressed separately. Auxiliary modalities cannot be directly combined because different sources observe different blocks and may represent the same underlying information in different coordinates. However, aligning these modalities using unadjusted source data can also be misleading because the cross-modal patterns in each source reflect its own outcome distribution rather than that of the target. Thus, the source-to-target shift must be taken into account before learning a target-relevant alignment, but the target labels needed for direct adjustment are not available. We study this problem when one reference modality (e.g., the codified and structured EHR features) is observed across all domains, auxiliary modalities are observed only in subsets of domains, and source gold-standard labels may be sparse. 

\subsection{Related Literature}
\label{sec:related}

Our work lies at the intersection of domain adaptation and data integration with blockwise missing covariates. Much of the statistical literature on domain adaptation begins with a covariate vector whose coordinates have the same meaning and are observed in both the source and target populations. Under covariate shift, importance-weighted likelihood and augmented estimating equations account for changes in the covariate distribution while retaining a stable conditional outcome model \citep{shimodaira2000improving,liu2023augmented}. More recent semiparametric work develops efficient and multiply robust estimators of target risks in broader forms of distributional shift \citep{qiu2024efficient}. Under label shift, the outcome distribution changes across populations while the covariate distribution conditional on the outcome remains stable. \citet{lee2025doubly} developed doubly flexible estimation of target parameters under this condition, and \citet{lee2025efficient} studied semiparametrically efficient inference based on the source-to-target outcome density ratio. Classification methods for label shift estimate target class proportions through expectation--maximization, moment matching, or likelihood-based calibration \citep{saerens2002adjusting,lipton2018labelshift,garg2020unified}. Multi-source methods have also combined label-shift correction with representation or distribution alignment. \citet{redko2019optimal} jointly estimate target class proportions and optimal-transport couplings, while \citet{li2019target} address target shift during adversarial representation learning and extend their construction to multiple domains. \citet{shui2021aggregating} aggregate multiple target-shifted sources according to similarities between their semantic conditional distributions. \citet{xu2025robust} develop robust multi-source label-shift adaptation under source contamination. These works show that label-shift correction need not be a post hoc calibration step after fixing a representation. Their setups nevertheless presume a common observed input or representation space across domains; they do not address source-specific absence of entire modality blocks or the resulting need to align incomparable auxiliary coordinates.

A separate statistical literature studies data integration with blockwise missing covariates. Structured matrix completion and integrative factor methods recover low-dimensional signals from overlapping data blocks \citep{cai2016structured,zhu2020generalized,zhou2023multisource}. Other methods construct prediction or inference procedures directly from observed blocks, using sparse prediction, multiple imputation, estimating equations, projections, or influence-function decompositions \citep{yu2020optimal,xue2021integrating,song2024semisupervised,berrett2024efficient,li2024adaptive}. Recent work further accommodates heterogeneous sources through multi-task learning or source-adaptive representation retrieval \citep{sui2026multisource,xu2025representation}. Recent representation-learning methods for blockwise missing data include \citet{liu2026representation}, whose anchor-based principal component method aligns group-specific subspaces under blockwise missingness and signal heterogeneity, and \citet{yu2026pattern}, whose MOSAIC framework learns shared and modality-specific representations, borrows across overlapping missingness patterns, and calibrates the resulting predictor using labeled subjects from the focal pattern. These works address representation recovery or pattern-specific supervised prediction rather than adaptation to an unlabeled target population under label shift. In particular, they do not use a target outcome mixture inferred from unlabeled target data to reweight the cross-modal alignment. Under label shift, a source cross-modal association averages over the source outcome distribution and can therefore differ from the corresponding target association. An alignment learned before correcting this difference may consequently emphasize source-specific rather than target-relevant relationships.

Multi-source multimodal domain adaptation itself has been studied by \citet{zhao2025multisource}, who jointly align modalities and domains under covariate shift. Their formulation places every domain in the same multimodal input space and does not allow whole modalities to be absent by source. \citet{fan2023incomplete} learn a common latent representation from incomplete views before estimating label-shift weights and fine-tuning the reweighted classifier; their formulation does not address source-specific absence of entire modality blocks or incomparable auxiliary coordinates across sources. \citet{sui2026multisource} impute data-source blocks missing for an entire task and learn shared and task-specific feature--response mappings from observed task responses. Their supervised multi-task objective does not include an unlabeled target outcome mixture or target-mixture weighting before cross-modal alignment. We study the simultaneous combination of an unlabeled target under label shift, source-specific absence of entire modality blocks, and alignment of incomparable auxiliary coordinates after reweighting source observations to the target outcome mixture. A reference modality shared across domains identifies the target outcome mixture and supports the target-directed alignment. The framework does not require either a shared multimodal input space or a single common latent representation.

\subsection{Contribution}\label{sec:contribution}

We develop a reference-anchored framework for multimodal domain adaptation under label shift and blockwise missing modalities. In this framework, a reference modality observed in every domain identifies a target-defined anchor. This reference modality is not required to be the most predictive data block. Its role is instead to provide the common observable structure needed to identify the target outcome distribution and to connect auxiliary modalities that cannot be compared directly across sources. Our contribution is threefold.

First, we introduce a target-directed identification strategy for cross-modal alignment. A reference modality observed in every domain provides an observable anchor, avoiding the requirement that all modalities share a single latent representation. Crucially, source observations are reweighted to the target outcome distribution before alignment because the label shift changes the marginal cross-modal associations used to construct the representation. This ``weight before alignment'' principle distinguishes our setting from methods that either correct label shift in a common input space or align incomplete modalities without using the target outcome mixture.

Second, we establish identification and consistency for the resulting target predictor. Although dimension-reduction and alignment matrices depend on arbitrary coordinate choices, we show that all target and source aligned features are recovered up to one common rotation. This removes the representation ambiguity; together with a separate aligned likelihood-ratio identification condition, it yields the target conditional outcome model.

Third, we extend the framework to settings with sparse gold-standard source labels by using surrogate variables for calibration and alignment, without replacing the gold-standard outcome as the estimand. Simulations and the renal-cell-carcinoma application demonstrate the benefits of target calibration and reference-anchored alignment under simultaneous label shift and blockwise modality availability.

\section{Model setup and notation}
\label{sec:model}

Suppose there is one target domain, indexed by $m=0$, and $M$ source domains, indexed by $m=1,\ldots,M$.  Domain $m$ contains subjects $i=1,\ldots,n_m$.  There are $K$ modalities, with modality 1 used as the reference modality and modalities $2,\ldots,K$ used as auxiliary modalities.  Let $\Acal_m\subseteq\{1,\ldots,K\}$ be the modalities observed in domain $m$, with $1\in\Acal_m$ for all $m$.  For the main formulation, the target domain observes the modalities under analysis, so $\Acal_0=\{1,\ldots,K\}$ after restricting to the selected modality set. Masked analyses are interpreted relative to this selected set. In this paper, bold lowercase symbols denote subject-level column vectors, and bold uppercase symbols denote their row-stacked matrices.

For subject $i$ in domain $m$, the observed feature vector for modality $k\in\Acal_m$ is $\mathbf x_{k,i}^{(m)}\in\R^{d_k}$, and the corresponding data matrix is $\mathbf X_k^{(m)}=(\mathbf x_{k,1}^{(m)},\ldots,\mathbf x_{k,n_m}^{(m)})^\T$.  The gold-standard outcome is $\mathbf y_i^{(m)}\in\Ycal=\{0,1\}^d$.  Source gold-standard outcomes are observed for subjects in $\Ical_m\subseteq\{1,\ldots,n_m\}$, whereas $\Ical_0=\varnothing$ during fitting.  Surrogate evidence, when available, is denoted by $\mathbf s_i^{(m)}$. Throughout, $\mathcal O(q)$ is the group of $q\times q$ orthogonal matrices, and $\|\cdot\|_2$ and $\|\cdot\|_{\mathrm F}$ denote Euclidean and Frobenius norms.

In domain $m$, the gold-standard outcome has distribution $\mathbf{y} ^ {(m)} \sim \rho_y(\cdot;\bm\theta^{(m)})$ on the finite state space $\Ycal$.  A working example is the Ising model \citep{wainwright2008graphical}:
\begin{equation}
\label{eq:ising}
\rho_y(\mathbf y;\bm\theta^{(m)})=
\frac{1}{C(\bm\theta^{(m)})}
\exp\left\{
\sum_{1\leq a<b\leq d}\theta_{ab}^{(m)}(2y_a-1)(2y_b-1)
+
\sum_{a=1}^d\theta_{aa}^{(m)}(2y_a-1)
\right\}.
\end{equation}
The parameter $\bm\theta^{(m)}$ may vary across domains.  We use \emph{label shift} for this variation in the distribution.  For $d=1$ this is ordinary class-prior shift; for $d>1$ it allows changes in both marginal outcome prevalence and dependence among outcome components.  Domain-specific modality loadings and missing modality blocks are not called label shift; they are handled by the representation and alignment components.

The working representation contains a latent reference factor $\widetilde{\mathbf u}_i^{(m)}\in\R^h$ and a latent auxiliary factor $\widetilde{\mathbf v}_i^{(m)}\in\R^h$.  Conditional on the outcome,
\begin{equation*}
\label{eq:latent_label}
\widetilde{\mathbf u}_i^{(m)}=\psi_u(\mathbf y_i^{(m)})+\bm\xi_{u,i}^{(m)},
\qquad
\widetilde{\mathbf v}_i^{(m)}=\psi_v(\mathbf y_i^{(m)})+\bm\xi_{v,i}^{(m)},
\end{equation*}
with noise independent of $\mathbf y_i^{(m)}$.  Auxiliary modality $k\geq2$ observes only coordinates in $\Hcal_k\subseteq\{1,\ldots,h\}$, and $\widetilde{\mathbf v}_{k,i}^{(m)}=\widetilde{\mathbf v}_{i,\Hcal_k}^{(m)}\in\R^{\ell_k}$, where $\ell_k=|\Hcal_k|$.  The observed features satisfy
\begin{equation*}
\label{eq:obs_model}
\mathbf x_{1,i}^{(m)}=\mathbf L_1\widetilde{\mathbf u}_i^{(m)}+\bm\zeta_{1,i}^{(m)},
\qquad
\mathbf x_{k,i}^{(m)}=\mathbf L_k^{(m)}\widetilde{\mathbf v}_{k,i}^{(m)}+\bm\zeta_{k,i}^{(m)},
\quad k=2,\ldots,K,
\end{equation*}
where $\mathbf L_1^\T\mathbf L_1=\mathbf I_h$ and $\{\mathbf L_k^{(m)}\}^\T\mathbf L_k^{(m)}=\mathbf I_{\ell_k}$.  The reference loading is shared across domains, whereas auxiliary loadings may be domain-specific; this is the loading heterogeneity addressed by the match-up component.

Surrogate variables are treated as evidence about the gold-standard outcome.  When they are used, the working bridge condition is
\begin{equation}
\label{eq:surrogate_independence}
\mathbf s_i^{(m)}\indep\{\mathbf x_{k,i}^{(m)}:k\in\Acal_m\}\mid
\mathbf y_i^{(m)}.
\end{equation}
The prediction target is the target-domain conditional distribution of the unobserved gold-standard outcome.  The theory below studies this distribution conditional on the retained aligned representation.  It coincides with the full-feature risk $P\{\mathbf y_i^{(0)}=\mathbf y\mid \mathbf x_{k,i}^{(0)}, k\in\Acal_0,\mathbf s_i^{(0)}\}$ when that representation is sufficient for the outcome.  Target gold-standard outcomes, if present for validation, are masked during fitting and used only for final evaluation.

\section{Methodology}
\label{sec:method}

\subsection{Overview: representation, match-up and transfer}
\label{sec:method_overview}

\begin{figure}[t]
\centering
\includegraphics[width=\linewidth]{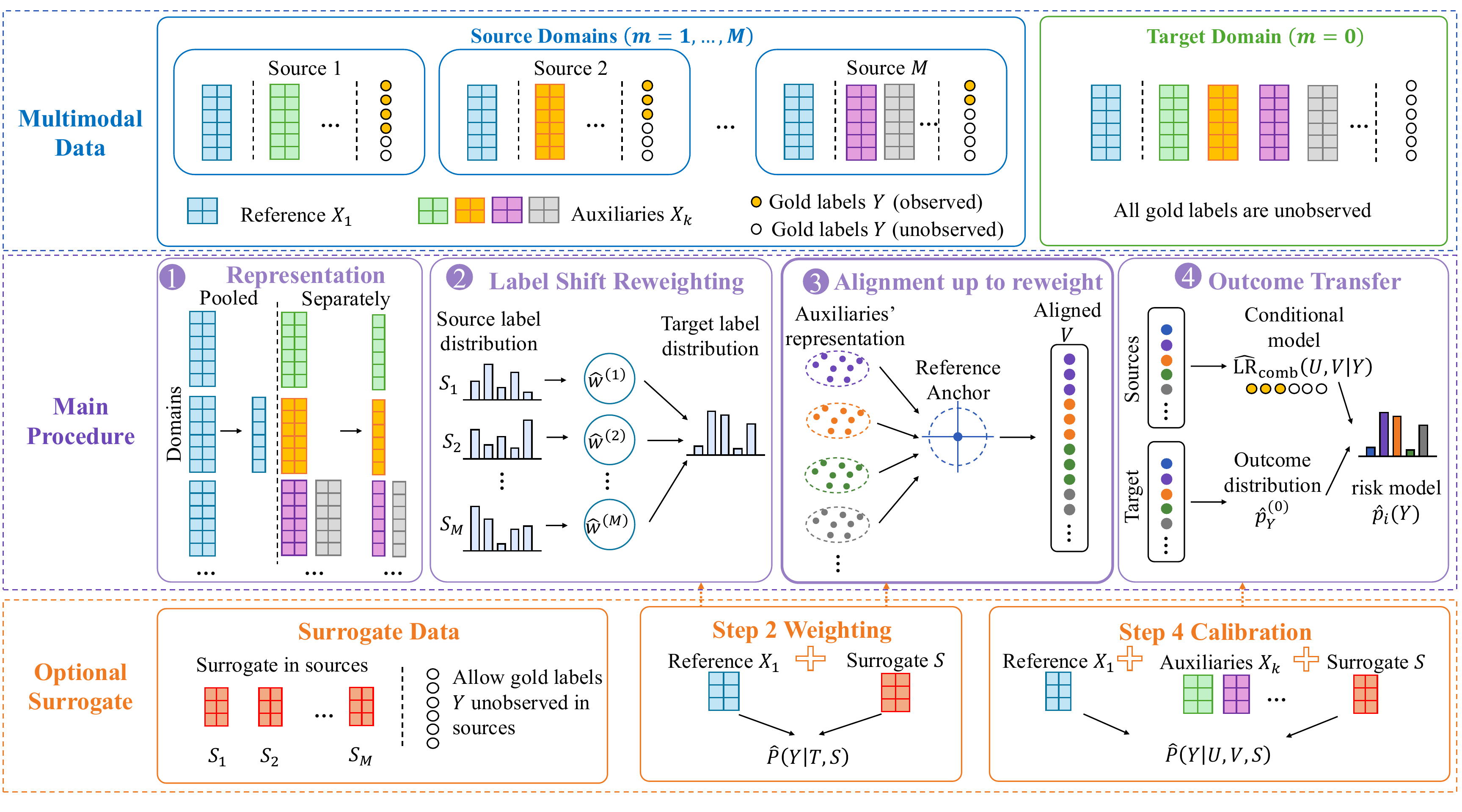}
\caption{Complete workflow of the proposed reference-anchored multimodal domain-adaptation procedure. }
\label{fig:method_flowchart}
\end{figure}

Our proposed procedure has four stages. First, a prespecified unsupervised representation-learning method is applied to each observed modality to obtain a low-dimensional score representation. Second, the reference scores observed in all domains are used to estimate the target outcome distribution and the corresponding label-shift weights. Third, these weights are incorporated before alignment: CCA fitted in the target domain defines a reference-side anchor, and weighted ridge regression maps the observed source auxiliary blocks into a common target-anchored coordinate system. Fourth, likelihood-ratio estimates obtained from the aligned coordinates are combined with the estimated target outcome distribution to estimate the target conditional outcome distribution. When gold-standard outcomes are sparsely observed or unavailable in the source domains, an optional surrogate component provides working outcome-state probabilities to construct alignment weights and estimate likelihood ratios. Target surrogate measurements, when available, may also be incorporated as additional outcome evidence. Figure~\ref{fig:method_flowchart} summarizes these four stages and the
optional surrogate-assisted pathway.

\subsection{Representation learning and reference calibration}
\label{sec:pca_method}
\label{sec:calibration_method}

In this section, we distinguish three stages of representation: raw backend scores $\widehat{\mathbf T}$, target-standardized scores $(\widehat{\mathbf R},\widehat{\mathbf Z})$, and final aligned coordinates $(\widehat{\mathbf U},\widehat{\mathbf V})$. This notation separates representation learning, target-weighted standardization, and cross-domain alignment. Each feature matrix is centered within its domain before representation learning. We then apply a prespecified unsupervised method to each observed modality to obtain a low-dimensional score representation.

For the reference modality, we pool the centered matrices $\widetilde{\mathbf X}_1^{(0)},\ldots,\widetilde{\mathbf X}_1^{(M)}$ by rows to estimate a common representation map. We then apply this map within each domain to obtain the raw reference-score matrices $\widehat{\mathbf T}_1^{(m)}\in\R^{n_m\times h}$, $m=0,\ldots,M$. For each domain $m$ and auxiliary modality $k\in\Acal_m\setminus\{1\}$, we fit a separate representation map and use it to obtain $\widehat{\mathbf T}_k^{(m)}\in\R^{n_m\times\ell_k}$. We denote the $i$th rows of $\widehat{\mathbf T}_1^{(m)}$ and $\widehat{\mathbf T}_k^{(m)}$ as column vectors $\widehat{\mathbf t}_{1,i}^{(m)}$ and $\widehat{\mathbf t}_{k,i}^{(m)}$, respectively. Thus, the reference scores share a common coordinate system across domains, whereas the auxiliary scores may have domain-specific coordinate systems to be mapped into a common aligned coordinate system in subsequent procedures.

Let $\mathbf0$ be the all-zero outcome vector. On the raw reference-score scale, we define the \textit{reference likelihood ratio} $\mathrm{LR}_{\mathrm{ref}}(\mathbf t\mid\mathbf y) = p(\mathbf t\mid \mathbf y)/p(\mathbf t\mid\mathbf0)$, where $\mathbf y\in\Ycal$. The ratio $\mathrm{LR}_{\mathrm{ref}}(\mathbf t\mid\mathbf y)$ is estimated from pooled labeled source pairs $\big\{
(\widehat{\mathbf t}_{1,i}^{(m)},\mathbf y_i^{(m)}):i\in\Ical_m,\ m=1,\ldots,M
\big\}$.  A probabilistic multiclass classifier can be converted into these baseline-referenced ratios through Bayes' rule, a standard classification-based route to density-ratio estimation \citep{sugiyama2012density}. The target outcome-distribution parameter is initialized from the target reference scores by
\begin{equation}
\label{eq:theta_init}
\widehat{\bm\theta}_{\mathrm{init}}^{(0)}
=
\argmax_{\bm\theta}
\sum_{i=1}^{n_0}
\log\left[
\sum_{\mathbf y\in\Ycal}
\rho_y(\mathbf y;\bm\theta)
\widehat{\mathrm{LR}}_{\mathrm{ref}}(\widehat{\mathbf t}_{1,i}^{(0)}\mid\mathbf y)
\right].
\end{equation}
For each source domain, estimate $\widehat{\bm\theta}^{(m)}$ by maximizing $\sum_{i\in\Ical_m}\log\rho_y(\mathbf y_i^{(m)};\bm\theta)$. Define the estimated outcome-state ratio and its subject-level evaluation by
\begin{equation}
\label{eq:source_weight}
\widehat\omega_m(\mathbf y)
=
\frac{\rho_y(\mathbf y;\widehat{\bm\theta}_{\mathrm{init}}^{(0)})}
{\rho_y(\mathbf y;\widehat{\bm\theta}^{(m)})},
\qquad
\widehat w_i^{(m)}
=
\widehat\omega_m(\mathbf y_i^{(m)}),
\quad i\in\Ical_m,\quad m=1,\ldots,M.
\end{equation}

Thus $\omega_m(\cdot)$ denotes an outcome-state ratio, whereas
$w_i^{(m)}=\omega_m\{\mathbf y_i^{(m)}\}$ denotes its evaluation for
subject $i$; hats indicate estimated quantities. We use $\Lcal_m$ for
the alignment index set. For the target domain, set
$\widehat w_i^{(0)}=1$ and
$\Lcal_0=\{1,\ldots,n_0\}$. For $m\geq1$, set
$\Lcal_m=\Ical_m$ in the gold-label procedure; when
surrogate-assisted weights are used, $\Lcal_m$ is enlarged to all
subjects with the required modalities and surrogate information.

\subsection{Target-calibrated reference-anchored match-up}
\label{sec:alignment_method}

For $k=1$ use $\widehat{\mathbf t}_{1,i}^{(m)}$, and for $k\geq2$ use $\widehat{\mathbf t}_{k,i}^{(m)}$. Define the target-weighted means and covariances
\begin{align*}
\widehat{\bm\mu}_{1}^{(m)}
&=
\frac{\sum_{i\in\Lcal_m}
\widehat w_i^{(m)}\widehat{\mathbf t}_{1,i}^{(m)}}
{\sum_{i\in\Lcal_m}\widehat w_i^{(m)}},
&
\widehat{\bm\Sigma}_{1}^{(m)}
&=
\frac{\sum_{i\in\Lcal_m}\widehat w_i^{(m)}
(\widehat{\mathbf t}_{1,i}^{(m)}-\widehat{\bm\mu}_{1}^{(m)})
(\widehat{\mathbf t}_{1,i}^{(m)}-\widehat{\bm\mu}_{1}^{(m)})^\T}
{\sum_{i\in\Lcal_m}\widehat w_i^{(m)}},
\\
\widehat{\bm\mu}_{k}^{(m)}
&=
\frac{\sum_{i\in\Lcal_m}
\widehat w_i^{(m)}\widehat{\mathbf t}_{k,i}^{(m)}}
{\sum_{i\in\Lcal_m}\widehat w_i^{(m)}},
&
\widehat{\bm\Sigma}_{k}^{(m)}
&=
\frac{\sum_{i\in\Lcal_m}\widehat w_i^{(m)}
(\widehat{\mathbf t}_{k,i}^{(m)}-\widehat{\bm\mu}_{k}^{(m)})
(\widehat{\mathbf t}_{k,i}^{(m)}-\widehat{\bm\mu}_{k}^{(m)})^\T}
{\sum_{i\in\Lcal_m}\widehat w_i^{(m)}}.
\end{align*}
The standardized pre-alignment scores are defined as
\begin{equation*}
\widehat{\mathbf r}_i^{(m)}
=
(\widehat{\bm\Sigma}_{1}^{(m)})^{-1/2}
(\widehat{\mathbf t}_{1,i}^{(m)}-\widehat{\bm\mu}_{1}^{(m)}),
\qquad
\widehat{\mathbf z}_{k,i}^{(m)}
=
(\widehat{\bm\Sigma}_{k}^{(m)})^{-1/2}
(\widehat{\mathbf t}_{k,i}^{(m)}-\widehat{\bm\mu}_{k}^{(m)}).
\end{equation*}
The inverse square root is the spectral inverse on the retained score subspace. Let $\widehat{\mathbf R}^{(m)}$ and $\widehat{\mathbf Z}_k^{(m)}$ be the corresponding row-stacked matrices. These are the empirical counterparts of the population pre-alignment scores $\mathbf R^{(m)}$ and $\mathbf Z_k^{(m)}$ used in Section~\ref{sec:theory}.

On the target domain, concatenate $\widehat{\mathbf z}_i^{(0)}= \operatorname{col} \big \{\widehat{\mathbf z}_{2,i}^{(0)},\ldots, \widehat{\mathbf z}_{K,i}^{(0)} \big\}\in\R^{d_z}$, where $d_z=\sum_{k=2}^K\ell_k$.  Let $\widehat{\bm\Sigma}_{rr}^{(0)}$, $\widehat{\bm\Sigma}_{zz}^{(0)}$ and $\widehat{\bm\Sigma}_{rz}^{(0)}$ be the empirical target covariance and cross-covariance matrices of $\widehat{\mathbf r}_i^{(0)}$ and $\widehat{\mathbf z}_i^{(0)}$. With regularization parameter $\lambda_z\geq0$ and rank $r_0\leq \min (h, d_z)$, the target anchor is obtained from canonical correlation analysis (CCA) \citep{hotelling1936relations}, which solves
\begin{equation}
\label{eq:target_cca}
(\widehat{\mathbf A}^{(0)},\widehat{\mathbf B}^{(0)}) = \argmax_{\mathbf A,\mathbf B} \tr\{\mathbf A^\T\widehat{\bm\Sigma}_{rz}^{(0)}\mathbf B\},
\end{equation}
subject to $\mathbf A^\T\widehat{\bm\Sigma}_{rr}^{(0)}\mathbf A=\mathbf I_{r_0}$ and $\mathbf B^\T(\widehat{\bm\Sigma}_{zz}^{(0)}+\lambda_z\mathbf I_{d_z})\mathbf B=\mathbf I_{r_0}$. Our CCA procedure is \textit{one-sided regularized}, as the auxiliary-side ridge term $\lambda_z$ stabilizes the joint covariance when different auxiliary modalities contain overlapping, thus nearly redundant information \citep{hardoon2004canonical,tenenhaus2011regularized}. Importantly, the paired left and right canonical bases are defined only up to one common rotation. For each source domain, the observed auxiliary blocks are mapped to the target reference-side anchor by weighted ridge regression. In particular, $\big \{\widehat{\mathbf G}_k^{(m)}: k\in\Acal_m\setminus\{1\} \big\}$ are obtained by minimizing the following weighted ridge objective:
\begin{equation*}
\sum_{i\in\Lcal_m}\widehat w_i^{(m)}
\left\| \big(\widehat{\mathbf A}^{(0)} \big)^\T\widehat{\mathbf r}_i^{(m)} - \sum_{k\in\Acal_m\setminus\{1\}} \mathbf G_k^\T\widehat{\mathbf z}_{k,i}^{(m)} \right\|_2^2 + \lambda_m \left(\sum_{i\in\Lcal_m}\widehat w_i^{(m)}\right) \sum_{k\in\Acal_m\setminus\{1\}} \|\mathbf G_k\|_{\mathrm F}^2.
\end{equation*}
The ridge penalty $\lambda_m$ is invariant to orthogonal changes of auxiliary score basis and keeps the stacked normal equations stable under collinearity or rank deficiency \citep{hastie2020ridge}. In experiments, we select $\lambda_m$ from a fixed positive grid using only source or intermediate-domain validation. We denote $\widehat{\mathbf B}^{(0)} =\operatorname{col}\{\widehat{\mathbf B}_2^{(0)},\ldots, \widehat{\mathbf B}_K^{(0)}\}$.  The final aligned coordinates are
\begin{equation} \label{eq:final_embeddings}
\widehat{\mathbf u}_i^{(m)}
=(\widehat{\mathbf A}^{(0)})^\T\widehat{\mathbf r}_i^{(m)},
\qquad
\widehat{\mathbf v}_{k,i}^{(m)}
=(\widehat{\mathbf G}_k^{(m)})^\T
\widehat{\mathbf z}_{k,i}^{(m)},
\qquad
\widehat{\mathbf v}_{k,i}^{(0)}
=(\widehat{\mathbf B}_k^{(0)})^\T
\widehat{\mathbf z}_{k,i}^{(0)}.
\end{equation}
The source ridge reconstruction uses the sum of the blockwise coordinates, whereas outcome transfer retains their ordered collection. If the empirical auxiliary score basis changes as $\widehat{\mathbf z}_{k,i}^{(m)}\mapsto{\mathbf Q_{z,k}^{(m)}}^{\T}\widehat{\mathbf z}_{k,i}^{(m)}$ and the common aligned basis is rotated by $\mathbf O \in \mathcal{O}(r_0)$, then the equivalent coefficient representation is $\widehat{\mathbf G}_k^{(m)}\mapsto{\mathbf Q_{z,k}^{(m)}}^{\T}\widehat{\mathbf G}_k^{(m)}\mathbf O$, yielding a fitted coordinate rotated by $\mathbf O^\T$. Thus, the entries of $\widehat{\mathbf G}_k^{(m)}$ depend on the nuisance score basis, whereas the fitted coordinate is preserved up to the common aligned-space rotation $\mathbf O$. Under conditional transportability, the weights in
\eqref{eq:source_weight} transport the source ridge normal equations to
the target outcome mixture. The weighted ridge maps then match the
observed source auxiliary blocks to the target CCA anchor. Thus, the
core methodological contribution is the target-calibrated,
reference-anchored match-up: calibration removes label-mixture bias
before alignment, whereas match-up resolves the domain-specific
auxiliary coordinate systems.

\subsection{Outcome transfer and surrogate bridge}
\label{sec:prediction_method}
\label{sec:surrogate_method}

Density ratios are estimated from pooled labeled sources on the aligned scale in \eqref{eq:final_embeddings}.  For a generic aligned reference coordinate $\mathbf U$ and blockwise aligned auxiliary coordinate $\mathbf V_k$, we define the likelihood ratio
\begin{equation}
\label{eq:aligned_ratios}
\mathrm{LR}_{\mathrm{U,V_k\mid Y}}(\mathbf u,\mathbf v\mid\mathbf y)
=
\frac{p(\mathbf U=\mathbf u,\mathbf V_k=\mathbf v\mid\mathbf y)}
{p(\mathbf U=\mathbf u,\mathbf V_k=\mathbf v\mid\mathbf0)},
\quad
\mathrm{LR}_{\mathrm U}(\mathbf u\mid\mathbf y)
=
\frac{p(\mathbf U=\mathbf u\mid\mathbf y)}
{p(\mathbf U=\mathbf u\mid\mathbf0)}.
\end{equation}
We define $\widehat{\mathrm{LR}}_{\mathrm{V_k\mid U,Y}} (\mathbf v\mid\mathbf u,\mathbf y) =\widehat{\mathrm{LR}}_{\mathrm{U,V_k\mid Y}}(\mathbf u,\mathbf v\mid\mathbf y)/\widehat{\mathrm{LR}}_{\mathrm U}(\mathbf u\mid\mathbf y)$, where $\widehat{\mathrm{LR}}_{\mathrm{U,V_k\mid Y}}$ and $\widehat{\mathrm{LR}}_{\mathrm U}$ denote the estimators of the ratios in \eqref{eq:aligned_ratios}. For a domain with observed auxiliary set
$\Acal_m\setminus\{1\}$, the aligned likelihood-ratio evidence is
combined as
\begin{equation}
\label{eq:combined_ratio}
\widehat{\mathrm{LR}}_{\mathrm{comb}}^{(m)}
\big(
\mathbf u;
\{\mathbf v_k\}_{k\in\Acal_m\setminus\{1\}}
\mid\mathbf y
\big)
=
\widehat{\mathrm{LR}}_{\mathrm U}(\mathbf u\mid\mathbf y)
\prod_{k\in\Acal_m\setminus\{1\}}
\widehat{\mathrm{LR}}_{\mathrm{V_k\mid U,Y}}
(\mathbf v_k\mid\mathbf u,\mathbf y).
\end{equation}

The product corresponds to the joint conditional likelihood ratio when the aligned auxiliary blocks are conditionally independent given $(\mathbf U,\mathbf Y)$. More generally, it represents a composite likelihood-ratio construction based on the transferable aligned
blocks. 

The target outcome-distribution parameter is estimated by the profile criterion, the finite-state likelihood-ratio form used in maximum-likelihood label-shift estimation \citep{garg2020unified,alexandari_maximum_2020},
\begin{equation}
\label{eq:target_profile}
\widehat{\bm\theta}^{(0)}
=
\argmax_{\bm\theta}
\frac{1}{n_0}\sum_{i=1}^{n_0}
\log\left[
\sum_{\mathbf y\in\Ycal}
\rho_y(\mathbf y;\bm\theta)
\widehat{\mathrm{LR}}_{\mathrm{comb}}^{(0)}
\big(\widehat{\mathbf u}_i^{(0)}; \widehat{\mathbf v}_{2,i}^{(0)},\ldots,
\widehat{\mathbf v}_{K,i}^{(0)}\mid\mathbf y\big)
\right].
\end{equation}
The estimated target conditional outcome distribution is
\begin{equation}
\label{eq:target_posterior}
\widehat p_i(\mathbf y)
=
\frac{
\rho_y(\mathbf y;\widehat{\bm\theta}^{(0)})
\widehat{\mathrm{LR}}_{\mathrm{comb}}^{(0)}
\big(\widehat{\mathbf u}_i^{(0)}; \widehat{\mathbf v}_{2,i}^{(0)},\ldots,
\widehat{\mathbf v}_{K,i}^{(0)}\mid\mathbf y \big)
}{
\sum_{\mathbf y'\in\Ycal}
\rho_y(\mathbf y';\widehat{\bm\theta}^{(0)})
\widehat{\mathrm{LR}}_{\mathrm{comb}}^{(0)}
\big(\widehat{\mathbf u}_i^{(0)}; \widehat{\mathbf v}_{2,i}^{(0)},\ldots,
\widehat{\mathbf v}_{K,i}^{(0)}\mid\mathbf y'\big)
}.
\end{equation}
Soft predictions are conditional means under $\widehat p_i$, and hard predictions use $\argmax_{\mathbf y\in\Ycal}\widehat p_i(\mathbf y)$.

We consider the source-only surrogate regime, in which gold-standard outcomes are unavailable in the source domains, surrogate outcomes $\mathbf s_i^{(m)}$ are observed for $m=1,\ldots,M$, and $\mathbf s_i^{(0)}$ is unavailable. In addition to \eqref{eq:surrogate_independence}, assume that the conditional surrogate distribution $p_S(\mathbf s\mid\mathbf y)$ is invariant across source domains and that the resulting surrogate bridge is identifiable. For a subject-level representation $\mathbf f$, let $\widehat{\mathrm{LR}}_{\mathrm{br},S}(\mathbf f\mid\mathbf y)$ denote a surrogate-bridge estimator of $p(\mathbf f\mid\mathbf y)/p(\mathbf f\mid\mathbf 0)$ estimated from source representations and surrogates. In the binary implementation, where $Y\in\{0,1\}$ and $S\in\R$, the bridge regresses $S$ on $\mathbf f$ and combines the fitted mean with an identifiable two-component model for $S\mid Y$. For reference calibration, we denote the raw reference score by $\mathbf f$ and denote the resulting ratio by $\widehat{\mathrm{LR}}_{\mathrm{ref},S}$. Replacing $\widehat{\mathrm{LR}}_{\mathrm{ref}}$ by $\widehat{\mathrm{LR}}_{\mathrm{ref},S}$ in \eqref{eq:theta_init} defines $\widehat{\bm\theta}_{\mathrm{init},S}^{(0)}$. For match-up, we define
\[
\widehat{\mathbf f}_{i,\mathrm{pre}}^{(m)}
=
\operatorname{col}
\left\{
\widehat{\mathbf t}_{1,i}^{(m)},
\widehat{\mathbf t}_{k,i}^{(m)}
:
k\in\Acal_m\setminus\{1\}
\right\},
\]
and let $\widehat q_{i,\mathrm{pre},S}^{(m)}(\mathbf y)$ denote the normalized working weight for the outcome state $\mathbf y$ obtained from $(\widehat{\mathbf f}_{i,\mathrm{pre}}^{(m)},\mathbf s_i^{(m)})$, with $\sum_{\mathbf y\in\Ycal} \widehat q_{i,\mathrm{pre},S}^{(m)}(\mathbf y)=1$. Because the auxiliary scores are not comparable before match-up, this bridge is fitted separately within each source domain. For
$m=1,\ldots,M$, define
\begin{align}
\widehat{\bm\theta}_{S}^{(m)}
&=
\argmax_{\bm\theta}
\sum_{i\in\Lcal_m}
\sum_{\mathbf y\in\Ycal}
\widehat q_{i,\mathrm{pre},S}^{(m)}(\mathbf y)
\log\rho_y(\mathbf y;\bm\theta),
\label{eq:source_theta_surrogate}\\
\widehat w_{i,S}^{(m)}
&=
\sum_{\mathbf y\in\Ycal}
\widehat q_{i,\mathrm{pre},S}^{(m)}(\mathbf y)
\frac{
\rho_y(\mathbf y;
\widehat{\bm\theta}_{\mathrm{init},S}^{(0)})
}{
\rho_y(\mathbf y;\widehat{\bm\theta}_{S}^{(m)})
},
\qquad i\in\Lcal_m.
\label{eq:surrogate_weight}
\end{align}
The surrogate-based weight in \eqref{eq:surrogate_weight} replaces the gold-label weight in \eqref{eq:source_weight}, while the subsequent alignment steps described in Section~\ref{sec:alignment_method} remain unchanged. For each domain $m=0,\ldots,M$, we construct a common-dimensional representation for surrogate-based outcome transfer by averaging the aligned auxiliary coordinates observed in that domain:
\[
\widehat{\mathbf v}_{\mathrm{agg},i}^{(m)}
=
\frac{1}{|\Acal_m\setminus\{1\}|}
\sum_{k\in\Acal_m\setminus\{1\}}
\widehat{\mathbf v}_{k,i}^{(m)},
\qquad
\widehat{\mathbf f}_{i,\mathrm{al}}^{(m)}
=
\operatorname{col}
\left\{
\widehat{\mathbf u}_i^{(m)},
\widehat{\mathbf v}_{\mathrm{agg},i}^{(m)}
\right\}.
\]
Let
$\widehat{\mathrm{LR}}_{\mathrm{comb},S}
(\mathbf f\mid\mathbf y)$
denote the aligned surrogate-bridge ratio estimated from the pooled source pairs $\left\{
\big(
\widehat{\mathbf f}_{i,\mathrm{al}}^{(m)},
\mathbf s_i^{(m)}
\big)
:
i\in\Lcal_m,\ m=1,\ldots,M
\right\}.$ Replacing the likelihood-ratio term in \eqref{eq:target_profile} with $\widehat{\mathrm{LR}}_{\mathrm{comb},S} (\widehat{\mathbf f}_{i,\mathrm{al}}^{(0)}\mid\mathbf y)$ defines $\widehat{\bm\theta}_{S}^{(0)}$. Using $\widehat{\bm\theta}_{S}^{(0)}$ and the same ratio in \eqref{eq:target_posterior} gives the estimated target conditional outcome
distribution. Surrogate information is used only to estimate source-side weights and likelihood ratios, while the estimand remains the conditional distribution of the target gold-standard outcome $\mathbf y_i^{(0)}$.

\section{Theory}
\label{sec:theory}

\subsection{Population coordinates and identification target}
\label{sec:theory_notation}

In this section, we consider the setting where PCA \citep{jolliffe2016principal} is used for representation learning and gold-standard outcome labels are observed for all source subjects, whereas target outcomes remain unobserved during model fitting. Throughout the section, $\mathbf T$ denotes raw backend scores, $(\mathbf R,\mathbf Z)$ denote target-standardized scores, and $(\mathbf U,\mathbf V)$ denote final aligned coordinates. We use unhatted symbols to denote population quantities and the hat symbol to denote their empirical counterparts. The domain samples are mutually independent, with independent subjects within each domain. For subject $i$ in domain $m$, let $\mathbf r_i^{(m)}\in\R^h$ be the population standardized reference score and let $\mathbf z_{k,i}^{(m)}\in\R^{\ell_k}$ be the population standardized score for auxiliary block $k\in\Acal_m\setminus\{1\}$. We set $\mathbf z_i^{(m)} = \operatorname{col} \big\{\mathbf z_{k,i}^{(m)}:k\in\Acal_m\setminus\{1\} \big\}$, and use $\mathbf R^{(m)}$, $\mathbf Z_k^{(m)}$, and $\mathbf Z^{(m)}$ for the corresponding row-stacked matrices for $\mathbf r_i^{(m)}$, $\mathbf z_{k,i}^{(m)}\in\R^{\ell_k}$, and $\mathbf z_i^{(m)}$, respectively. Let $P_m$ and $E_m$ denote probability and expectation in domain $m$, and define the population label-shift ratio $\omega_m(\mathbf y) = \rho_y(\mathbf y;\bm\theta^{(0)}) / \rho_y(\mathbf y;\bm\theta^{(m)})$, where $m=1,\ldots,M$.

For a labeled source subject, we define the subject-specific weight $w_i^{(m)}=\omega_m(\mathbf y_i^{(m)})$. In the target domain, we set $w_i^{(0)}=1$. We reserve $\omega_m(\cdot)$ for the outcome-state ratio and $w_i^{(m)}$ for its evaluation at a subject's outcome. Let $(\mathbf A^{(0)},\mathbf B^{(0)})$ denote a paired population solution of the regularized target CCA problem corresponding to \eqref{eq:target_cca}. For each source domain $m$, define the population ridge alignment map as the optimizer of the penalized population objective
\begin{equation*}
\mathbf G^{(m)}
=
\argmin_{\mathbf G}
\left\{
E_m\left[
\omega_m(\mathbf y^{(m)})
\left\|
\big(\mathbf A^{(0)}\big)^{\T}\mathbf r^{(m)}
-
\mathbf G^{\T}\mathbf z^{(m)}
\right\|_2^2
\right]
+
\lambda_m\|\mathbf G\|_{\mathrm F}^2
\right\}.
\end{equation*}
Define the auxiliary-score covariance $\bm\Sigma_{zz}^{(m)}$ and the auxiliary--reference cross-moment $\bm\Sigma_{zr}^{(m)}$ under the outcome-reweighted source distribution by
\begin{equation*}
\bm\Sigma_{zz}^{(m)}
=
E_m\left[
\omega_m(\mathbf y^{(m)})
\mathbf z^{(m)}\mathbf z^{(m)\T}
\right],
\qquad
\bm\Sigma_{zr}^{(m)}
=
E_m\left[
\omega_m(\mathbf y^{(m)})
\mathbf z^{(m)}\mathbf r^{(m)\T}
\right].
\end{equation*}
Because $E_m\big[\omega_m(\mathbf y^{(m)})\big]=1$, these are moments under the outcome-reweighted source distribution. For $\lambda_m>0$, $\bm\Sigma_{zz}^{(m)}+\lambda_m\mathbf I$ is positive definite, where $\mathbf I$ is the identity matrix of dimension $\dim(\mathbf{z}^{(m)})$. Hence the penalized population optimizer is unique and satisfies
\[
\mathbf G^{(m)}
=
\left(\bm\Sigma_{zz}^{(m)}+\lambda_m\mathbf I\right)^{-1}
\bm\Sigma_{zr}^{(m)}\mathbf A^{(0)}.
\]
Thus, $\mathbf G^{(m)}$ is the population counterpart of the empirical ridge alignment map $\widehat{\mathbf G}^{(m)}$ subject to the same penalty parameter $\lambda_m$. Positive ridge regularization ensures uniqueness and stability even when the stacked auxiliary scores are collinear or rank deficient \citep{hastie2020ridge}. We then partition $\mathbf G^{(m)}=\operatorname{col}\big\{\mathbf G_k^{(m)}:k\in\Acal_m\setminus\{1\} \big\}$ and $\mathbf B^{(0)}=\operatorname{col} \big\{\mathbf B_2^{(0)},\ldots,\mathbf B_K^{(0)} \big\}$, where the population aligned coordinates are
\begin{equation*}
\label{eq:population_aligned_coordinates}
\begin{aligned}
\mathbf u_i^{(m)}
&=
\{\mathbf A^{(0)}\}^{\T}\mathbf r_i^{(m)} \in \R ^ {r_0},
&&m=0,\ldots,M,\\
\mathbf v_{k,i}^{(0)}
&=
\{\mathbf B_k^{(0)}\}^{\T}\mathbf z_{k,i}^{(0)} \in \R ^ {r_0},
&&k=2,\ldots,K,\\
\mathbf v_{k,i}^{(m)}
&=
\{\mathbf G_k^{(m)}\}^{\T}\mathbf z_{k,i}^{(m)} \in \R ^ {r_0},
&&k\in\Acal_m\setminus\{1\},\quad m\geq1.
\end{aligned}
\end{equation*}
We use $\mathbf v_i^{(m)} = \operatorname{col}\big\{\mathbf v_{k,i}^{(m)}:k\in\Acal_m\setminus\{1\} \big\}$ for the ordered collection of the observed aligned auxiliary blocks in domain $m$, and $\mathbf U^{(m)}$, $\mathbf V_k^{(m)}$, and $\mathbf V^{(m)}$ for the corresponding row-stacked reference coordinates, auxiliary blocks, and ordered auxiliary collection, respectively.

We distinguish two types of \textit{nuisance rotation} of the score bases. We use $\mathbf Q_r\in\mathcal O(h)$ to represent a change of basis for the common reference scores, and $\mathbf Q_{z,k}^{(m)}\in\mathcal O(\ell_k)$ to represent a \textit{domain-modality-specific} change of basis for auxiliary modality $k$ in domain $m$. In addition, $\mathbf O\in\mathcal O(r_0)$ represents the single residual rotational symmetry of the common aligned space. If the subject-level score vectors are re-expressed as
\begin{equation*}
    \mathbf r_i^{(m)}
\mapsto
\mathbf Q_r^{\T}\mathbf r_i^{(m)},
\qquad
\mathbf z_{k,i}^{(m)}
\mapsto
\big(\mathbf Q_{z,k}^{(m)}\big)^{\T}
\mathbf z_{k,i}^{(m)},
\end{equation*}
then the equivalent coefficient representations under the common aligned-space rotation $\mathbf O$ are
\begin{equation*}
\mathbf A^{(0)}
\mapsto
\mathbf Q_r^{\T}\mathbf A^{(0)}\mathbf O,
\qquad
\mathbf B_k^{(0)}
\mapsto
\big(\mathbf Q_{z,k}^{(0)}\big)^{\T}
\mathbf B_k^{(0)}\mathbf O,
\qquad
\mathbf G_k^{(m)}
\mapsto
\big(\mathbf Q_{z,k}^{(m)}\big)^{\T}
\mathbf G_k^{(m)}\mathbf O.
\end{equation*}
Consequently, every fitted row-stacked block transforms as $\mathbf Z_k^{(m)}\mathbf G_k^{(m)}\mapsto\mathbf Z_k^{(m)}\mathbf G_k^{(m)}\mathbf O$, and the same $\mathbf O$ is applied to each observed aligned auxiliary block. For a row-stacked matrix $\mathbf H$ whose $i$th row is $\mathbf h_i^{\T}$, we define the weighted matrix norm $\|\mathbf H\|_{\mathrm F,m}^2 = \frac{ \sum_i w_i^{(m)}\|\mathbf h_i\|_2^2 }{ \sum_i w_i^{(m)}}$. We use a single deterministic sequence $\Delta_n\to0$ for the primitive standardized-score and second-moment errors. Target CCA, source ridge, and fitted-coordinate errors are derived from this sequence rather than included in its definition. For the PCA backend, the Supplementary Material gives an explicit $\Delta_n$ obtained from the worst PCA eigenspace rate, the outcome-weight rate, and $n_{\min}^{-1/2}$ sampling variation.

\subsection{Assumptions}
\label{sec:theory_assumptions}

Assumption~\ref{ass:theory_transport} is the standard label-shift and overlap condition: outcome prevalences may differ across domains, whereas the class-conditional distribution of the retained scores remains stable; it is used to express target moments as weighted source moments \citep{lipton2018labelshift,garg2020unified}.

\begin{assumption}[Label shift and overlap]
\label{ass:theory_transport}
For every source domain $m$ and outcome state $\mathbf y$,
\begin{equation*}
\begin{split}
    &\rho_y(\mathbf y;\bm\theta^{(0)})>0
\quad\Longrightarrow\quad
\rho_y(\mathbf y;\bm\theta^{(m)})>0, \\
& (\mathbf r^{(m)},\mathbf z^{(m)})
\mid
\{\mathbf y^{(m)}=\mathbf y\}
\stackrel{d}{=}
(\mathbf r^{(0)},\mathbf z^{(m,0)})
\mid
\{\mathbf y^{(0)}=\mathbf y\},
\end{split}
\end{equation*}
where $\mathbf z^{(m,0)}$ contains the target auxiliary blocks observed in source domain $m$ and expressed in the corresponding fixed population score orientation. The empirical score bases are related to these fixed population representatives through the rotations $\mathbf Q_{z,k}^{(m)}$. Moreover, $\sup_{m,\mathbf y}\omega_m(\mathbf y)<\infty$, where the supremum is over states with positive target probability.
\end{assumption}

Let $\bm\Gamma^{(0)}
=
\{\bm\Sigma_{rr}^{(0)}\}^{-1/2}
\bm\Sigma_{rz}^{(0)}
\{\bm\Sigma_{zz}^{(0)}+\lambda_z\mathbf I\}^{-1/2}$ be the population regularized target CCA operator, and let $\widehat{\bm\Gamma}^{(0)}$ be its empirical counterpart. Assumption~\ref{ass:theory_alignment} collects high-level conditions for covariance concentration, subspace recovery, and ridge stability. Standard references provide these individual ingredients \citep{koltchinskii2017covariance,yu2015davis,wedin1972perturbation,hastie2020ridge}; the Supplementary Material proves their method-specific assembly into one common rotation.

\begin{assumption}[Stable regularized alignment]
\label{ass:theory_alignment}
The number of domains, modalities, and retained ranks is fixed. The relevant population second and fourth moments are bounded, and the covariance matrices used for standardization and regularized CCA have eigenvalues bounded away from zero on their retained subspaces. The target CCA operator satisfies $\sigma_{r_0}\{\bm\Gamma^{(0)}\}
-
\sigma_{r_0+1}\{\bm\Gamma^{(0)}\}
\geq
\delta_{\mathrm{CCA}}>0.$
The source ridge parameters satisfy $0<\underline\lambda\leq\lambda_m\leq\overline\lambda<\infty$. For suitable nuisance score rotations, the standardized-score errors, the three target CCA moment errors, and the source auxiliary covariance and auxiliary--reference cross-moment errors are uniformly $O_p(\Delta_n)$. Any tuning parameter selected from a finite candidate set satisfies these conditions uniformly.
\end{assumption}

The PCA-backend result in the Supplementary Material verifies Assumption~\ref{ass:theory_alignment} and propagates its explicit rate through CCA and ridge regression. A different representation backend may replace PCA by verifying the same primitive $O_p(\Delta_n)$ moment bounds.

Let $\mathbf 0$ denote the baseline outcome state and define the target aligned likelihood ratio
\begin{equation}
\label{eq:population_aligned_ratio}
\mathrm{LR}_{\mathrm{U,V\mid Y}}^{(0)}
(\mathbf u,\mathbf v\mid\mathbf y)
=
\frac{
p_0(\mathbf u,\mathbf v\mid\mathbf y)
}{
p_0(\mathbf u,\mathbf v\mid\mathbf 0)
}.
\end{equation}
Bayes' rule gives
\begin{equation}
\label{eq:population_target_posterior}
p_0(\mathbf y\mid\mathbf u,\mathbf v)
=
\frac{
\rho_y(\mathbf y;\bm\theta^{(0)})
\mathrm{LR}_{\mathrm{U,V\mid Y}}^{(0)}
(\mathbf u,\mathbf v\mid\mathbf y)
}{
\sum_{\mathbf y'\in\Ycal}
\rho_y(\mathbf y';\bm\theta^{(0)})
\mathrm{LR}_{\mathrm{U,V\mid Y}}^{(0)}
(\mathbf u,\mathbf v\mid\mathbf y')
}.
\end{equation}

Assumption~\ref{ass:theory_outcome} gives standard consistency and identification conditions for the aligned likelihood-ratio estimator and target profile criterion; after the common rotation is established, it yields consistency of the target outcome parameter and posterior by standard plug-in and argmax arguments \citep{sugiyama2012density,van2000asymptotic}.

\begin{assumption}[Aligned ratio consistency and profile identification]
\label{ass:theory_outcome}
For every $\mathbf y\in\Ycal$ with positive target probability, the conditional distribution of $(\mathbf u^{(0)},\mathbf v^{(0)})$ given $\mathbf y^{(0)}=\mathbf y$ is absolutely continuous with respect to its conditional distribution given $\mathbf y^{(0)}=\mathbf 0$. After expressing the estimated coordinates in the common orientation of Lemma~\ref{lem:unified_rotation_matchup}, we denote
\[
\mathrm{LR}_i(\mathbf y)
=
\mathrm{LR}_{\mathrm{U,V\mid Y}}^{(0)}
\{\mathbf u_i^{(0)},\mathbf v_i^{(0)}\mid\mathbf y\},
\]
and define the corresponding estimated ratio by
\[
\widehat{\mathrm{LR}}_i(\mathbf y)
=
\widehat{\mathrm{LR}}_{\mathrm{comb}}^{(0)}
\big(
\widehat{\mathbf u}_i^{(0)};
\widehat{\mathbf v}_{2,i}^{(0)},\ldots,
\widehat{\mathbf v}_{K,i}^{(0)}
\mid\mathbf y
\big).
\]
The ratios are nonnegative and satisfy $\mathrm{LR}_i(\mathbf 0)=\widehat{\mathrm{LR}}_i(\mathbf 0)=1$. Let $\kappa_n\to0$ denote the intrinsic likelihood-ratio estimation-and-transfer rate at the population aligned coordinates. After evaluation at the estimated coordinates, suppose
\begin{equation}
\label{eq:average_ratio_consistency}
\frac{1}{n_0}
\sum_{i=1}^{n_0}
\max_{\mathbf y\in\Ycal}
\left|
\widehat{\mathrm{LR}}_i(\mathbf y)
-
\mathrm{LR}_i(\mathbf y)
\right|
=
O_p(\kappa_n+\Delta_n),
\qquad
\frac{1}{n_0}
\sum_{i=1}^{n_0}
\max_{\mathbf y\in\Ycal}
\left\{
\mathrm{LR}_i(\mathbf y)
+
\widehat{\mathrm{LR}}_i(\mathbf y)
\right\}
=
O_p(1).
\end{equation}
The parameter space for $\bm\theta$ is compact, $\rho_y(\mathbf y;\bm\theta)$ is continuous in $\bm\theta$ uniformly over the finite outcome space $\Ycal$, and $\inf_{\bm\theta}\rho_y(\mathbf 0;\bm\theta)>0$. The oracle profile criterion obtained from \eqref{eq:target_profile} by replacing $\widehat{\mathrm{LR}}_i$ with $\mathrm{LR}_i$ converges uniformly to a deterministic population criterion whose unique maximizer is $\bm\theta^{(0)}$. For the rate conclusion below, $\bm\theta^{(0)}$ is an interior point, $\rho_y$ is twice continuously differentiable with bounded derivatives in a neighbourhood of $\bm\theta^{(0)}$, the negative population profile Hessian is positive definite there, the oracle empirical Hessian converges uniformly to its population counterpart in that neighbourhood, and the oracle empirical score at $\bm\theta^{(0)}$ is $O_p(n_0^{-1/2})$.
\end{assumption}

\subsection{Unified-rotation alignment and posterior consistency}
\label{sec:main_theory}

\begin{lemma}[Target-distribution transport]
\label{lem:target_distribution_transport}
Under Assumption~\ref{ass:theory_transport}, for every source domain $m$ and every square-integrable measurable $f$,
\begin{equation}
\label{eq:transport_identity}
E_m\left[
\omega_m\{\mathbf y^{(m)}\}
f\{\mathbf r^{(m)},\mathbf z^{(m)}\}
\right]
=
E_0\left[
f\{\mathbf r^{(0)},\mathbf z^{(m,0)}\}
\right].
\end{equation}
\end{lemma}

\begin{lemma}[One-common-rotation recovery]
\label{lem:unified_rotation_matchup}
Under Assumptions~\ref{ass:theory_transport} and \ref{ass:theory_alignment}, there exist nuisance basis rotations $\mathbf Q_r \in\mathcal O(h)$ and $\mathbf Q_{z,k}^{(m)} \in\mathcal O(\ell_k)$ and one common $\mathbf O\in\mathcal O(r_0)$ such that
\begin{equation}
\label{eq:unified_parameter_matchup}
\begin{aligned}
&
\left\|
\widehat{\mathbf A}^{(0)}
-
\mathbf Q_r^{\T}\mathbf A^{(0)}\mathbf O
\right\|_{\mathrm F}
+
\max_{2\leq k\leq K}
\left\|
\widehat{\mathbf B}_k^{(0)}
-
\{\mathbf Q_{z,k}^{(0)}\}^{\T}
\mathbf B_k^{(0)}\mathbf O
\right\|_{\mathrm F}
\\
&\qquad
+
\max_{1\leq m\leq M}
\max_{k\in\Acal_m\setminus\{1\}}
\left\|
\widehat{\mathbf G}_k^{(m)}
-
\{\mathbf Q_{z,k}^{(m)}\}^{\T}
\mathbf G_k^{(m)}\mathbf O
\right\|_{\mathrm F}
=
O_p(\Delta_n),
\end{aligned}
\end{equation}
and
\begin{equation}
\label{eq:unified_coordinate_matchup}
\max_{0\leq m\leq M}
\left\{
\left\|
\widehat{\mathbf U}^{(m)}
-
\mathbf U^{(m)}\mathbf O
\right\|_{\mathrm F,m}
+
\sum_{k\in\Acal_m\setminus\{1\}}
\left\|
\widehat{\mathbf V}_k^{(m)}
-
\mathbf V_k^{(m)}\mathbf O
\right\|_{\mathrm F,m}
\right\}
=
O_p(\Delta_n).
\end{equation}
The same $\mathbf O$ appears in the target CCA pair, every source ridge block, and every fitted aligned coordinate.
\end{lemma}

Lemma~\ref{lem:unified_rotation_matchup} establishes recovery of each fitted coordinate relative to its own population map in one common orientation. It does not assert that the target CCA map $\mathbf B^{(0)}$ equals a source ridge map $\mathbf G^{(m)}$, nor does common orientation alone imply source-to-target likelihood-ratio transport. The latter is the separate nuisance condition in Assumption~\ref{ass:theory_outcome}.

The raw coefficient matrices $\mathbf G_k^{(m)}$ depend on the nuisance score bases and are not intrinsically identifiable. In contrast, each fitted block and the ordered collection of aligned coordinates are identified up to the same common rotation. For generic subject-level aligned coordinates $\mathbf u$ and $\mathbf v_k$, let $\mathbf v=\operatorname{col}\{\mathbf v_2,\ldots,\mathbf v_K\}$. The transformation
\[
\left(
\mathbf u,
\operatorname{col}\{\mathbf v_2,\ldots,\mathbf v_K\}
\right)
\longmapsto
\left(
\mathbf O^{\T}\mathbf u,
\operatorname{col}\{\mathbf O^{\T} \mathbf v_2,\ldots,\mathbf O^{\T} \mathbf v_K\}
\right)
\]
is measurable and one-to-one, so the transformed coordinates generate the same sigma-field as $(\mathbf u,\mathbf v)$. If $p_0^{\mathbf O}$ denotes the conditional distribution expressed in the transformed coordinates, then
\begin{equation}
\label{eq:rotation_invariant_posterior}
p_0^{\mathbf O}
\left(
\mathbf y
\mid
\mathbf O^{\T}\mathbf u,
\operatorname{col}\{\mathbf O^{\T} \mathbf v_2,\ldots,\mathbf O^{\T} \mathbf v_K\}
\right)
=
p_0(\mathbf y\mid\mathbf u,\mathbf v)
\quad\text{almost surely}.
\end{equation}

\begin{theorem}[Consistency and rate of the target conditional outcome distribution]
\label{thm:outcome_transfer}
Suppose Assumptions~\ref{ass:theory_transport}, \ref{ass:theory_alignment}, and
\ref{ass:theory_outcome} hold. If the target profile criterion in
\eqref{eq:target_profile} is maximized up to $o_p(1)$, then
\[
\widehat{\bm\theta}^{(0)}\xrightarrow{p}\bm\theta^{(0)}
\]
and
\begin{equation}
\label{eq:posterior_consistency}
\frac{1}{n_0}
\sum_{i=1}^{n_0}
\sum_{\mathbf y\in\Ycal}
\left|
\widehat p_i(\mathbf y)
-
p_0\big(
\mathbf y
\mid
\mathbf u_i^{(0)},
\mathbf v_i^{(0)}
\big)
\right|
\xrightarrow{p}
0.
\end{equation}
If the local differentiability, interiority, and nonsingular-Hessian conditions in Assumption~\ref{ass:theory_outcome} hold and the fitted criterion satisfies its first-order condition, then
\begin{equation}
\label{eq:posterior_rate}
\left\|\widehat{\bm\theta}^{(0)}-\bm\theta^{(0)}\right\|_2
+
\frac{1}{n_0}
\sum_{i=1}^{n_0}
\sum_{\mathbf y\in\Ycal}
\left|
\widehat p_i(\mathbf y)
-
p_0\big(\mathbf y\mid\mathbf u_i^{(0)},\mathbf v_i^{(0)}\big)
\right|
=
O_p\big(n_0^{-1/2}+\kappa_n+\Delta_n\big).
\end{equation}
A locally Lipschitz ratio-evaluation map with an integrable envelope is one sufficient route from intrinsic rate $\kappa_n$ and the coordinate rate $\Delta_n$ to the assumed average ratio bound. Thus no additional error-rate symbol is required.
\end{theorem}

The theorem concerns the exact target likelihood ratio in \eqref{eq:population_aligned_ratio}. If the composite construction in \eqref{eq:combined_ratio} is not an exact joint ratio, the same argument gives consistency and the corresponding rate for its working conditional distribution rather than for the exact target posterior.

\subsection{Proof sketch}
\label{sec:proof_sketch}

The main nonstandard step is to show that a single rotation, rather than separate domain-specific rotations, aligns all estimated coordinates. The remaining transport identity, covariance concentration, singular-subspace perturbation, ridge perturbation, and profile-estimation arguments are standard procedures. We establish these steps, with appropriate adaptations, in the Supplementary Material.

Standard perturbation of the symmetric dilation of $\widehat{\bm\Gamma}^{(0)}$ around $\bm\Gamma^{(0)}$ yields a single $\mathbf O\in\mathcal O(r_0)$ for both the left and right target CCA bases, because the paired canonical vectors span a joint invariant subspace \citep{wedin1972perturbation,yu2015davis}. The target reference coordinate therefore carries this same $\mathbf O$ into every source ridge problem.

Orthogonal equivariance of the ridge normal equations then gives $\widehat{\mathbf G}_k^{(m)}=\{\mathbf Q_{z,k}^{(m)}\}^{\T}\mathbf G_k^{(m)}\mathbf O+O_p(\Delta_n)$ for every observed source block. Multiplication by the estimated score matrices cancels the nuisance $\mathbf Q$ rotations, so $\widehat{\mathbf U}^{(m)}=\mathbf U^{(m)}\mathbf O+O_p(\Delta_n)$ and $\widehat{\mathbf V}_k^{(m)}=\mathbf V_k^{(m)}\mathbf O+O_p(\Delta_n)$ in every domain. Thus, blockwise differences in modality availability determine which aligned blocks are present but do not change their common orientation. Assumption~\ref{ass:theory_outcome} separately supplies the intrinsic likelihood-ratio rate $\kappa_n$. A first-order expansion of the profile score then yields the parameter and posterior rate in \eqref{eq:posterior_rate}; the Supplementary Material gives the explicit PCA-to-CCA-to-ridge contribution to $\Delta_n$.

\section{Simulation studies}
\label{sec:simulation}

The simulation studies evaluate the proposed method under three main research questions (RQs). \textbf{RQ1}: Can target-distribution calibration together with target-anchored CCA and regularized source maps handle source--target label shift and blockwise modality heterogeneity? \textbf{RQ2}: When gold-standard labels are sparse, can the surrogate bridge recover most of the accuracy of a full-label procedure, and does adding a small fraction of gold-standard labels accelerate convergence to the full-label oracle? \textbf{RQ3}: Do the alignment and target-calibration components make complementary contributions, rather than one component compensating for instability introduced by the other? We therefore report a main multivariate simulation, a surrogate-assisted simulation, and an ablation study.

\subsection{Simulation study settings}
\label{sec:simulation_settings}

For the main multivariate simulation, each replication contained one unlabeled target domain and five source domains, with 400 subjects per domain. The target outcome was a three-dimensional binary vector generated from the Ising model in \eqref{eq:ising}, with marginal prevalences $(0.22,0.36,0.40)$ and pairwise interactions $(0.45,0.35,0.30)$. Source outcome distributions were perturbed away from the target distribution, while the conditional latent representation given the outcome was kept transportable. Each subject had one shared reference modality and four rank-deficient auxiliary modalities generated from an $h=5$ latent factor.

We varied one stress factor at a time: the joint label-shift magnitude $\delta$, auxiliary loading rotation $\varepsilon_{\mathrm{rot}}$, source gold-label missingness $p_{\mathrm{lab}}$, and blockwise source-modality pattern index $p_{\mathrm{mod}}$. The default values were $\delta=0.24$, $\varepsilon_{\mathrm{rot}}=0.50$, $p_{\mathrm{lab}}=0.35$, and $p_{\mathrm{mod}}=0$.  Target gold-standard outcomes were masked during fitting and used only for evaluation. We compared the proposed method with direct source transfer, orthogonal Procrustes transfer \citep{schonemann1966generalized}, RGCCA \citep{tenenhaus2011regularized}, BONMI \citep{zhou2023multisource}, target-only PCA--EM \citep{jolliffe2016principal,dempster1977maximum}, and two raw-feature learners: XGBoost with missing-block indicators \citep{chen2016xgboost}, which can capture nonlinear feature effects and interactions, and LASSO \citep{tibshirani1996regression}, which provides a sparse linear benchmark. Together, the two learners distinguish the performance available from flexible nonlinear prediction from that available under regularized linear prediction. Performance was averaged over 100 Monte Carlo replications and summarized by target macro-AUC and target mean squared error.

For the surrogate-assisted simulation, we used a binary version of the same domain-adaptation problem. The target prevalence was 0.22, the label-shift magnitude was fixed at $\delta=0.24$, and a continuous source-domain surrogate with signal level $q=0.80$ was observed. The per-domain sample size ranged from 25 to 6400. We compared the surrogate-assisted method with a reference-only surrogate learner, direct and orthogonal full-label transfer rules, the proposed full-label oracle using all source gold-standard labels, and a semi-supervised variant that additionally used 20\% of the source gold-standard labels for calibration. For \textbf{RQ3}, we performed a crossed ablation at $\varepsilon_{\mathrm{rot}}=1$, with the other perturbation parameters fixed at their defaults. The four variants crossed the target CCA/ridge alignment component with the stabilized target update.

\subsection{Main simulation}
\label{sec:simulation_main}

\begin{figure}[t]
\centering
\includegraphics[width=.82\linewidth]{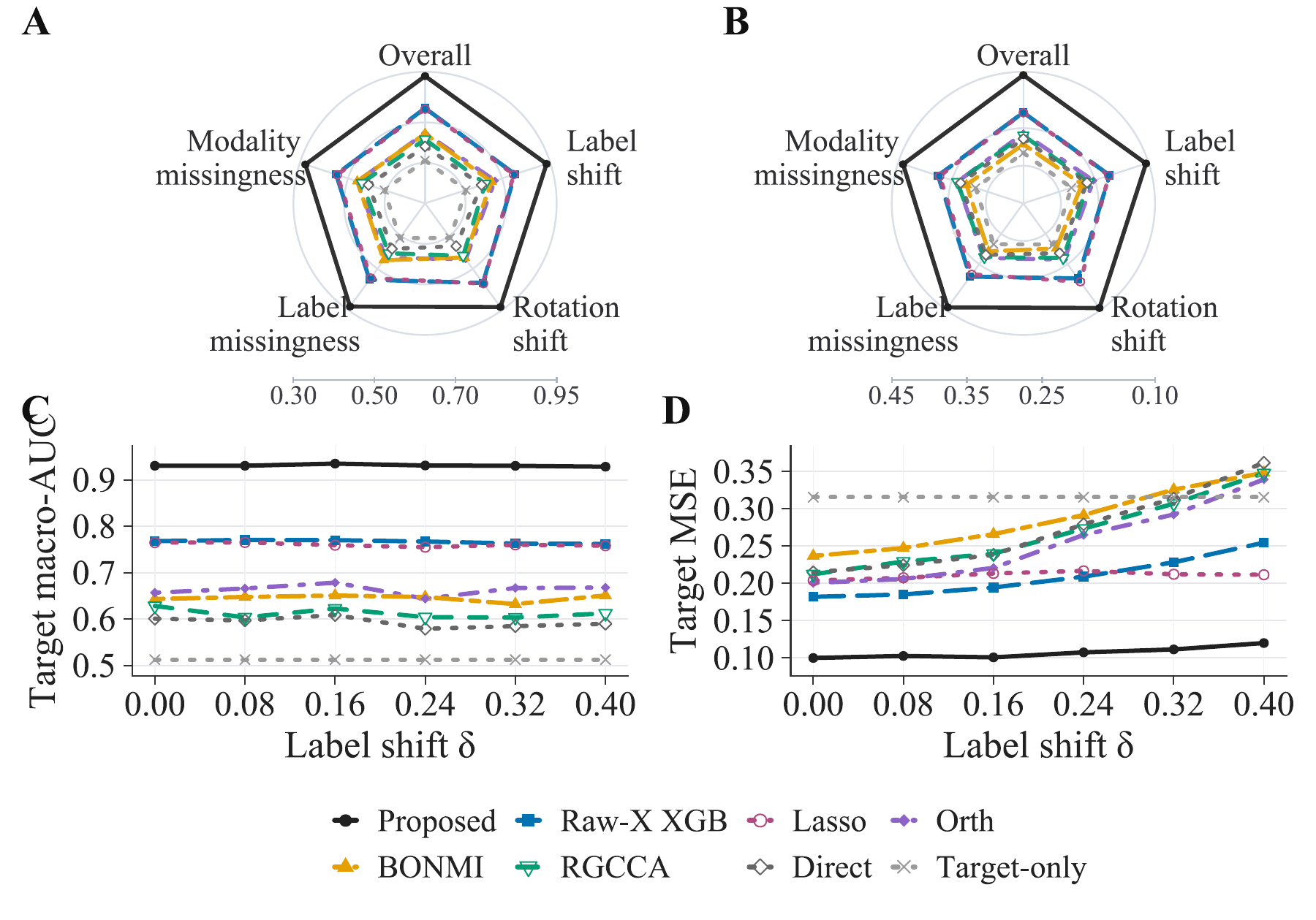}
\caption{Main simulation. Panels A and B summarize performance across the four perturbation sweeps and overall, whereas Panels C and D display the full label-shift trajectory. Panels A and C report target macro-AUC, for which higher values are preferred; Panels B and D report target MSE, for which lower values are preferred. The radial scale in Panel B is reversed so that farther from the center indicates better performance in both radar plots. Robustness to increasing label shift is therefore read jointly as a high, flat trajectory in Panel C and a low, flat trajectory in Panel D. Results are Monte Carlo means over 100 replications.}
\label{fig:sim_main}
\end{figure}

To answer \textbf{RQ1}, Figure~\ref{fig:sim_main} visualizes target macro-AUC and MSE over the four perturbation sweeps; detailed setting-level results are reported in the Supplementary Material. Averaged over all settings, the proposed method achieved macro-AUC 0.9269 and MSE 0.1003, corresponding to a 20.3\% relative increase in macro-AUC and a 51.8\% relative reduction in MSE compared with Raw-X XGBoost. Under the largest label shift, $\delta=0.40$, the corresponding improvements were 21.7\% and 59.5\%, while the proposed trajectories remained stable across rotation, source-label missingness and the nonnested modality-pattern configurations. These results answer \textbf{RQ1}: target-distribution calibration addresses source--target outcome shift, and the target-anchor CCA/ridge match-up supplies a common coordinate system for heterogeneous auxiliary blocks.

\subsection{Surrogate-assisted simulation}
\label{sec:simulation_surrogate}

\begin{figure}[t]
\centering
\includegraphics[width=.70\linewidth]{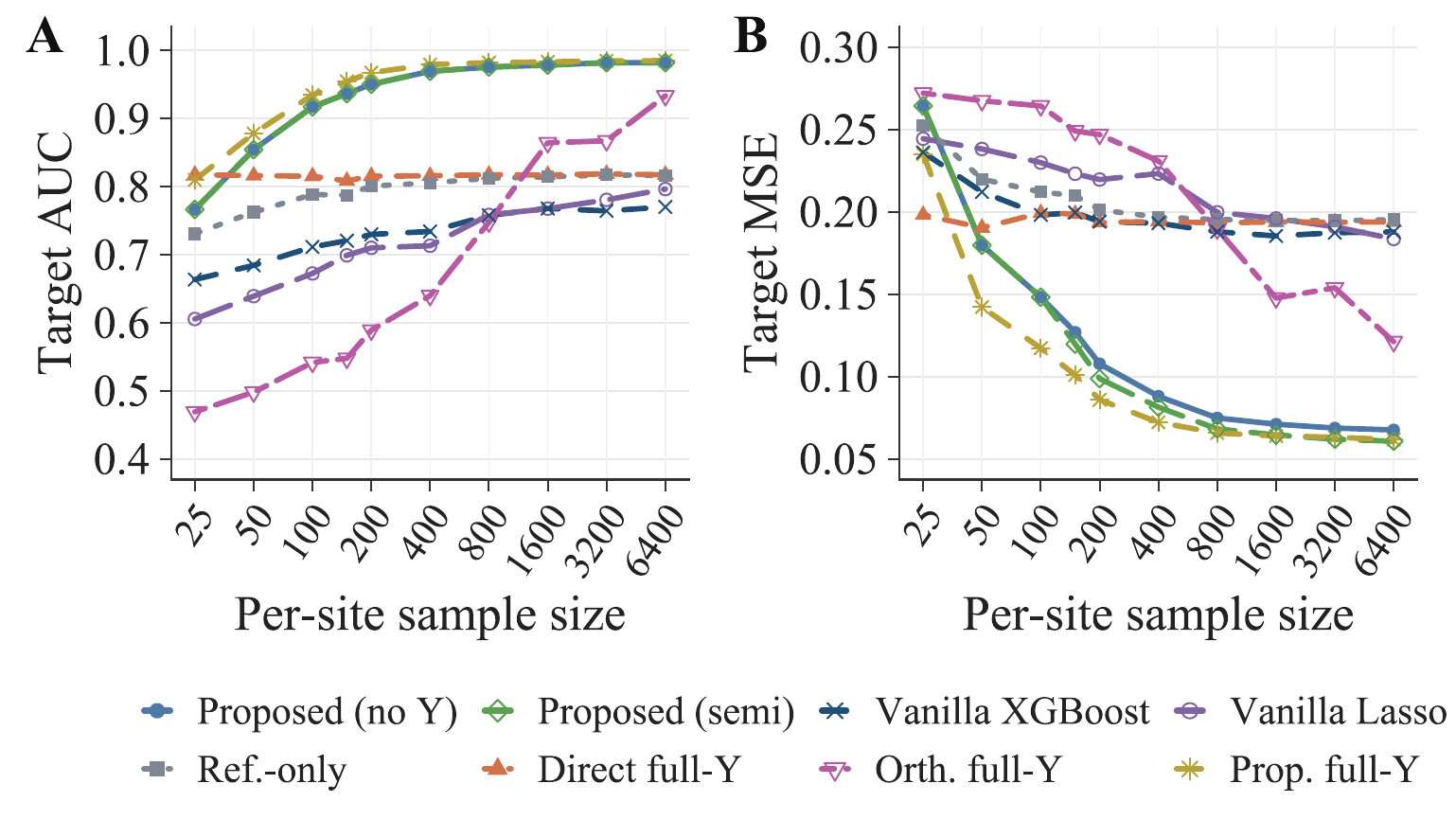}
\caption{Surrogate-assisted simulation. Panel A reports target AUC, for which higher values are preferred, and Panel B reports target MSE, for which lower values are preferred; sample size is measured per source domain. The key trajectories are \textit{Proposed (no $Y$)}, \textit{Proposed (semi)}, and \textit{Prop. full-$Y$}. The shrinking vertical gap between the first two curves and the full-label oracle as sample size increases shows how much oracle performance is recovered from surrogate information alone and from additionally using 20\% of the source gold-standard labels. Vanilla XGBoost and LASSO provide nonlinear and sparse linear benchmarks without the proposed target adaptation.}

\label{fig:sim_surrogate}
\end{figure}

To answer \textbf{RQ2}, Figure~\ref{fig:sim_surrogate} visualizes target AUC and MSE as functions of the per-domain sample size. The surrogate bridge recovered most of the full-label benchmark. The principal comparison is among the surrogate-only proposed method, the semi-supervised variant, and the proposed full-label oracle: as the sample size increased, the first two trajectories rapidly approached the oracle in both ranking and probability accuracy. Beyond the smallest-sample regime, vanilla XGBoost and LASSO improved more slowly and remained clearly separated from the proposed trajectories, showing that generic nonlinear or sparse linear fitting does not replace target-calibrated multimodal transfer. At $n=400$ per domain, the surrogate-only method increased AUC by 20.4\% and reduced MSE by 55.3\% relative to reference-only surrogate learning; relative to direct full-label transfer, the corresponding improvements were 18.8\% and 54.6\%. As the sample size increased, the surrogate-only curve approached the full-label oracle, reaching within 0.3\% of the oracle AUC at $n=6400$. The semi-supervised surrogate variant further narrowed the gap to the oracle, especially for probability accuracy: at $n=6400$, with 20\% source gold-standard calibration, its MSE was 0.061, essentially the same as the full-label oracle value of 0.062. Thus, the surrogate alone provides most of the ranking information, while a small subset of gold-standard labels calibrates the scale and lets the procedure approach the full-label oracle more rapidly. These findings answer \textbf{RQ2}: surrogate evidence is effective when it is incorporated into the same target-calibrated multimodal transfer pathway as gold-standard labels.

\subsection{Ablation of alignment and calibration}
\label{sec:simulation_ablation}

To answer \textbf{RQ3}, the crossed ablation separates the additional target-anchor CCA/ridge match-up from the stabilized target update; the preliminary outcome-reweighted alignment remains active when the CCA switch is off. Removing the target-anchor component produced the larger deterioration, decreasing macro-AUC from 0.9297 to 0.8635 and increasing MSE from 0.0983 to 0.1361. Relative to the no-update variant, the full method reduced MSE by 0.0069 but also had a small macro-AUC decrease of 0.0030. Removing both components gave the largest MSE and classification error. These trends answer \textbf{RQ3}: the target-anchor CCA/ridge maps provide the main discrimination gain under loading rotation, whereas the stabilized target update primarily refines the posterior probability scale. The two components therefore serve distinct and complementary roles.

\section{Real-data analysis}
\label{sec:real_data}

\subsection{Patient cohort and modalities}
\label{sec:real_cohort}

In this section, we evaluate the proposed method for predicting renal cell carcinoma (RCC) recurrence within 12 months after nephrectomy. The analysis assesses predictive performance under three prominent data challenges: (i) a temporal shift between the source and target cohorts, (ii) domain-specific heterogeneity in the availability of modalities, and (iii) limited availability of physician-adjudicated recurrence outcomes.

In our evaluation, the cohort includes 7,713 patients diagnosed with RCC who underwent nephrectomy in the Mass General Brigham (MGB) healthcare system between 2000 and 2022, who had a recurrence event observed within 12 months after nephrectomy, or at least 12 months of follow-up records were available after nephrectomy \citep{hou2026rcc}. The time of patient enrollment is defined as the calendar year of the initial ICD code mapped to {\tt PheCode 189.11} (\textit{malignant neoplasm of the kidney, except the renal pelvis}). Patients enrolled during 2000-2016 form the source cohort, whereas those enrolled during 2017-2022 form the target cohort. 

The primary analysis includes all 5,674 source patients, grouped according to their observed modality patterns, and 379 target patients with all modalities available. The remaining 1,660 target patients have one or more missing auxiliary modalities and are included in supplementary analyses. Among the 6,053 patients in the primary analysis, 196 have physician-adjudicated 12-month recurrence labels. For the remaining 5,857 patients, the recurrence status is represented by a surrogate label derived using LATTE (\underline{LA}bel-efficien\underline{T} inciden\underline{T} ph\underline{E}notyping) \citep{wen2024latte,hou2026rcc}. Our analysis uses physician-adjudicated recurrence labels when available and LATTE-derived surrogate labels otherwise.

We collect data from five data modalities. (i) \textit{Modality 1} contains $d_1=14$ demographic and NICE-curated clinicopathologic attributes, including TNM stage, pathology-based grade, and histologic subtype \citep{hou2026rcc}. (ii) \textit{Modality 2} contains $d_2=87$ healthcare-utilization and codified EHR count features, including {\tt PheCode}, {\tt RxNorm}, {\tt CCS/PCS}, and standardized UMLS concept unique identifiers (CUIs) \citep{gronsbell2025pehrt}. The EHR codes relevant to RCC recurrence were selected using the ONCE feature-selection procedure\citep{xiong2023knowledge}. (iii) \textit{Modality 3} contains $d_3=14$ UMLS CUIs extracted from preoperative CT radiology reports and selected using NILE \citep{yu2013nile}, focusing on anatomic characteristics of the tumor, including tumor encapsulation, lymph-node invasion, and necrosis. (iv) \textit{Modality 4} contains $d_4=27$ quantitative features obtained from raw CT images using nnU-Net for tumor segmentation \citep{isensee2021nnunet} and PyRadiomics for quantitative feature extraction \citep{vangriethuysen2017pyradiomics}. (v) \textit{Modality 5} contains $d_5=56$ residual image--text features intended to represent CT information not explained by the report-derived semantic embedding \citep{radford2021clip,bannur2023biovilt}. Among these modalities, Modality 1 serves as the reference modality, while the other modalities are auxiliary.

The patients enrolled during 2000-2016 are grouped into three source domains according to modality availability: $n_1=265$ patients with Modalities 1 and 2, $n_2=4,248$ patients with Modalities 1, 2, and 3, and $n_3=1,161$ patients with Modalities 1, 4, and 5. The $n_0 = 379$ patients enrolled during 2017-2022 with complete modalities form the target domain. Their recurrence labels for the target patients are masked throughout preprocessing, model fitting, tuning, and target adaptation and are revealed only for final evaluation. The patient cohort and modality composition are summarized in Table~\ref{tab:rcc_primary_domains}. Other real-data analysis configurations are examined in the Supplementary Material.

\begin{table}[h]
\centering
\caption{Patient cohort and modality composition in the primary MGB-RCC analysis. Events are defined using physician-adjudicated 12-month recurrence when available and the LATTE-derived surrogate otherwise.}
\label{tab:rcc_primary_domains}
\small
\setlength{\tabcolsep}{3pt}
\renewcommand{\arraystretch}{1.10}

\begin{tabular*}{\linewidth}
{@{\extracolsep{\fill}}lccrrrrr@{}}
\toprule
\textbf{Domain}
& \textbf{Cohort years}
& \textbf{Observed modalities}
& $n_m$
& $d_{\mathcal A_m}$
& \textbf{Events}
& \textbf{Gold labels}
& \textbf{Surrogate labels} \\
\midrule

\textbf{Source}, $m=1$
& 2000--2016
& $\{1,2\}$
& 265
& 101
& 56
& 5
& 260 \\

\textbf{Source}, $m=2$
& 2000--2016
& $\{1,2,3\}$
& 4,248
& 115
& 805
& 130
& 4,118 \\

\textbf{Source}, $m=3$
& 2000--2016
& $\{1,4,5\}$
& 1,161
& 97
& 172
& 43
& 1,118 \\

\midrule

\textbf{Target}, $m=0$
& 2017--2022
& $\{1,2,3,4,5\}$
& 379
& 198
& 100
& 18
& 361 \\

\midrule

\textbf{Pooled source}
& 2000--2016
& Heterogeneous
& 5,674
& --
& 1,033
& 178
& 5,496 \\

\textbf{Total}
& 2000--2022
& --
& 6,053
& --
& 1,133
& 196
& 5,857 \\

\bottomrule
\end{tabular*}
\end{table}

\subsection{Tasks, baselines, and evaluation protocol}
\label{sec:real_protocol}

\noindent\textbf{Task and Baseline Models.} The primary task is to predict 12-month RCC recurrence in the complete late-target cohort using the heterogeneous source domains summarized in Table~\ref{tab:rcc_primary_domains}. This design evaluates transport across calendar time and modality patterns under limited physician-adjudicated supervision. The target CCA rank is fixed at $r_0 = 10$ before evaluation to prevent data leakage from the target labels. We compare our method with XGBoost, a nonlinear gradient-boosted tree model \citep{chen2016xgboost}; category-specific logistic LASSO, which fits a separate sparse linear model for each modality configuration \citep{tibshirani1996regression}; RGCCA, which learns regularized multiblock components \citep{tenenhaus2011regularized}; and BONMI, which recovers a shared low-rank representation from blockwise-overlapping data \citep{zhou2023multisource}. All comparators are trained using only the data from source domains with their corresponding recurrence labels and applied directly to the target data.

\noindent\textbf{Evaluation Protocols.} We use the Brier skill score (BSS) and the absolute calibration gap as our primary metrics. The BSS quantifies improvement in probabilistic prediction relative to the constant predictor equal to the target event prevalence \citep{brier1950verification,mason2004climatology}. Such a prevalence-based metric prevents a model from being favorable solely because it predicts the majority outcome. The absolute calibration gap measures the absolute difference between the mean predicted risk and the observed target event prevalence \citep{vancalster2016hierarchy}. This metric is particularly informative under label shift because it directly evaluates whether the predicted risks recover the target event prevalence, whereas AUC measures ranking performance and may remain unchanged despite systematic miscalibration. We additionally report AUC and average precision (AP). We obtain the standard errors from 1,000 event-stratified patient-level bootstrap resamples of the fixed target predictions without model refitting, following prior work such as \citep{robin2011proc, collins2016sample}. Additional details on hyperparameter tuning, bootstrap details, and ablation studies are provided in the Supplementary Material.

\begin{table}[h]
\centering
\caption{BSS, calibration gap (Cal. gap), AUC and AP in the complete target cohort. Parentheses give standard errors from bootstrap resamples. Higher BSS, AUC and AP and lower calibration gap are preferred. The best results are shown in \textbf{bold}.}
\label{tab:rcc_main}
\small
\setlength{\tabcolsep}{4pt}
\renewcommand{\arraystretch}{1}

\begin{tabular*}{\linewidth}
{@{\extracolsep{\fill}}lcccc@{}}
\toprule
\textbf{Method}
& \textbf{BSS}
& \textbf{Cal. gap}
& \textbf{AUC}
& \textbf{AP} \\
\midrule

Proposed
& \textbf{0.4194 (0.0485)}
& \textbf{0.0015 (0.0070)}
& \textbf{0.9027 (0.0153)}
& 0.7057 (0.0465) \\

XGBoost
& 0.3457 (0.0446)
& 0.0846 (0.0100)
& 0.9008 (0.0166)
& \textbf{0.7564 (0.0363)} \\

Category-specific LASSO
& 0.3021 (0.0497)
& 0.0752 (0.0105)
& 0.8749 (0.0205)
& 0.6911 (0.0451) \\

RGCCA
& 0.1419 (0.0339)
& 0.1064 (0.0071)
& 0.8619 (0.0181)
& 0.6048 (0.0430) \\

BONMI
& -0.0347 (0.0098)
& 0.0860 (0.0021)
& 0.5329 (0.0348)
& 0.3030 (0.0296) \\

\bottomrule
\end{tabular*}
\end{table}

\subsection{Results}
\label{sec:real_results}

We present the BSS, calibration gap, AUC, and AP results for our proposed method and the baselines in Table~\ref{tab:rcc_main}. We summarize the key observations below: (i) From the perspective of overall probabilistic accuracy, being the highest in BSS indicates that our method estimates patient-level recurrence probabilities more accurately than the baseline methods despite the label imbalance in the target cohort with $26.39\%$ prevalence. (ii) From the perspective of marginal calibration, having the smallest calibration gap shows that the average predicted risk of the proposed method \textit{most closely matches} the observed target event rate under the temporal shift in recurrence prevalence from $18.21\%$ in the source cohort to $26.39\%$ in the target cohort. As we mentioned in Section \ref{sec:real_protocol}, this aspect of performance is not measured by AUC. (iii) From the perspective of cross-metric consistency, our method ranks first among BSS, calibration gap, and AUC, while remaining competitive in AP. These metrics evaluate complementary aspects of model performance, indicating that our advantage is not confined to a single evaluation criterion. These findings support the proposed method as a robust approach to 12-month RCC recurrence prediction under temporal prevalence shift, heterogeneous modality availability, and limited physician-adjudicated supervision.

\clearpage

\setcounter{section}{0}
\setcounter{equation}{0}
\setcounter{table}{0}
\setcounter{figure}{0}
\renewcommand{\thesection}{S\arabic{section}}
\makeatletter
\@addtoreset{equation}{section}
\makeatother
\renewcommand{\theequation}{\thesection.\arabic{equation}}
\renewcommand{\thetable}{S\arabic{table}}
\renewcommand{\thefigure}{S\arabic{figure}}
\renewcommand{\theHsection}{supp.\arabic{section}}
\renewcommand{\theHtable}{supp.\arabic{table}}
\renewcommand{\theHfigure}{supp.\arabic{figure}}
\renewcommand{\theHequation}{supp.\arabic{section}.\arabic{equation}}

\newcounter{suppclaim}[section]
\renewcommand{\thesuppclaim}{\thesection.\arabic{suppclaim}}
\newenvironment{supplemma}[1][]{%
  \refstepcounter{suppclaim}\par\medskip\noindent
  {\bf Lemma~\thesuppclaim\if\relax\detokenize{#1}\relax\else\ (#1)\fi.}\ \itshape
}{\par\medskip}
\newenvironment{supptheorem}[1][]{%
  \refstepcounter{suppclaim}\par\medskip\noindent
  {\bf Theorem~\thesuppclaim\if\relax\detokenize{#1}\relax\else\ (#1)\fi.}\ \itshape
}{\par\medskip}
\newenvironment{suppcorollary}[1][]{%
  \refstepcounter{suppclaim}\par\medskip\noindent
  {\bf Corollary~\thesuppclaim\if\relax\detokenize{#1}\relax\else\ (#1)\fi.}\ \itshape
}{\par\medskip}
\newenvironment{suppproof}[1][Proof]{%
  \par\smallskip\noindent{\it #1.}\ }{\hfill$\square$\par\smallskip}
\newenvironment{suppremark}[1][]{%
  \par\smallskip\noindent{\it Remark\if\relax\detokenize{#1}\relax\else\ (#1)\fi.}\ }{\par\smallskip}

\clearpage
\begin{center}
{\Large\bfseries Supplementary material for ``Multimodal domain adaptation under label shift and blockwise missing modalities''\par}
\vspace{1em}
{\normalsize Z. Wang, Z. Dou, M. Liu and T. Cai\par}
\end{center}

\begin{abstract}
This supplement contains the proofs for one-common-rotation alignment and target conditional-outcome estimation, complete implementation and numerical details for the simulation studies, and the pre-specified protocol and supporting results for the renal-cell-carcinoma recurrence analysis.  The theory uses one primitive representation rate and propagates it explicitly through PCA, target CCA, source ridge maps and fitted coordinates.  Likelihood-ratio estimation and transfer remain a separate nuisance component; under local profile curvature, the two components yield an end-to-end posterior rate.  The simulation appendix documents the main multivariate, surrogate-assisted and crossed-ablation experiments, including the raw-feature and vanilla XGBoost and LASSO benchmarks and the numerical summaries underlying Figures~2 and~3 of the main paper.  The real-data appendix documents cohort and label composition, leakage controls, the primary complete-target analysis, additional late-target configurations, the complete-domain matched-modality sensitivity analysis, and the crossed CCA--target-EM ablation.

\end{abstract}

\begin{keywords}
Ablation study; Domain adaptation; Multimodal data fusion; Renal cell carcinoma; Supplementary proof; Surrogate outcomes.
\end{keywords}

\section{Proofs for one-common-rotation alignment and target conditional-outcome consistency}
\label{sec:supp_proof}

\subsection{Proof roadmap and population notation}
\label{subsec:supp_notation}

The proof has two branches.  Its core representation result verifies one primitive rate
$\Delta_n$, propagates it through target CCA and the source ridge maps, and recovers all fitted
coordinates under one common rotation.  The outcome branch combines this coordinate error with the
intrinsic rate $\kappa_n$ of the final aligned likelihood-ratio learner.  The dependency order is
\begin{equation*}
\begin{aligned}
\text{PCA + weights + sampling}
&\Longrightarrow \Delta_n
\Longrightarrow \delta_{\mathrm{CCA}}^{-1}\Delta_n\ \text{(CCA)}\\
&\Longrightarrow
\{\underline\lambda^{-1}(1+\delta_{\mathrm{CCA}}^{-1})+
\underline\lambda^{-2}\}\Delta_n\ \text{(ridge)}\\
&\Longrightarrow \text{common-orientation fitted coordinates},\\[2pt]
\text{coordinate error}+\kappa_n
&\Longrightarrow \text{evaluated ratio error}
\Longrightarrow \widehat{\bm\theta}^{(0)}-\bm\theta^{(0)}
\Longrightarrow \widehat p-p_0.
\end{aligned}
\end{equation*}
At the high-level theorem, $\Delta_n$ is one envelope for the pre-CCA score and moment events.
For the PCA backend it is assembled from only four controllable inputs: outcome-weight error,
reference-PCA error, auxiliary-PCA error and ordinary sampling fluctuation.  Initial reference
calibration enters only through the outcome-weight error; plug-in standardization and empirical
moments are consequences of those four inputs and are not added again.  The subsequent lemmas
propagate the same envelope through CCA, ridge and the fitted coordinates.  In contrast,
$\kappa_n$ concerns the final likelihood-ratio learner when population aligned coordinates are
available; stability to replacing them by fitted coordinates consumes $\Delta_n$ once.  No
independence or relative ordering is assumed.
The representation statements do not imply likelihood-ratio transport.  In particular, common
orientation does not make the target CCA map equal to a source ridge map.  The separate ratio
condition is therefore retained in Assumption~\ref{ass:supp_outcome}.  This separation also lets
another representation backend replace PCA by verifying the same primitive $O_p(\Delta_n)$
score and moment events.

The target domain is indexed by $0$ and sources by $m=1,\ldots,M$.  Write $n_m$ for a
domain sample size, $n_{\min}=\min_{0\leq m\leq M}n_m$ for the smallest domain sample
size, $N_s=\sum_{m=1}^M n_m$ for the total source sample size, and
$N=n_0+N_s=\sum_{m=0}^M n_m$ for the total sample size.  The outcome takes values in finite $\Ycal$.  Before
target-distribution standardization, let
$\mathbf t_1^{(m)}\in\R^h$ be the population raw reference score and
$\mathbf t_k^{(m)}\in\R^{\ell_k}$ the raw auxiliary score for
$k\in\Acal_m\setminus\{1\}$, and set
$\mathbf t^{(m)}=\operatorname{col}\{\mathbf t_k^{(m)}:
k\in\Acal_m\setminus\{1\}\}$.  Their empirical row-stacked matrices are
$\widehat{\mathbf T}_1^{(m)}$ and $\widehat{\mathbf T}_k^{(m)}$, with population
counterparts $\mathbf T_1^{(m)}$ and $\mathbf T_k^{(m)}$.

Set $\omega_0(\mathbf y)\equiv1$ and, for every domain, define the target-mixture expectation
\begin{equation}
E_m^{\mathrm{tar}}\{a\}
=E_m\left[\omega_m(\mathbf y^{(m)})a\right]
=\sum_{\mathbf y\in\Ycal}\rho_y(\mathbf y;\bm\theta^{(0)})
E_m(a\mid\mathbf y^{(m)}=\mathbf y)
\label{eq:supp_reweighted_expectation}
\end{equation}
for $m\geq1$, and let $E_0^{\mathrm{tar}}=E_0$. In particular, $E_m\{\omega_m(\mathbf y^{(m)})\}=1$. For every available
score block, define target-distribution means and covariances
\begin{align*}
\bm\mu_{1,m}&=E_m^{\mathrm{tar}}(\mathbf t_1^{(m)}),&
\mathbf S_{1,m}&=E_m^{\mathrm{tar}}\{(\mathbf t_1^{(m)}-\bm\mu_{1,m})
(\mathbf t_1^{(m)}-\bm\mu_{1,m})^\T\},\\
\bm\mu_{k,m}&=E_m^{\mathrm{tar}}(\mathbf t_k^{(m)}),&
\mathbf S_{k,m}&=E_m^{\mathrm{tar}}\{(\mathbf t_k^{(m)}-\bm\mu_{k,m})
(\mathbf t_k^{(m)}-\bm\mu_{k,m})^\T\}.
\end{align*}
The population standardized scores are
\begin{equation*}
\mathbf r^{(m)}=\mathbf S_{1,m}^{-1/2}(\mathbf t_1^{(m)}-\bm\mu_{1,m}),
\qquad
\mathbf z_k^{(m)}=\mathbf S_{k,m}^{-1/2}(\mathbf t_k^{(m)}-\bm\mu_{k,m}),
\end{equation*}
and
$\mathbf z^{(m)}=\operatorname{col}\{\mathbf z_k^{(m)}:
k\in\Acal_m\setminus\{1\}\}$.  Their row-stacked matrices are
$\mathbf R^{(m)}$ and $\mathbf Z^{(m)}$; hats denote the plug-in standardized matrices.

For the target, write
\begin{align*}
\bm\Sigma_{rr}^{(0)}&=E_0(\mathbf r^{(0)}\mathbf r^{(0)\T}),&
\bm\Sigma_{zz}^{(0)}&=E_0(\mathbf z^{(0)}\mathbf z^{(0)\T}),&
\bm\Sigma_{rz}^{(0)}&=E_0(\mathbf r^{(0)}\mathbf z^{(0)\T}).
\end{align*}
The population regularized CCA operator is
\begin{equation}
\bm\Gamma^{(0)}
=\{\bm\Sigma_{rr}^{(0)}\}^{-1/2}\bm\Sigma_{rz}^{(0)}
(\bm\Sigma_{zz}^{(0)}+\lambda_z\mathbf I)^{-1/2}
=\mathbf P_{\Gamma}^{(0)}\mathbf D_{\Gamma}^{(0)}\{\mathbf Q_{\Gamma}^{(0)}\}^\T,
\label{eq:supp_population_cca}
\end{equation}
where $\mathbf P_{\Gamma}^{(0)}$ and $\mathbf Q_{\Gamma}^{(0)}$ contain the first $r_0$ left
and right singular vectors.  The empirical operator $\widehat{\bm\Gamma}^{(0)}$ is obtained by replacing the three population covariance matrices with their empirical counterparts.  Define
\begin{equation*}
\mathbf A^{(0)}=\{\bm\Sigma_{rr}^{(0)}\}^{-1/2}\mathbf P_{\Gamma}^{(0)},
\qquad
\mathbf B^{(0)}=(\bm\Sigma_{zz}^{(0)}+\lambda_z\mathbf I)^{-1/2}
\mathbf Q_{\Gamma}^{(0)}.
\end{equation*}
Partition $\mathbf B^{(0)}=\operatorname{col}\{\mathbf B_2^{(0)},\ldots,\mathbf B_K^{(0)}\}$.
For source $m$, define the regularized population alignment map by the same penalized criterion used in the main paper,
\begin{equation}
\mathbf G^{(m)}
=\argmin_{\mathbf G}
\left\{
E_m\left[
\omega_m(\mathbf y^{(m)})
\left\|
\{\mathbf A^{(0)}\}^{\T}\mathbf r^{(m)}
-\mathbf G^{\T}\mathbf z^{(m)}
\right\|_2^2
\right]
+\lambda_m\|\mathbf G\|_{\mathrm F}^2
\right\}.
\label{eq:supp_population_ridge}
\end{equation}
Let $d_{z,m}=\dim(\mathbf z^{(m)})=\sum_{k\in\Acal_m\setminus\{1\}}\ell_k$, and define
\begin{equation}
\bm\Sigma_{zz}^{(m)}
=E_m^{\mathrm{tar}}\{\mathbf z^{(m)}\mathbf z^{(m)\T}\},
\qquad
\bm\Sigma_{zr}^{(m)}
=E_m^{\mathrm{tar}}\{\mathbf z^{(m)}\mathbf r^{(m)\T}\}.
\label{eq:supp_ridge_moments}
\end{equation}
The standardized scores have zero mean under $E_m^{\mathrm{tar}}$, so these are the auxiliary-score covariance and auxiliary--reference cross-moment under the outcome-reweighted source distribution.  If $\lambda_m>0$, then $\bm\Sigma_{zz}^{(m)}+\lambda_m\mathbf I_{d_{z,m}}$ is positive definite and the unique optimizer of \eqref{eq:supp_population_ridge} is
\begin{equation}
\mathbf G^{(m)}
=\left\{\bm\Sigma_{zz}^{(m)}+\lambda_m\mathbf I_{d_{z,m}}\right\}^{-1}
\bm\Sigma_{zr}^{(m)}\mathbf A^{(0)}.
\label{eq:supp_population_ridge_closed}
\end{equation}
Partition $\mathbf G^{(m)}=\operatorname{col}\{\mathbf G_k^{(m)}:
k\in\Acal_m\setminus\{1\}\}$.

The aligned subject-level coordinates are
\begin{equation*}
\begin{aligned}
\mathbf u_i^{(m)}&=\{\mathbf A^{(0)}\}^{\T}\mathbf r_i^{(m)}\in\R^{r_0},&&0\leq m\leq M,\\
\mathbf v_{k,i}^{(0)}&=\{\mathbf B_k^{(0)}\}^{\T}\mathbf z_{k,i}^{(0)}\in\R^{r_0},&&k=2,\ldots,K,\\
\mathbf v_{k,i}^{(m)}&=\{\mathbf G_k^{(m)}\}^{\T}\mathbf z_{k,i}^{(m)}\in\R^{r_0},&&k\in\Acal_m\setminus\{1\},\quad m\geq1.
\end{aligned}
\end{equation*}
For each domain, let $\mathbf v_i^{(m)}=\operatorname{col}\{\mathbf v_{k,i}^{(m)}:k\in\Acal_m\setminus\{1\}\}$ in increasing modality order.  We use $\mathbf U^{(m)}$, $\mathbf V_k^{(m)}$, and $\mathbf V^{(m)}$ for the row-stacked matrices whose $i$th rows are $\mathbf u_i^{(m)\T}$, $\mathbf v_{k,i}^{(m)\T}$, and $\mathbf v_i^{(m)\T}$, respectively.
Set $q_0=K-1$.  For each source, write the observed auxiliary indices in increasing order as
$k_{m,1}<\cdots<k_{m,q_m}$, where $q_m=|\Acal_m\setminus\{1\}|$.
Their empirical row-stacked counterparts, used later in the perturbation bounds, are defined
explicitly by
\begin{equation}
\begin{aligned}
\widehat{\mathbf U}^{(m)}
&=\widehat{\mathbf R}^{(m)}\widehat{\mathbf A}^{(0)},\qquad 0\leq m\leq M,\\
\widehat{\mathbf V}_k^{(0)}
&=\widehat{\mathbf Z}_k^{(0)}\widehat{\mathbf B}_k^{(0)},\qquad k=2,\ldots,K,\\
\widehat{\mathbf V}_k^{(m)}
&=\widehat{\mathbf Z}_k^{(m)}\widehat{\mathbf G}_k^{(m)},\qquad k\in\Acal_m\setminus\{1\},\quad m\geq1,\\
\widehat{\mathbf V}^{(0)}
&=\big[\widehat{\mathbf Z}_2^{(0)}\widehat{\mathbf B}_2^{(0)}
\ \cdots\ 
\widehat{\mathbf Z}_K^{(0)}\widehat{\mathbf B}_K^{(0)}\big],\\
\widehat{\mathbf V}^{(m)}
&=\big[\widehat{\mathbf Z}_{k_{m,1}}^{(m)}\widehat{\mathbf G}_{k_{m,1}}^{(m)}
\ \cdots\ 
\widehat{\mathbf Z}_{k_{m,q_m}}^{(m)}\widehat{\mathbf G}_{k_{m,q_m}}^{(m)}\big],
\qquad m\geq1.
\end{aligned}
\label{eq:supp_empirical_aligned_coordinates}
\end{equation}
The horizontal concatenation in the last two rows follows increasing modality order.  Thus each subject-level $\mathbf u_i^{(m)}$ and $\mathbf v_{k,i}^{(m)}$ has dimension $r_0$, while $\mathbf U^{(m)}$ and $\mathbf V_k^{(m)}$ are $n_m\times r_0$ row-stacked matrices and $\mathbf V^{(m)}$ is an $n_m\times q_mr_0$ ordered block matrix;
the source ridge reconstruction itself is the sum of the $q_m$ fitted blocks.

Let $\mathbf Q_r\in\mathcal O(h)$ and
$\mathbf Q_z^{(m)}=\operatorname{blockdiag}\{\mathbf Q_{z,k}^{(m)}:
k\in\Acal_m\setminus\{1\}\}$ denote pre-alignment score-basis rotations.  The common
aligned-space rotation is $\mathbf O\in\mathcal O(r_0)$.  For a row-stacked matrix $\mathbf H$, write
\begin{equation*}
\|\mathbf H\|_{\F,m}^2
=\frac{\sum_{i\in\Lcal_m}w_i^{(m)}\|\mathbf h_i\|_2^2}
{\sum_{j\in\Lcal_m}w_j^{(m)}}.
\end{equation*}
Here $w_i^{(m)}=\omega_m(\mathbf y_i^{(m)})$ for a labeled source subject, while $w_i^{(0)}=1$ and $\Lcal_0=\{1,\ldots,n_0\}$.  Let
\begin{equation*}
\mathbf W_m=\frac{\operatorname{diag}\{w_i^{(m)}:i\in\Lcal_m\}}
{\sum_{j\in\Lcal_m}w_j^{(m)}},
\qquad
\mathbf W_0=n_0^{-1}\mathbf I,
\end{equation*}
and define $\widehat{\mathbf W}_m$ analogously from $\widehat w_i^{(m)}$.  The empirical ridge
moments are
\begin{equation*}
\widehat{\bm\Sigma}_{zz}^{(m)}
=\widehat{\mathbf Z}^{(m)\T}\widehat{\mathbf W}_m
\widehat{\mathbf Z}^{(m)},
\qquad
\widehat{\bm\Sigma}_{zr}^{(m)}
=\widehat{\mathbf Z}^{(m)\T}\widehat{\mathbf W}_m
\widehat{\mathbf R}^{(m)}.
\end{equation*}
Dividing the empirical ridge objective in the main paper by $\sum_{i\in\Lcal_m}\widehat w_i^{(m)}$ does not change its minimizer.  The divided objective has the normalized moment matrices above and penalty $\lambda_m$, so its normal equations are
\[
\widehat{\mathbf G}^{(m)}
=
\{\widehat{\bm\Sigma}_{zz}^{(m)}+\lambda_m\mathbf I_{d_{z,m}}\}^{-1}
\widehat{\bm\Sigma}_{zr}^{(m)}\widehat{\mathbf A}^{(0)}.
\]
Thus the empirical and population maps have exactly the same penalty scale.  A single sequence $\Delta_n$ will control the initial weight, standardized-score and second-moment errors; CCA, ridge, and fitted-coordinate errors are derived from it.

\medskip\noindent\textbf{Notation at a glance.}\quad
Bold lower-case symbols denote subject-level column vectors and bold upper-case symbols denote
row-stacked matrices; a hat always denotes an estimator.  The operator $\operatorname{col}$ stacks
subject-level blocks vertically, whereas square brackets in
\eqref{eq:supp_empirical_aligned_coordinates} concatenate row-matrix blocks horizontally.
\begin{center}
\small
\begin{tabular}{@{}p{0.28\linewidth}p{0.65\linewidth}@{}}
\toprule
Symbol & Role \\
\midrule
$\mathbf T_1^{(m)},\mathbf T_k^{(m)}$ & Raw backend score matrices before target-distribution standardization. \\
$\mathbf R^{(m)},\mathbf Z^{(m)}$ & Target-distribution standardized, pre-alignment reference and stacked auxiliary scores. \\
$\mathbf A^{(0)},\mathbf B^{(0)},\mathbf G^{(m)}$ & Population target CCA maps and the regularized source alignment map; $\mathbf G_k^{(m)}$ is its block for modality $k$. \\
$\bm\Sigma_{zz}^{(m)},\bm\Sigma_{zr}^{(m)}$ & Outcome-reweighted auxiliary-score covariance and auxiliary--reference cross-moment defining the penalized population ridge target. \\
$\mathbf u_i^{(m)},\mathbf v_{k,i}^{(m)},\mathbf v_i^{(m)}$ & Subject-level aligned reference vector, aligned auxiliary-block vector, and ordered auxiliary collection. \\
$\mathbf U^{(m)},\mathbf V_k^{(m)},\mathbf V^{(m)}$ & Corresponding row-stacked reference matrix, auxiliary-block matrix, and ordered auxiliary matrix. \\
$\mathbf Q_r,\mathbf Q_{z,k}^{(m)}$ & Arbitrary rotations of the raw score bases induced by PCA or another representation backend. \\
$\mathbf O$ & One common aligned-space rotation. \\
$\omega_m(\mathbf y),w_i^{(m)}$ & Outcome-state ratio and its subject-level evaluation; hats denote their estimators. \\
$n_m,n_{\min},N_s,N$ & Domain size, smallest domain size, total source size and total size, respectively. \\
$\LR_{\mathrm{U,V\mid Y}}^{(0)}(\mathbf u,\mathbf v\mid\mathbf y),\widehat{\LR}_i(\mathbf y)$ & Population target aligned likelihood ratio and its fitted-coordinate estimate; $\widehat{\LR}_i^{\circ}$ denotes the counterfactual population-coordinate output. \\
$\Delta_n$ & Single envelope for the pre-CCA score and moment events; for PCA its four inputs are weight, reference-PCA, auxiliary-PCA and sampling errors in \eqref{eq:supp_delta_rate}. \\
$\kappa_n$ & Intrinsic rate of the final aligned likelihood-ratio learner at population coordinates; perturbing those coordinates is excluded. \\
\bottomrule
\end{tabular}
\end{center}

\medskip\noindent\textbf{Rate provenance and accounting.}\quad
The following ledger distinguishes externally supplied inequalities, method-specific deductions and
learner assumptions.  It introduces no further rate sequence.
\begin{enumerate}[leftmargin=1.7em,itemsep=1pt,topsep=2pt]
\small
\item \emph{Inputs to $\Delta_n$.}  These are the maximum outcome-weight error, reference-PCA
error, auxiliary-PCA error and $n_{\min}^{-1/2}$ sampling fluctuation.  Initial reference
calibration affects alignment only through the weight error.  \citet[Theorem~3]{garg2020unified}
bounds a likelihood-based label-shift estimate by a root-target-sample term plus calibration error;
when the latter is also root-order, a smooth transformation gives a root-order weight error.
Covariance concentration is supplied by
\citet[Theorem~4.7.1 and Exercise~4.7.3]{vershynin2018hdp}, and the population-gap
Procrustes factor by \citet[Theorem~2]{yu2015davis}.
\item \emph{Verified from $\Delta_n$.}  Lemma~\ref{lem:supp_pca_raw} derives standardized-score
recovery and the five moment events from those inputs, fixed retained dimensions and bounded
overlap.  They are not additional summands.
\item \emph{Derived alignment errors.}  Lemma~\ref{lem:supp_paired_cca} obtains
$O_p\{(1+\delta_{\mathrm{CCA}}^{-1})\Delta_n\}$ by applying the cited subspace theorem to one symmetric
dilation; the common rotation is the deduction made here.  Lemma~\ref{lem:supp_ridge} then uses
normal equations, a resolvent identity and product decompositions to control ridge maps and fitted
coordinates without another rate input.
\item \emph{Learner assumptions.}  Assumption~\ref{ass:supp_outcome} assigns $\kappa_n$ to the
population-coordinate likelihood-ratio learner and $\Delta_n$ to stability of the complete
fit-and-evaluation map under coordinate perturbation.  Regular parametric $M$-estimation gives a
smooth finite-dimensional route \citep[Chap.~5]{van2000asymptotic}.  For boosted trees, both
clauses remain assumed: \citet{chen2016xgboost} specifies the algorithm, not a statistical rate.
\item \emph{Derived target rate.}  Theorem~\ref{thm:supp_conditional} adds the oracle target-score
term $n_0^{-1/2}$ and obtains $O_p\{n_0^{-1/2}+\kappa_n+\Delta_n\}$.
\end{enumerate}
Thus $\Delta_n$ is a common upper envelope, not a sum of its inputs, and neither $\Delta_n$ nor
$\kappa_n$ is charged twice later in the proof.

Throughout this proof section, all source gold-standard outcomes are observed, so
$\Lcal_m=\{1,\ldots,n_m\}$ for $m\geq1$; $\Lcal_m$ denotes these subjects when forming the
alignment moments.  The optional surrogate bridge is not covered by the rate theorem below.

\subsection{Regularity conditions}
\label{subsec:supp_assumptions}

\begin{assumption}[Label shift and overlap]
\label{ass:supp_transport}
For every source $m$ and outcome state $\mathbf y$ with positive target probability,
there exist fixed orthonormal bases of the retained population score subspaces for which
\begin{equation}
(\mathbf t_1^{(m)},\mathbf t^{(m)})\mid\{\mathbf y^{(m)}=\mathbf y\}
\stackrel d=
(\mathbf t_1^{(0)},\mathbf t^{(m,0)})\mid\{\mathbf y^{(0)}=\mathbf y\},
\label{eq:supp_transportability}
\end{equation}
and
$\omega_m(\mathbf y)=\rho_y(\mathbf y;\bm\theta^{(0)})/
\rho_y(\mathbf y;\bm\theta^{(m)})$ is uniformly bounded.  Here
$\mathbf t^{(m,0)}$ denotes the target auxiliary latent-score representative restricted to the
blocks observed in source $m$ and expressed in the corresponding fixed population
orientation.  This equality is a population transport assumption, not a claim that numerical
sample PCA coordinates coincide across domains.  The sample bases are matched to these
representatives by the rotations in Lemma~\ref{lem:supp_pca_raw}.  By
Lemma~\ref{lem:supp_transport}, the transported source means and covariances equal the
target-mixture means and covariances, so the same equality holds after the deterministic
population standardization defining $(\mathbf r,\mathbf z)$.
\end{assumption}

\begin{assumption}[Stable score, CCA and ridge moments]
\label{ass:supp_alignment}
The number of domains, modality blocks and retained ranks is fixed.  Relevant population
second and fourth moments are bounded.  The raw-score covariance eigenvalues retained for
standardization are bounded below by $c_0>0$.  Also
$\lambda_{\min}(\bm\Sigma_{rr}^{(0)})\geq c_0$ and
$\lambda_{\min}(\bm\Sigma_{zz}^{(0)}+\lambda_z\mathbf I)\geq c_0$.
The singular-value gap of \eqref{eq:supp_population_cca} satisfies
$\sigma_{r_0}(\bm\Gamma^{(0)})-\sigma_{r_0+1}(\bm\Gamma^{(0)})\geq\delta_{\mathrm{CCA}}>0$.
CCA and ridge tuning parameters are fixed, or selected from finite grids over which the
bounds hold uniformly, and
$0<\underline\lambda\leq\lambda_m\leq\overline\lambda<\infty$.

For a deterministic sequence $\Delta_n\to0$ and suitable score rotations, the following primitive errors are uniformly $O_p(\Delta_n)$:
\begin{align}
\max_m\|\widehat{\mathbf R}^{(m)}-\mathbf R^{(m)}\mathbf Q_r\|_{\F,m}
+\max_m\|\widehat{\mathbf Z}^{(m)}-\mathbf Z^{(m)}\mathbf Q_z^{(m)}\|_{\F,m}
&=O_p(\Delta_n),
\label{eq:supp_score_event}\\
\|\widehat{\bm\Sigma}_{rr}^{(0)}-\mathbf Q_r^\T\bm\Sigma_{rr}^{(0)}\mathbf Q_r\|_{\op}
+\|\widehat{\bm\Sigma}_{zz}^{(0)}-\mathbf Q_z^{(0)\T}\bm\Sigma_{zz}^{(0)}\mathbf Q_z^{(0)}\|_{\op}
&=O_p(\Delta_n),
\label{eq:supp_target_cov_event}\\
\|\widehat{\bm\Sigma}_{rz}^{(0)}-\mathbf Q_r^\T\bm\Sigma_{rz}^{(0)}\mathbf Q_z^{(0)}\|_{\op}
&=O_p(\Delta_n).
\label{eq:supp_target_cross_event}
\end{align}
For the source ridge moments,
\begin{align}
\max_m\|\widehat{\bm\Sigma}_{zz}^{(m)}-
\mathbf Q_z^{(m)\T}\bm\Sigma_{zz}^{(m)}\mathbf Q_z^{(m)}\|_{\op}
&=O_p(\Delta_n),
\label{eq:supp_source_cov_event}\\
\max_m\|\widehat{\bm\Sigma}_{zr}^{(m)}-
\mathbf Q_z^{(m)\T}\bm\Sigma_{zr}^{(m)}\mathbf Q_r\|_{\op}
&=O_p(\Delta_n).
\label{eq:supp_source_zr_event}
\end{align}
The target CCA, source ridge, and fitted-coordinate errors are not part of $\Delta_n$; they are conclusions derived below.
\end{assumption}

\begin{assumption}[Target ratio and profile criterion]
\label{ass:supp_outcome}
Let
\begin{equation*}
\LR_{\mathrm{U,V\mid Y}}^{(0)}(\mathbf u,\mathbf v\mid\mathbf y)
=\frac{p_0(\mathbf u,\mathbf v\mid\mathbf y)}
{p_0(\mathbf u,\mathbf v\mid\mathbf0)}.
\end{equation*}
After expressing estimated coordinates in the common population orientation, write
$\LR_i(\mathbf y)=\LR_{\mathrm{U,V\mid Y}}^{(0)}\{\mathbf u_i^{(0)},\mathbf v_i^{(0)}\mid\mathbf y\}$ and let
$\widehat{\LR}_i(\mathbf y)$ denote the estimated ratio at the estimated target coordinates.
To expose the two sources of ratio error without introducing another rate, let
$\widehat{\LR}_i^{\circ}(\mathbf y)$ denote the counterfactual output of the same ratio-learning
procedure when both its source training inputs and target evaluation inputs are the population
aligned coordinates.  All three ratios are nonnegative and normalized to one at $\mathbf0$.
Let the deterministic sequence $\kappa_n\to0$ denote the intrinsic estimation-and-transfer rate
of this population-coordinate learner.  It excludes the outcome-weight error already absorbed by
$\Delta_n$ and excludes all coordinate estimation.  Suppose
\begin{align}
\frac1{n_0}\sum_{i=1}^{n_0}\max_{\mathbf y\in\Ycal}
|\widehat{\LR}_i^{\circ}(\mathbf y)-\LR_i(\mathbf y)|
&=O_p(\kappa_n),
\label{eq:supp_oracle_coordinate_ratio_rate}\\
\frac1{n_0}\sum_{i=1}^{n_0}\max_{\mathbf y\in\Ycal}
|\widehat{\LR}_i(\mathbf y)-\widehat{\LR}_i^{\circ}(\mathbf y)|
&=O_p(\Delta_n),
\label{eq:supp_ratio_coordinate_stability}\\
\frac1{n_0}\sum_{i=1}^{n_0}\max_{\mathbf y\in\Ycal}
|\widehat{\LR}_i(\mathbf y)-\LR_i(\mathbf y)|
&=O_p(\kappa_n+\Delta_n),
\label{eq:supp_average_ratio_rate}\\
\frac1{n_0}\sum_{i=1}^{n_0}\max_{\mathbf y\in\Ycal}
\{\LR_i(\mathbf y)+\widehat{\LR}_i(\mathbf y)\}&=O_p(1).
\label{eq:supp_ratio_envelope}
\end{align}
The third line follows from the first two by the triangle inequality and is displayed because it is
the quantity consumed by the profile theorem.  The first line is learner-specific.  The second is
a stability condition for the complete fit-and-evaluation map, not merely for target-point
evaluation.
The parameter space $\Theta$ is compact,
$\rho_y(\mathbf y;\bm\theta)$ is continuous in $\bm\theta$ uniformly over
finite $\Ycal$, and $\inf_{\bm\theta\in\Theta}\rho_y(\mathbf0;\bm\theta)>0$.  The oracle
criterion class
\begin{equation*}
\ell_{\bm\theta}(\mathbf u,\mathbf v)
=\log\left\{
\sum_{\mathbf y\in\Ycal}\rho_y(\mathbf y;\bm\theta)
\LR_{\mathrm{U,V\mid Y}}^{(0)}(\mathbf u,\mathbf v\mid\mathbf y)
\right\}
\end{equation*}
is Glivenko--Cantelli under $P_0$.  Its population criterion
\begin{equation}
\mathcal M(\bm\theta)
=E_0\ell_{\bm\theta}(\mathbf u^{(0)},\mathbf v^{(0)})
\label{eq:supp_population_profile}
\end{equation}
has the unique maximizer $\bm\theta^{(0)}$.

For the rate conclusion, $\bm\theta^{(0)}$ is an interior point, $\rho_y$ is twice
continuously differentiable with bounded derivatives in a neighbourhood of
$\bm\theta^{(0)}$, the negative population profile Hessian is positive definite there, the
oracle empirical Hessian converges uniformly to its population counterpart in that
neighbourhood, the oracle empirical profile score at $\bm\theta^{(0)}$ is
$O_p(n_0^{-1/2})$, and the fitted criterion satisfies its first-order condition.  For a smooth
finite-dimensional ratio learner, local strong curvature and Lipschitz score contributions give one
route from \eqref{eq:supp_unified_coordinates} to
\eqref{eq:supp_ratio_coordinate_stability} by the usual $M$-estimator perturbation argument
\citep[Chap.~5]{van2000asymptotic}.  For a flexible boosted-tree learner, both its intrinsic rate in
\eqref{eq:supp_oracle_coordinate_ratio_rate} and this coordinate-stability property are retained as
high-level conditions.
\end{assumption}

\begin{suppremark}[A primitive route to averaged ratio consistency]
A sufficient route to \eqref{eq:supp_average_ratio_rate}--\eqref{eq:supp_ratio_envelope}
is a standard support-and-tail condition.  After the common rotation, suppose there are sets
$\mathcal S_n$ with target probability tending to one such that the ratio estimator is
uniformly consistent on a vanishing neighbourhood of $\mathcal S_n$, the population ratios
are uniformly continuous and bounded there, and
\begin{equation*}
\frac1{n_0}\sum_{i=1}^{n_0}
\mathbf 1\{(\mathbf u_i^{(0)},\mathbf v_i^{(0)})\notin\mathcal S_n\}
\max_{\mathbf y\in\Ycal}\{\LR_i(\mathbf y)+\widehat{\LR}_i(\mathbf y)\}=o_p(1).
\end{equation*}
Together with the coordinate error in \eqref{eq:supp_unified_coordinates}, a
modulus-of-continuity argument yields \eqref{eq:supp_average_ratio_rate}; boundedness on
$\mathcal S_n$ and the displayed tail condition give \eqref{eq:supp_ratio_envelope}.
\end{suppremark}

\subsection{Transport and PCA-backend verification}
\label{subsec:supp_pca}

\begin{supplemma}[Target-moment transport]
\label{lem:supp_transport}
Under Assumption~\ref{ass:supp_transport}, for every square-integrable measurable $f$,
\begin{equation*}
E_m\left[\omega_m(\mathbf y^{(m)})
 f\{\mathbf t_1^{(m)},\mathbf t^{(m)}\}\right]
=E_0\left[f\{\mathbf t_1^{(0)},\mathbf t^{(m,0)}\}\right].
\end{equation*}
The same identity holds for measurable functions of the standardized scores
$(\mathbf r,\mathbf z)$.
\end{supplemma}

\begin{suppproof}
Condition on the finite outcome state.  The factor $\omega_m(\mathbf y)$ cancels the source
outcome probability and replaces it with the target probability; then use
\eqref{eq:supp_transportability}.  This is the usual importance-weighting identity under
label shift \citep{lipton2018labelshift,garg2020unified}.
\end{suppproof}

\begin{supplemma}[PCA-backend verification and primitive rate]
\label{lem:supp_pca_raw}
For each domain and observed block, let the population-centred feature rows be independent and
identically distributed.  Uniformly over domains, blocks and deterministic vectors $\mathbf a$,
suppose
\begin{equation*}
\|\langle\mathbf x,\mathbf a\rangle\|_{\psi_2}
\leq C_{\mathrm{sg}}\{\operatorname{var}(\langle\mathbf x,\mathbf a\rangle)\}^{1/2}
\end{equation*}
for a fixed $C_{\mathrm{sg}}$, and suppose the covariance operator norms are uniformly bounded.  Write $d_1$
and $d_k$ for the raw reference- and auxiliary-block
dimensions, and define
$\bm\Sigma_{x,1}^{(m)}=\operatorname{var}(\mathbf x_1^{(m)})$ and
$\bm\Sigma_{x,k}^{(m)}=\operatorname{var}(\mathbf x_k^{(m)})$.
Suppose $n_m/N\to\pi_m\in(0,1)$.  If
$\widehat{\bm\Sigma}_{x,1}^{(m)}$ denotes the within-domain sample covariance, set
$\widehat{\bm\Sigma}_1=\sum_{m=0}^M(n_m/N)
\widehat{\bm\Sigma}_{x,1}^{(m)}$.  For the pooled reference block define
$\overline{\bm\Sigma}_1=\sum_{m=0}^M\pi_m\bm\Sigma_{x,1}^{(m)}$, let
$\mathbf D_1$ be a fixed orthonormal basis of its leading $h$-dimensional eigenspace, and
suppose its eigengap at $h$ is $\delta_1>0$.  For auxiliary block $(m,k)$, let
$\mathbf D_k^{(m)}$ be a fixed orthonormal basis of the leading $\ell_k$-dimensional
eigenspace of $\bm\Sigma_{x,k}^{(m)}$ and suppose the corresponding eigengap is
$\delta_{k,m}>0$.  The bases are chosen to represent the population score orientations in
Assumption~\ref{ass:supp_transport}; changing those bases only changes the rotations below.

Suppose the population standardization covariances have minimum eigenvalue at least $c_0>0$.
Choose one deterministic $\Delta_n\to0$ such that
\begin{align}
&\max_{1\leq m\leq M}\max_{\mathbf y\in\Ycal}
|\widehat\omega_m(\mathbf y)-\omega_m(\mathbf y)|=O_p(\Delta_n),
\notag\\
&\frac{\sqrt h}{\delta_1}
\left[
\max_{0\leq m\leq M}
\left\{\sqrt{\frac{d_1}{n_m}}+\frac{d_1}{n_m}\right\}
+\max_{0\leq m\leq M}\left|\frac{n_m}{N}-\pi_m\right|
\right]
=O(\Delta_n),
\notag\\
&\max_{0\leq m\leq M}
\max_{k\in\Acal_m\setminus\{1\}}
\frac{\sqrt{\ell_k}}{\delta_{k,m}}
\left\{\sqrt{\frac{d_k}{n_m}}+\frac{d_k}{n_m}\right\}
=O(\Delta_n),
\qquad n_{\min}^{-1/2}=O(\Delta_n).
\label{eq:supp_delta_rate}
\end{align}
Then there are a common reference rotation $\mathbf Q_r$ and domain--block rotations
$\mathbf Q_{z,k}^{(m)}$ such that
\begin{equation}
\max_m n_m^{-1/2}
\|\widehat{\mathbf T}_1^{(m)}-\mathbf T_1^{(m)}\mathbf Q_r\|_{\F}
+
\max_{m,k}n_m^{-1/2}
\|\widehat{\mathbf T}_k^{(m)}-
\mathbf T_k^{(m)}\mathbf Q_{z,k}^{(m)}\|_{\F}
=O_p(\Delta_n).
\label{eq:supp_pca_raw_score}
\end{equation}
After plug-in target-mixture standardization, every primitive event
\eqref{eq:supp_score_event}--\eqref{eq:supp_source_zr_event} holds with the single rate
$O_p(\Delta_n)$.  Thus the PCA backend verifies the stochastic part of
Assumption~\ref{ass:supp_alignment} whenever $\Delta_n\to0$.
Equivalently, controllability requires the weight error to be $o_p(1)$ and each deterministic
PCA, domain-proportion and sampling expression in \eqref{eq:supp_delta_rate} to vanish; no
downstream lemma imposes another stochastic requirement on $\Delta_n$.
\end{supplemma}

\begin{suppproof}
Within each domain, sub-Gaussian covariance concentration gives
\begin{equation*}
\|\widehat{\bm\Sigma}_x-\bm\Sigma_x\|_{\op}
=O_p\{\sqrt{d/n}+d/n\},
\end{equation*}
by \citet[Theorem~4.7.1 and Exercise~4.7.3]{vershynin2018hdp}.  Decomposing
$\widehat{\bm\Sigma}_1-\overline{\bm\Sigma}_1$ adds the displayed domain-proportion term.
The population-gap Procrustes bound in \citet[Theorem~2]{yu2015davis} then gives a basis error
of order $\sqrt h/\delta_1$ times the pooled covariance error, and analogously gives
$\sqrt{\ell_k}/\delta_{k,m}$ times the auxiliary covariance error.  For a
population-centred feature matrix $\mathbf X_c$,
\begin{equation*}
n^{-1}\|\mathbf X_c(\widehat{\mathbf D}-\mathbf D\mathbf Q)\|_{\F}^2
=\tr\{(\widehat{\mathbf D}-\mathbf D\mathbf Q)^\T
\widehat{\bm\Sigma}_x
(\widehat{\mathbf D}-\mathbf D\mathbf Q)\},
\end{equation*}
so the basis rate gives \eqref{eq:supp_pca_raw_score}; sample centering has the same or a
smaller order.

Bounded overlap makes each normalized source weight $O(n_m^{-1})$.  Because $\Ycal$ is
finite, the first line of \eqref{eq:supp_delta_rate} perturbs a normalized weighted first or
second moment by $O_p(\Delta_n)$.  Raw-score recovery and empirical fluctuation are also
$O_p(\Delta_n)$ by the remaining lines; for the fixed retained score dimensions, bounded fourth
moments give the unperturbed empirical first- and second-moment fluctuation
$O_p(n_{\min}^{-1/2})$.  If symmetric positive-definite $\mathbf C$ and
$\widetilde{\mathbf C}$ have minimum eigenvalue at least $c>0$, matrix-square-root functional
calculus gives
\begin{equation*}
\|\widetilde{\mathbf C}^{-1/2}-\mathbf C^{-1/2}\|_{\op}
\leq \frac{1}{2c^{3/2}}
\|\widetilde{\mathbf C}-\mathbf C\|_{\op};
\end{equation*}
this follows from the resolvent integral for the inverse square root; see
\citet[Chap.~6]{higham2008functions}.  This applies with probability tending to one because
the corresponding population eigenvalues are at least $c_0$.  Expanding the standardized scores,
target CCA moments and source matrices $\widehat{\bm\Sigma}_{zz}^{(m)}$ and
$\widehat{\bm\Sigma}_{zr}^{(m)}$ therefore gives all five primitive events at rate
$O_p(\Delta_n)$.  Standardization and moment estimation have thus been verified, not added as
new terms.  This is a backend verification, not a new PCA theorem.
\end{suppproof}

\begin{suppremark}[Scope of the backend rate]
Lemma~\ref{lem:supp_pca_raw} gives sufficient backend conditions.  A robust covariance
estimator, an effective-rank covariance bound under its own distributional conditions
\citep{koltchinskii2017covariance}, or a non-PCA backend may instead verify the same
$O_p(\Delta_n)$ events.  The weight condition in \eqref{eq:supp_delta_rate} is separate
from PCA.  In particular, the MLLS bound of \citet[Theorem~3]{garg2020unified} separates a
root-target-sample term from calibration error.  Under its curvature conditions, root-order
calibration error, root-order estimation of the source outcome probabilities and a smooth
transformation to $\omega_m$ verify the weight row of \eqref{eq:supp_delta_rate}.  The practical calibration study
of \citet{alexandari_maximum_2020} motivates an implementation but is not used as a rate theorem.
\end{suppremark}

\subsection{Paired CCA, source moments and ridge maps}
\label{subsec:supp_alignment_lemmas}

\begin{supplemma}[Paired regularized-CCA perturbation]
\label{lem:supp_paired_cca}
Under the structural conditions of Assumption~\ref{ass:supp_alignment} and the target moment
events \eqref{eq:supp_target_cov_event}--\eqref{eq:supp_target_cross_event}, there is one
$\mathbf O\in\mathcal O(r_0)$ such that
\begin{equation}
\|\widehat{\mathbf A}^{(0)}-\mathbf Q_r^\T\mathbf A^{(0)}\mathbf O\|_{\F}
+\|\widehat{\mathbf B}^{(0)}-\mathbf Q_z^{(0)\T}\mathbf B^{(0)}\mathbf O\|_{\F}
=O_p\{(1+\delta_{\mathrm{CCA}}^{-1})\Delta_n\}.
\label{eq:supp_paired_cca_rate}
\end{equation}
\end{supplemma}

\begin{suppproof}
The inverse-square-root perturbation bound \citep{higham2008functions} and
\eqref{eq:supp_target_cov_event}--\eqref{eq:supp_target_cross_event} imply
\begin{equation*}
\|\widehat{\bm\Gamma}^{(0)}-
\mathbf Q_r^\T\bm\Gamma^{(0)}\mathbf Q_z^{(0)}\|_{\op}
=O_p(\Delta_n).
\end{equation*}
Apply the population-gap eigenspace bound of \citet[Theorem~2]{yu2015davis} to the symmetric dilation
$\left(\begin{smallmatrix}0&\bm\Gamma^{(0)}\\\bm\Gamma^{(0)\T}&0\end{smallmatrix}\right)$.
Its leading positive eigenvectors are the stacked pairs
$2^{-1/2}\operatorname{col}(\mathbf p_j,\mathbf q_j)$, so the theorem supplies one Procrustes
matrix $\mathbf O$ for the joint eigenspace and therefore the same $\mathbf O$ on both CCA
sides.  This common-rotation conclusion comes from applying the perturbation theorem to one
dilation rather than perturbing the two singular subspaces separately.  Transforming back through
the inverse square roots gives \eqref{eq:supp_paired_cca_rate}.  Wedin's theorem provides the
classical singular-subspace background \citep{wedin1972perturbation}, but the common rotation used
here is the joint-dilation deduction just given.
\end{suppproof}

\begin{supplemma}[Ridge propagation of one common rotation]
\label{lem:supp_ridge}
Under Assumptions~\ref{ass:supp_transport} and \ref{ass:supp_alignment}, let $\mathbf O$
be the rotation in Lemma~\ref{lem:supp_paired_cca}.  The population ridge map is equivariant:
\begin{equation}
\left[
\mathbf Q_z^{(m)\T}\bm\Sigma_{zz}^{(m)}\mathbf Q_z^{(m)}
+\lambda_m\mathbf I_{d_{z,m}}
\right]^{-1}
\mathbf Q_z^{(m)\T}\bm\Sigma_{zr}^{(m)}\mathbf A^{(0)}\mathbf O
=
\mathbf Q_z^{(m)\T}\mathbf G^{(m)}\mathbf O.
\label{eq:supp_ridge_equivariance}
\end{equation}
Moreover, uniformly over the fixed source collection,
\begin{equation}
\max_m\|\widehat{\mathbf G}^{(m)}-
\mathbf Q_z^{(m)\T}\mathbf G^{(m)}\mathbf O\|_{\F}
=O_p\!\left(
\left[\underline\lambda^{-1}(1+\delta_{\mathrm{CCA}}^{-1})
+\underline\lambda^{-2}\right]\Delta_n
\right).
\label{eq:supp_ridge_rate}
\end{equation}
The same $\mathbf O$ satisfies
\begin{align}
&\|\widehat{\mathbf A}^{(0)}-\mathbf Q_r^\T\mathbf A^{(0)}\mathbf O\|_{\F}
+\max_{2\leq k\leq K}
\|\widehat{\mathbf B}_k^{(0)}-
\mathbf Q_{z,k}^{(0)\T}\mathbf B_k^{(0)}\mathbf O\|_{\F}
\notag\\
&\qquad+
\max_m\max_{k\in\Acal_m\setminus\{1\}}
\|\widehat{\mathbf G}_k^{(m)}-
\mathbf Q_{z,k}^{(m)\T}\mathbf G_k^{(m)}\mathbf O\|_{\F}
=O_p(\Delta_n),
\label{eq:supp_unified_parameters}\\
&\max_{0\leq m\leq M}
\left\{
\|\widehat{\mathbf U}^{(m)}-\mathbf U^{(m)}\mathbf O\|_{\F,m}
+\sum_{k\in\Acal_m\setminus\{1\}}
\|\widehat{\mathbf V}_k^{(m)}-
\mathbf V_k^{(m)}\mathbf O\|_{\F,m}
\right\}
=O_p(\Delta_n).
\label{eq:supp_unified_coordinates}
\end{align}
Here and below, the simpler $O_p(\Delta_n)$ form uses the fixed positive CCA gap and fixed
ridge lower bound in Assumption~\ref{ass:supp_alignment}; equation~\eqref{eq:supp_ridge_rate}
shows their explicit contribution.
\end{supplemma}

\begin{suppproof}
Differentiating the population criterion in \eqref{eq:supp_population_ridge} gives
\begin{equation}
\{\bm\Sigma_{zz}^{(m)}+\lambda_m\mathbf I_{d_{z,m}}\}\mathbf G^{(m)}
=\bm\Sigma_{zr}^{(m)}\mathbf A^{(0)}.
\label{eq:supp_ridge_population_normal}
\end{equation}
Its Hessian is $2\mathbf I_{r_0}\otimes\{\bm\Sigma_{zz}^{(m)}+\lambda_m\mathbf I_{d_{z,m}}\}$, whose smallest eigenvalue is at least $2\underline\lambda$.  Hence the criterion is strictly convex and its unique minimizer is \eqref{eq:supp_population_ridge_closed}.  If the auxiliary scores are re-expressed through $\mathbf Q_z^{(m)}$ and the target anchor through $\mathbf O$, then the normal-equation moments become
\begin{equation*}
\bm\Sigma_{zz}^{(m)}\mapsto\mathbf Q_z^{(m)\T}\bm\Sigma_{zz}^{(m)}\mathbf Q_z^{(m)},
\qquad
\bm\Sigma_{zr}^{(m)}\mathbf A^{(0)}
\mapsto\mathbf Q_z^{(m)\T}\bm\Sigma_{zr}^{(m)}\mathbf A^{(0)}\mathbf O.
\end{equation*}
Substituting these two expressions into the closed-form solution and using $\mathbf Q_z^{(m)}\mathbf Q_z^{(m)\T}=\mathbf I_{d_{z,m}}$ gives
\begin{align*}
&\left[
\mathbf Q_z^{(m)\T}\bm\Sigma_{zz}^{(m)}\mathbf Q_z^{(m)}
+\lambda_m\mathbf I_{d_{z,m}}
\right]^{-1}
\mathbf Q_z^{(m)\T}\bm\Sigma_{zr}^{(m)}\mathbf A^{(0)}\mathbf O\\
&\qquad=
\mathbf Q_z^{(m)\T}
\{\bm\Sigma_{zz}^{(m)}+\lambda_m\mathbf I_{d_{z,m}}\}^{-1}
\bm\Sigma_{zr}^{(m)}\mathbf A^{(0)}\mathbf O
=\mathbf Q_z^{(m)\T}\mathbf G^{(m)}\mathbf O,
\end{align*}
which proves \eqref{eq:supp_ridge_equivariance}.  Thus ridge regularization preserves, rather than creates, the common right rotation inherited from the target CCA anchor.

For the empirical perturbation, set
\begin{equation*}
\widehat{\mathbf M}
=\widehat{\bm\Sigma}_{zz}^{(m)}+\lambda_m\mathbf I_{d_{z,m}},
\qquad
\mathbf M
=\mathbf Q_z^{(m)\T}
\{\bm\Sigma_{zz}^{(m)}+\lambda_m\mathbf I_{d_{z,m}}\}
\mathbf Q_z^{(m)},
\end{equation*}
Both inverse operator norms are at most $\underline\lambda^{-1}$.  By
\eqref{eq:supp_source_zr_event} and Lemma~\ref{lem:supp_paired_cca},
\begin{equation}
\max_m\left\|
\widehat{\bm\Sigma}_{zr}^{(m)}\widehat{\mathbf A}^{(0)}-
\mathbf Q_z^{(m)\T}\bm\Sigma_{zr}^{(m)}\mathbf A^{(0)}\mathbf O
\right\|_{\F}
=O_p\{(1+\delta_{\mathrm{CCA}}^{-1})\Delta_n\}.
\label{eq:supp_ridge_rhs_rate}
\end{equation}
Using \eqref{eq:supp_ridge_equivariance} and adding and subtracting the population right-hand
side gives
\begin{equation}
\widehat{\mathbf G}^{(m)}-\mathbf Q_z^{(m)\T}\mathbf G^{(m)}\mathbf O
=
\widehat{\mathbf M}^{-1}
\left\{\widehat{\bm\Sigma}_{zr}^{(m)}\widehat{\mathbf A}^{(0)}-
\mathbf Q_z^{(m)\T}\bm\Sigma_{zr}^{(m)}\mathbf A^{(0)}\mathbf O\right\}
+(\widehat{\mathbf M}^{-1}-\mathbf M^{-1})
\mathbf Q_z^{(m)\T}\bm\Sigma_{zr}^{(m)}\mathbf A^{(0)}\mathbf O.
\label{eq:supp_ridge_exact_decomposition}
\end{equation}
The resolvent identity $\widehat{\mathbf M}^{-1}-\mathbf M^{-1}=\widehat{\mathbf M}^{-1}(\mathbf M-\widehat{\mathbf M})\mathbf M^{-1}$ gives
\begin{equation}
\|\widehat{\mathbf M}^{-1}-\mathbf M^{-1}\|_{\op}
\leq\underline\lambda^{-2}
\|\widehat{\bm\Sigma}_{zz}^{(m)}-
\mathbf Q_z^{(m)\T}\bm\Sigma_{zz}^{(m)}\mathbf Q_z^{(m)}\|_{\op}.
\label{eq:supp_ridge_resolvent}
\end{equation}
Equations~\eqref{eq:supp_ridge_rhs_rate} and \eqref{eq:supp_ridge_resolvent}, bounded
population moments and the exact decomposition prove
\eqref{eq:supp_ridge_rate}.  Taking row blocks gives the coefficient bound in
\eqref{eq:supp_unified_parameters}; the target blocks follow from
Lemma~\ref{lem:supp_paired_cca}.

For fitted coordinates, the two relevant decompositions are
\begin{align*}
\widehat{\mathbf U}^{(m)}-\mathbf U^{(m)}\mathbf O
&=(\widehat{\mathbf R}^{(m)}-\mathbf R^{(m)}\mathbf Q_r)
\widehat{\mathbf A}^{(0)}
+\mathbf R^{(m)}\mathbf Q_r
(\widehat{\mathbf A}^{(0)}-\mathbf Q_r^\T\mathbf A^{(0)}\mathbf O),\\
\widehat{\mathbf V}_k^{(m)}-\mathbf V_k^{(m)}\mathbf O
&=(\widehat{\mathbf Z}_k^{(m)}-\mathbf Z_k^{(m)}\mathbf Q_{z,k}^{(m)})
\widehat{\mathbf G}_k^{(m)}
+\mathbf Z_k^{(m)}\mathbf Q_{z,k}^{(m)}
(\widehat{\mathbf G}_k^{(m)}-
\mathbf Q_{z,k}^{(m)\T}\mathbf G_k^{(m)}\mathbf O),
\end{align*}
with $\mathbf G_k^{(m)}$ replaced by $\mathbf B_k^{(0)}$ for a target auxiliary block.
The primitive score event, bounded second moments and the coefficient bounds make each term
$O_p(\Delta_n)$ in $\|\cdot\|_{\F,m}$.  Summing over the fixed number of blocks proves
\eqref{eq:supp_unified_coordinates}.  The nuisance $\mathbf Q$ rotations cancel in the
products, while the one right rotation $\mathbf O$ remains.  The displayed resolvent identity
is the only additional stability argument needed for ridge regression.
\end{suppproof}

\begin{suppremark}[What common orientation does not prove]
Equation~\eqref{eq:supp_unified_coordinates} compares each fitted block with the population
block defined by its own map.  It neither asserts $\mathbf B_k^{(0)}=\mathbf G_k^{(m)}$ nor
transports a source class-conditional density to the target.  That distributional step is the
separate likelihood-ratio condition in Assumption~\ref{ass:supp_outcome}.
\end{suppremark}

\begin{supplemma}[Rotation invariance of the conditional distribution]
\label{lem:supp_rotation}
For generic target subject-level coordinates, write
$\mathbf c=\operatorname{col}(\mathbf u,\mathbf v_2,\ldots,\mathbf v_K)$ and define
\begin{equation}
\mathbf c^{\mathbf O}
=\operatorname{col}(\mathbf O^\T\mathbf u,
\mathbf O^\T\mathbf v_2,\ldots,\mathbf O^\T\mathbf v_K).
\label{eq:supp_rotation_map}
\end{equation}
Then
\begin{equation}
\sigma(\mathbf c)=\sigma(\mathbf c^{\mathbf O}),
\qquad
p_0^{\mathbf O}(\mathbf y\mid\mathbf c^{\mathbf O})
=p_0(\mathbf y\mid\mathbf c)\quad\text{almost surely},
\label{eq:supp_rotation_conditional}
\end{equation}
where $p_0^{\mathbf O}$ denotes the same target law expressed in the rotated coordinates.  If $f_{\mathbf y}$ and $f_{\mathbf0}$ are target conditional Lebesgue densities of $\mathbf c$ given outcome states $\mathbf y$ and $\mathbf0$, respectively, then their rotated-coordinate likelihood ratio satisfies
\begin{equation}
\LR_{\mathrm{U,V\mid Y}}^{(0),\mathbf O}
(\mathbf c^{\mathbf O}\mid\mathbf y)
=
\LR_{\mathrm{U,V\mid Y}}^{(0)}(\mathbf c\mid\mathbf y).
\label{eq:supp_rotation_ratio}
\end{equation}
Consequently, the population profile criterion and the target posterior are unchanged by the residual common rotation.
\end{supplemma}

\begin{suppproof}
Applying $\mathbf O$ blockwise is an invertible Borel map, so $\mathbf c$ and
$\mathbf c^{\mathbf O}$ generate the same sigma-field.  This proves
\eqref{eq:supp_rotation_conditional}.  Under a change of variables, the numerator and
denominator densities in the likelihood ratio acquire the same Jacobian, which cancels and
gives \eqref{eq:supp_rotation_ratio}.  Substitution into the profile criterion and Bayes
formula proves the final claim.
\end{suppproof}

\subsection{Profile and target conditional-outcome consistency}
\label{subsec:supp_conditional}

\begin{suppremark}[A sufficient identification condition]
Let $f_{\mathbf y}$ be the density of the target aligned coordinates conditional on outcome
$\mathbf y$.  If $f_{\mathbf0}>0$, the finite collection $\{f_{\mathbf y}\}$ is linearly
independent up to null sets, and $\bm\theta\mapsto\{\rho_y(\mathbf y;\bm\theta)\}$ is
injective, then the unique-maximizer clause in Assumption~\ref{ass:supp_outcome} holds.  Indeed,
with $f_{\bm\theta}=\sum_{\mathbf y}\rho_y(\mathbf y;\bm\theta)f_{\mathbf y}$,
\begin{equation*}
\mathcal M(\bm\theta)-\mathcal M(\bm\theta^{(0)})
=-\operatorname{KL}(f_{\bm\theta^{(0)}},f_{\bm\theta})\leq0,
\end{equation*}
and equality identifies first the mixture probabilities and then $\bm\theta^{(0)}$.  Linear
independence is sufficient, not necessary.
\end{suppremark}

Define
\begin{align*}
\mathcal M_n^0(\bm\theta)
&=\frac1{n_0}\sum_{i=1}^{n_0}
\log\left\{\sum_{\mathbf y\in\Ycal}
\rho_y(\mathbf y;\bm\theta)\LR_i(\mathbf y)\right\},\\
\widehat{\mathcal M}_n(\bm\theta)
&=\frac1{n_0}\sum_{i=1}^{n_0}
\log\left\{\sum_{\mathbf y\in\Ycal}
\rho_y(\mathbf y;\bm\theta)\widehat{\LR}_i(\mathbf y)\right\}.
\end{align*}

Define the target conditional outcome distribution
\begin{equation*}
p_0(\mathbf y\mid\mathbf u,\mathbf v)
=\frac{\rho_y(\mathbf y;\bm\theta^{(0)})
\LR_{\mathrm{U,V\mid Y}}^{(0)}(\mathbf u,\mathbf v\mid\mathbf y)}
{\sum_{\mathbf y'\in\Ycal}\rho_y(\mathbf y';\bm\theta^{(0)})
\LR_{\mathrm{U,V\mid Y}}^{(0)}(\mathbf u,\mathbf v\mid\mathbf y')}.
\end{equation*}

\begin{supptheorem}[Target profile and conditional-outcome rate]
\label{thm:supp_conditional}
Under Assumptions~\ref{ass:supp_transport}--\ref{ass:supp_outcome}, if
$\widehat{\mathcal M}_n$ is maximized up to $o_p(1)$, then
$\widehat{\bm\theta}^{(0)}\to_p\bm\theta^{(0)}$ and
\begin{equation}
\frac1{n_0}\sum_{i=1}^{n_0}\sum_{\mathbf y\in\Ycal}
\left|
\widehat p_i(\mathbf y)-
p_0\{\mathbf y\mid\mathbf u_i^{(0)},\mathbf v_i^{(0)}\}
\right|
\xrightarrow{p}0.
\label{eq:supp_conditional_consistency}
\end{equation}
Under the local rate conditions in Assumption~\ref{ass:supp_outcome},
\begin{align}
&\|\widehat{\bm\theta}^{(0)}-\bm\theta^{(0)}\|_2
+\frac1{n_0}\sum_{i=1}^{n_0}\sum_{\mathbf y\in\Ycal}
\left|
\widehat p_i(\mathbf y)-
p_0\{\mathbf y\mid\mathbf u_i^{(0)},\mathbf v_i^{(0)}\}
\right|
\notag\\
&\hspace{8em}=O_p\{n_0^{-1/2}+\kappa_n+\Delta_n\}.
\label{eq:supp_conditional_rate}
\end{align}
No relative ordering is required.  If $\kappa_n=O(\Delta_n)$, alignment is the slower nuisance
component; if $\Delta_n=O(\kappa_n)$, likelihood-ratio learning is slower.  The sum is retained to
show provenance, but $\kappa_n+\Delta_n$ has the same order as
$\max(\kappa_n,\Delta_n)$.  The benchmark is
inverse square root: root-sample-size rates mean
$\sqrt{n_{\min}}\Delta_n=O(1)$ and $\sqrt{N_s}\kappa_n=O(1)$; divergence of either
scaled sequence means slower convergence, whereas convergence to zero means faster convergence.
For example, when $d_k/n_m\to0$, the leading auxiliary-PCA term in
\eqref{eq:supp_delta_rate} is $\sqrt{\ell_kd_k}/(\delta_{k,m}\sqrt{n_m})$, so growing
dimension or a shrinking eigengap can make $\Delta_n$ larger than $n_m^{-1/2}$.
For the PCA backend, the admissible $\Delta_n$ is given explicitly by
\eqref{eq:supp_delta_rate}.  In the fixed-dimensional case with
fixed eigengaps, root-$n_{\min}$ weight and domain-proportion rates give
$\Delta_n=O(n_{\min}^{-1/2})$; the weight-rate route is described after
Lemma~\ref{lem:supp_pca_raw}.  If, in addition, a correctly specified smooth
finite-dimensional ratio procedure has $\kappa_n=O(N_s^{-1/2})$ by regular parametric
$M$-estimation \citep[Theorem~5.23]{van2000asymptotic}, the rate in
\eqref{eq:supp_conditional_rate} becomes
\begin{equation}
O_p\{n_0^{-1/2}+N_s^{-1/2}+n_{\min}^{-1/2}\}.
\label{eq:supp_parametric_rate}
\end{equation}
Because $n_{\min}\leq n_0$ and $n_{\min}\leq N_s$, this is also
$O_p(n_{\min}^{-1/2})$.  Under the positive limiting domain proportions imposed in
Lemma~\ref{lem:supp_pca_raw}, $n_{\min}\asymp n_0\asymp N_s\asymp N$, so the same
specialization can be written $O_p(N^{-1/2})$.  This specialization is conditional and is not an XGBoost rate:
for boosted trees, $\kappa_n$ and coordinate stability remain the two learner-specific clauses in
Assumption~\ref{ass:supp_outcome}.  The displayed orders do not determine a numerical constant;
that constant depends on covariance and sub-Gaussian scales, inverse eigengaps, the ridge lower
bound, ratio envelopes and profile curvature.
\end{supptheorem}

\begin{suppproof}
The baseline ratio equals one and its outcome probability is uniformly positive, so every
mixture inside the logarithm is bounded away from zero.  Hence
\eqref{eq:supp_average_ratio_rate} implies
\begin{equation*}
\sup_{\bm\theta\in\Theta}
|\widehat{\mathcal M}_n(\bm\theta)-\mathcal M_n^0(\bm\theta)|
=O_p(\kappa_n+\Delta_n)=o_p(1).
\end{equation*}
The Glivenko--Cantelli and unique-maximizer clauses and the argmax theorem give parameter
consistency \citep[Theorem~5.7]{van2000asymptotic}.

For the rate, bounded first and second derivatives of $\rho_y$ make the profile score and
Hessian locally Lipschitz in the finite ratio vector.  Thus the fitted score at
$\bm\theta^{(0)}$ is the oracle score plus $O_p(\kappa_n+\Delta_n)$.  A Taylor expansion
of the fitted first-order condition around $\bm\theta^{(0)}$, together with the positive
population curvature, yields
\begin{equation*}
\|\widehat{\bm\theta}^{(0)}-\bm\theta^{(0)}\|_2
=O_p\{n_0^{-1/2}+\kappa_n+\Delta_n\}.
\end{equation*}
The finite-state Bayes normalization is also Lipschitz.  Its average error is bounded by a
constant times
\begin{equation*}
\|\widehat{\bm\theta}^{(0)}-\bm\theta^{(0)}\|_2
\left\{\frac1{n_0}\sum_i\max_{\mathbf y\in\Ycal}\widehat{\LR}_i(\mathbf y)\right\}
+\frac1{n_0}\sum_i\max_{\mathbf y\in\Ycal}
|\widehat{\LR}_i(\mathbf y)-\LR_i(\mathbf y)|.
\end{equation*}
Equations~\eqref{eq:supp_average_ratio_rate}--\eqref{eq:supp_ratio_envelope} prove
\eqref{eq:supp_conditional_rate}; consistency follows as a special case.  Lemma
\ref{lem:supp_rotation} removes the common rotation.  The final two specializations follow
by substituting the displayed PCA and ratio rates.
\end{suppproof}

\begin{suppremark}[Exact and working ratios]
The theorem concerns the true target distribution given the aligned coordinates.  If the tempered
composite ratio used in practice is not the exact joint ratio, the identical proof yields
consistency for the corresponding working conditional distribution.  This distinction is stated explicitly so
that a working composite model is not presented as an exact likelihood-ratio identity.
\end{suppremark}

\section{Additional simulation details and results}
\label{sec:supp_simulation}

\subsection{Main multivariate simulation}
\label{subsec:supp_main_simulation}

Each replication contained one target domain, indexed by 0, and five source domains, indexed by $m=1,\ldots,5$, with 400 subjects per domain.  The target observed all five modalities.  Each source observed the shared reference modality and a prespecified subset of the four auxiliary modalities.  Target gold-standard outcomes were hidden from all preprocessing, fitting and tuning steps and were merged back only after the predictions had been fixed.

The target outcome was a three-dimensional binary vector.  Its marginal probabilities were
\begin{equation*}
\bm\pi^{(0)}=(0.22,0.36,0.40)^\T,
\end{equation*}
and the Ising pairwise interaction parameters were $(0.45,0.35,0.30)$.  For source $m$, the marginal probabilities were
\begin{equation}
\bm\pi^{(m)}=\operatorname{clip}\{\bm\pi^{(0)}+\delta\mathbf c_m,10^{-4},1-10^{-4}\},
\label{eq:supp_source_prevalence}
\end{equation}
where clipping was componentwise and
\begin{equation*}
\begin{pmatrix}
\mathbf c_1^\T\\ \mathbf c_2^\T\\ \mathbf c_3^\T\\ \mathbf c_4^\T\\ \mathbf c_5^\T
\end{pmatrix}
=
\begin{pmatrix}
0.60&0.48&0.36\\
0.80&0.64&0.48\\
1.00&0.80&0.60\\
1.20&0.96&0.72\\
1.40&1.12&0.84
\end{pmatrix}.
\end{equation*}
For each source, the Ising main effects were solved numerically to attain the marginals in \eqref{eq:supp_source_prevalence}, while the pairwise interactions were held fixed.  Hence $\delta$ changed the joint source outcome distribution but not the conditional feature law given the outcome.

Let $\mathbf e_{\mathbf y}=2\mathbf y-\mathbf1$ and
\begin{equation*}
\mathbf q(\mathbf y)=(e_{y1}e_{y2},e_{y1}e_{y3},e_{y2}e_{y3})^\T.
\end{equation*}
Independently generated matrices $\mathbf P,\mathbf Q\in\mathbb R^{5\times3}$ with orthonormal columns were shared by the reference and auxiliary latent factors.  State-specific Gaussian template matrices were rescaled across the eight outcome states to Frobenius norms 0.04 and 0.13.  The conditional latent means were
\begin{align*}
\bm\psi_u(\mathbf y)
&=\mathbf P\operatorname{diag}(0.48,0.43,0.39)\mathbf e_{\mathbf y}
 +\mathbf Q\operatorname{diag}(0.03,0.03,0.03)\mathbf q(\mathbf y)
 +\mathbf g_u(\mathbf y),\\
\bm\psi_v(\mathbf y)
&=\mathbf P\operatorname{diag}(1.34,1.20,1.07)\mathbf e_{\mathbf y}
 +\mathbf Q\operatorname{diag}(0.33,0.33,0.33)\mathbf q(\mathbf y)
 +\mathbf g_v(\mathbf y).
\end{align*}
Conditional on $\mathbf y$, the two latent factors were independent:
\begin{equation*}
\widetilde{\mathbf u}=\bm\psi_u(\mathbf y)+\bm\xi_u,
\qquad
\widetilde{\mathbf v}=\bm\psi_v(\mathbf y)+\bm\xi_v,
\qquad
\bm\xi_u,\bm\xi_v\sim N_5(\mathbf0,0.70^2\mathbf I).
\end{equation*}

The observed dimensions were $(d_1,\ldots,d_5)=(180,150,150,130,130)$.  Auxiliary modalities 2, 3, 4 and 5 retained latent coordinates $\{1,2,3\}$, $\{2,3,4\}$, $\{3,4,5\}$ and $\{1,4,5\}$, respectively.  With $\mathbf M_k$ denoting the corresponding diagonal mask,
\begin{equation*}
\mathbf x_1^{(m)}=\mathbf L_1\widetilde{\mathbf u}+\bm\epsilon_1^{(m)},
\qquad
\mathbf x_k^{(m)}=\mathbf L_k^{(m)}\mathbf M_k\widetilde{\mathbf v}+\bm\epsilon_k^{(m)},
\quad k=2,\ldots,5.
\end{equation*}
The reference loading $\mathbf L_1$ was exactly shared across domains.  The measurement-error standard deviations were 0.26 for the reference modality and 0.41 for each auxiliary modality.

For auxiliary modality $k$, let $\mathbf L_k$ be a shared orthonormal loading and $\widetilde{\mathbf L}_k^{(m)}$ an independently generated orthonormal loading.  From a Gaussian matrix $\mathbf A_k^{(m)}$, define
\begin{equation*}
\mathbf R_k^{(m)}(\varepsilon_{\mathrm{rot}})
=\operatorname{qf}\left\{\mathbf I+
\frac{\varepsilon_{\mathrm{rot}}}{\sqrt5}
(\mathbf A_k^{(m)}-\mathbf A_k^{(m)\T})\right\},
\end{equation*}
and set
\begin{equation*}
\mathbf L_k^{(m)}
=\operatorname{qf}\left\{
0.5\mathbf L_k\mathbf R_k^{(m)}(\varepsilon_{\mathrm{rot}})
+0.5\widetilde{\mathbf L}_k^{(m)}
\right\}.
\end{equation*}
Here $\operatorname{qf}$ denotes the orthonormal factor from a QR decomposition.  No loading signs were reversed.  Thus $\varepsilon_{\mathrm{rot}}$ controlled the rotation of the shared loading component, while the independent subspace component retained weight 0.5.

For the label-shift, rotation and source-label sweeps, the source modality sets were
\begin{equation*}
\Acal_1=\{1,2\},\quad
\Acal_2=\{1,3\},\quad
\Acal_3=\{1,3,4,5\},\quad
\Acal_4=\{1,2,5\},\quad
\Acal_5=\{1,4,5\}.
\end{equation*}
A source gold-standard outcome was independently hidden with probability $p_{\mathrm{lab}}$, with a deterministic repair only if a source otherwise contained no observed outcome.  The quantity $p_{\mathrm{mod}}$ indexed the following five fixed source-pattern configurations rather than independent modality deletion:
\begin{center}
\small
\begin{tabular}{c@{\qquad}ccccc}
\toprule
$p_{\mathrm{mod}}$ & $\Acal_1$ & $\Acal_2$ & $\Acal_3$ & $\Acal_4$ & $\Acal_5$\\
\midrule
0.0 & $\{1,2\}$ & $\{1,3\}$ & $\{1,3,4,5\}$ & $\{1,2,5\}$ & $\{1,4,5\}$\\
0.1 & $\{1,2\}$ & $\{1,3\}$ & $\{1,3,4,5\}$ & $\{1,2,5\}$ & $\{1,4\}$\\
0.2 & $\{1,2\}$ & $\{1,3\}$ & $\{1,3,4,5\}$ & $\{1,2\}$ & $\{1,4,5\}$\\
0.3 & $\{1,2\}$ & $\{1,3\}$ & $\{1,3,4\}$ & $\{1,2,5\}$ & $\{1,4,5\}$\\
0.4 & $\{1,2\}$ & $\{1,3\}$ & $\{1,3,4\}$ & $\{1,2,5\}$ & $\{1,4\}$\\
\bottomrule
\end{tabular}
\end{center}
Because these configurations are nonnested, increasing $p_{\mathrm{mod}}$ does not necessarily remove more information monotonically.

The four one-factor-at-a-time sweeps were
\begin{align*}
\delta&\in\{0,0.08,0.16,0.24,0.32,0.40\},&
\varepsilon_{\mathrm{rot}}&\in\{0,0.2,0.4,0.6,0.8,1.0\},\\
p_{\mathrm{lab}}&\in\{0,0.1,0.2,0.3,0.4,0.5\},&
p_{\mathrm{mod}}&\in\{0,0.1,0.2,0.3,0.4\}.
\end{align*}
The default values were $\delta=0.24$, $\varepsilon_{\mathrm{rot}}=0.50$, $p_{\mathrm{lab}}=0.35$ and $p_{\mathrm{mod}}=0$.  Only one quantity varied in each sweep.  Within a setting, every method was evaluated on the same generated replication; the seed schedule was also matched across the settings of a sweep.

The proposed implementation centered every feature block within domain, used rank-five pooled PCA for the reference modality and rank-three within-domain PCA for each auxiliary modality, and used the uncentered reference projection only in the initial target outcome-distribution calibration.  After calibration, source observations were reweighted to the estimated target outcome distribution.  The code then performed its outcome-reweighted blockwise alignment followed by a rank-five single-joint target CCA anchor and the corresponding source ridge regression maps.  The CCA and source regression penalties were both $10^{-4}$.  Source-overlap augmentation was enabled: a source pattern $\Acal_m$ could borrow records from a donor source whose observed pattern contained $\Acal_m$, after the donor record was restricted to the requested blocks.

The reference density-ratio model used XGBoost with 200 trees, depth four, learning rate 0.05, row and column sampling fractions 0.9, and ridge penalty one \citep{chen2016xgboost}.  The joint ratio model was multinomial logistic regression.  Its penalty parameter was selected from $\{0.2,0.5,1,2\}$ by threefold group validation with source domains as folds.  The auxiliary contribution was retained only if it improved validation log loss and was then multiplied by 0.5.

The final target update used the likelihood-based label-shift formulation \citep{saerens2002adjusting,alexandari_maximum_2020,garg2020unified}.  It allowed at most 120 iterations, tolerance $10^{-5}$ and initial damping 0.4, with damping reduced as far as 0.05 to preserve a nondecreasing profile likelihood.  If the iteration failed to converge or moved the estimated target marginal probabilities by more than 0.15 in $L_1$ distance from the initial reference-based estimates, the final posterior was exponentially tilted back to those initial marginals.  This target-outcome-free safeguard was activated in 94.8\%, 94.2\%, 96.0\% and 95.0\% of the replications in the label-shift, rotation, label-missingness and modality-pattern sweeps, respectively.  The reported results therefore characterize the stabilized target update rather than unrestricted EM.

The comparator set was as follows.
\begin{itemize}[leftmargin=2.0em,itemsep=2pt,topsep=3pt]
\item \textit{Raw-X XGBoost} pooled labeled source rows, concatenated the five raw modality blocks, represented an absent source block by missing values, and appended four block-availability indicators.  It used 50 trees, depth three, learning rate 0.05, row and column sampling fractions 0.9, and ridge penalty one.  It performed no source reweighting, alignment or target adaptation \citep{chen2016xgboost}.
\item \textit{Raw-X LASSO} used the same pooled labeled source rows and raw block layout.  For each of the three binary outcomes, all-source-missing columns were removed, the remaining missing entries were median-imputed using source data with automatically generated missingness indicators, and the features were standardized.  An $L_1$-penalized logistic model selected $C$ from 12 logarithmically spaced values between $10^{-3}$ and $10^2$ by up to fivefold stratified cross-validation using negative Brier score, followed by SAGA fitting with 4000 maximum iterations and tolerance $10^{-3}$.  The three marginal probabilities were combined through their independent product to obtain the eight-state distribution.  No target label or adaptation step was used \citep{tibshirani1996regression}.
\item \textit{Direct transfer} used local PCA scores, a pooled source outcome model and no cross-domain alignment or target update.  \textit{Orthogonal transfer} used local reference PCA and subset-unweighted Procrustes alignment on labeled source observations, without target calibration \citep{schonemann1966generalized}.
\item \textit{RGCCA} used rank five, $\tau=0.5$ and ridge $10^{-4}$ \citep{tenenhaus2011regularized}; \textit{BONMI} used rank 12 \citep{zhou2023multisource}.  Neither procedure used target outcomes or the proposed target update.
\item \textit{Target-only PCA--EM} fitted PCA in the target and separate two-component Gaussian mixtures to its first three principal-component scores \citep{jolliffe2016principal,dempster1977maximum}.  It used no source data and no target outcome.
\end{itemize}

Macro-AUC was the average of the three componentwise AUCs.  Target MSE was
\begin{equation*}
\operatorname{MSE}
=\frac{1}{3n_0}\sum_{i=1}^{n_0}\sum_{a=1}^3
\{\widehat p_{ia}^{(0)}-y_{ia}^{(0)}\}^2.
\end{equation*}
Higher macro-AUC and lower MSE were preferred.

\subsection{Main-simulation results}
\label{subsubsec:supp_main_results}

Across the 23 settings, the proposed method attained average macro-AUC 0.9269 and average MSE 0.1003.  Relative to Raw-X XGBoost, this corresponds to a 20.3\% increase in macro-AUC and a 51.8\% reduction in MSE.  Relative to Raw-X LASSO, the corresponding changes were 21.0\% and 51.9\%.  Raw-X XGBoost and Raw-X LASSO were nearly tied overall: the seed-clustered paired XGBoost-minus-LASSO AUC difference was 0.0045 (95\% Monte Carlo interval $-0.0066$ to 0.0157), and the LASSO-minus-XGBoost MSE difference was 0.0006 ($-0.0103$ to 0.0114).  Thus, the experiment does not support a stable overall advantage of the nonlinear tree learner over the sparse linear learner.

The proposed method had the largest macro-AUC and smallest MSE in 22 of the 23 settings.  The one exception is informative: at $\varepsilon_{\mathrm{rot}}=0$, Raw-X LASSO achieved macro-AUC 0.9557 and MSE 0.0795, compared with 0.9369 and 0.0943 for the proposed method.  When the source auxiliary coordinate systems were not rotated, a sparse raw-feature model could exploit the nearly shared linear structure directly.  Its performance deteriorated with rotation, reaching macro-AUC 0.7329 and MSE 0.2288 at $\varepsilon_{\mathrm{rot}}=1$, whereas the proposed method remained at 0.9297 and 0.0983.  This exception therefore sharpens, rather than weakens, the interpretation of the alignment experiment.

The label-shift trajectory most clearly separates ranking from probability calibration.  The proposed macro-AUC remained between 0.9270 and 0.9323, while Raw-X XGBoost MSE increased from 0.1820 at $\delta=0$ to 0.2546 at $\delta=0.40$.  At the largest shift, the proposed method had 21.7\% higher macro-AUC and 59.5\% lower MSE than Raw-X XGBoost.  Across the rotation sweep, the proposed macro-AUC ranged from 0.9245 to 0.9369 and its MSE from 0.0943 to 0.1035.  Under source-label missingness, degradation was gradual; at $p_{\mathrm{lab}}=0.50$, macro-AUC was 0.9198 and MSE was 0.1054.  The modality-pattern results were nonmonotone because $p_{\mathrm{mod}}$ labels the nonnested patterns listed above rather than an ordered deletion probability.

\begin{table}[h]
\centering
\caption{Setting-equal summaries corresponding to the radar panels of Figure~2 in the main paper. Each sweep column averages its setting-level Monte Carlo means, and the Overall column averages all 23 settings. Higher macro-AUC and lower MSE are preferred; the best point estimate in each column is shown in bold.}
\label{tab:supp_main_aggregate}
\scriptsize
\setlength{\tabcolsep}{3.0pt}
\renewcommand{\arraystretch}{1.05}
\resizebox{\linewidth}{!}{%
\begin{tabular}{@{}lccccc@{}}
\toprule
Method & Overall & Label shift & Rotation & Label missingness & Modality pattern\\
\midrule
\multicolumn{6}{@{}l}{\textit{Panel A: Target macro-AUC}}\\
Proposed & \textbf{0.9269} & \textbf{0.9287} & \textbf{0.9299} & \textbf{0.9289} & \textbf{0.9187}\\
Raw-X XGBoost & 0.7707 & 0.7667 & 0.7862 & 0.7666 & 0.7617\\
Raw-X LASSO & 0.7661 & 0.7603 & 0.7902 & 0.7560 & 0.7563\\
Orthogonal & 0.6469 & 0.6633 & 0.6411 & 0.6324 & 0.6517\\
BONMI & 0.6437 & 0.6453 & 0.6313 & 0.6472 & 0.6521\\
RGCCA & 0.6157 & 0.6124 & 0.6196 & 0.6049 & 0.6278\\
Direct & 0.5809 & 0.5933 & 0.5612 & 0.5778 & 0.5935\\
Target-only & 0.5119 & 0.5126 & 0.5110 & 0.5126 & 0.5112\\
\midrule
\multicolumn{6}{@{}l}{\textit{Panel B: Target MSE}}\\
Proposed & \textbf{0.1003} & \textbf{0.0996} & \textbf{0.0988} & \textbf{0.0986} & \textbf{0.1051}\\
Raw-X XGBoost & 0.2080 & 0.2090 & 0.2033 & 0.2090 & 0.2114\\
Raw-X LASSO & 0.2086 & 0.2109 & 0.1922 & 0.2166 & 0.2158\\
Orthogonal & 0.2636 & 0.2540 & 0.2650 & 0.2705 & 0.2651\\
BONMI & 0.2928 & 0.2860 & 0.3017 & 0.2923 & 0.2908\\
RGCCA & 0.2694 & 0.2682 & 0.2700 & 0.2734 & 0.2653\\
Direct & 0.2785 & 0.2721 & 0.2862 & 0.2804 & 0.2745\\
Target-only & 0.3156 & 0.3156 & 0.3149 & 0.3156 & 0.3164\\
\bottomrule
\end{tabular}%
}
\end{table}

\begin{table}[h]
\centering
\caption{Joint label shift. Entries are Monte Carlo means with Monte Carlo standard errors in parentheses from 100 replications at each setting. Higher macro-AUC and lower MSE are preferred; the best point estimate in each column is shown in bold.}
\label{tab:supp_label_shift}
\scriptsize
\setlength{\tabcolsep}{2.4pt}
\renewcommand{\arraystretch}{1.05}
\resizebox{\linewidth}{!}{%
\begin{tabular}{@{}lcccccc@{}}
\toprule
Method & $\delta=0$ & $\delta=0.08$ & $\delta=0.16$ & $\delta=0.24$ & $\delta=0.32$ & $\delta=0.4$\\
\midrule
\multicolumn{7}{@{}l}{\textit{Panel A: Target macro-AUC}}\\
Proposed & \textbf{0.9283 (0.0035)} & \textbf{0.9280 (0.0030)} & \textbf{0.9323 (0.0028)} & \textbf{0.9273 (0.0031)} & \textbf{0.9292 (0.0032)} & \textbf{0.9270 (0.0033)}\\
Raw-X XGBoost & 0.7678 (0.0095) & 0.7705 (0.0091) & 0.7700 (0.0091) & 0.7670 (0.0093) & 0.7628 (0.0096) & 0.7617 (0.0090)\\
Raw-X LASSO & 0.7648 (0.0122) & 0.7651 (0.0119) & 0.7591 (0.0125) & 0.7549 (0.0126) & 0.7600 (0.0124) & 0.7580 (0.0124)\\
Orthogonal & 0.6568 (0.0266) & 0.6659 (0.0266) & 0.6784 (0.0269) & 0.6438 (0.0269) & 0.6669 (0.0277) & 0.6678 (0.0282)\\
BONMI & 0.6432 (0.0248) & 0.6475 (0.0244) & 0.6507 (0.0247) & 0.6475 (0.0229) & 0.6323 (0.0240) & 0.6508 (0.0239)\\
RGCCA & 0.6283 (0.0231) & 0.6036 (0.0228) & 0.6227 (0.0223) & 0.6041 (0.0219) & 0.6033 (0.0230) & 0.6122 (0.0233)\\
Direct & 0.6007 (0.0255) & 0.5970 (0.0256) & 0.6086 (0.0248) & 0.5791 (0.0257) & 0.5847 (0.0259) & 0.5898 (0.0250)\\
Target-only & 0.5126 (0.0129) & 0.5126 (0.0129) & 0.5126 (0.0129) & 0.5126 (0.0129) & 0.5126 (0.0129) & 0.5126 (0.0129)\\
\midrule
\multicolumn{7}{@{}l}{\textit{Panel B: Target MSE}}\\
Proposed & \textbf{0.0992 (0.0022)} & \textbf{0.0991 (0.0022)} & \textbf{0.0956 (0.0022)} & \textbf{0.1008 (0.0024)} & \textbf{0.0997 (0.0023)} & \textbf{0.1031 (0.0025)}\\
Raw-X XGBoost & 0.1820 (0.0018) & 0.1852 (0.0018) & 0.1946 (0.0019) & 0.2091 (0.0021) & 0.2283 (0.0022) & 0.2546 (0.0023)\\
Raw-X LASSO & 0.2039 (0.0073) & 0.2075 (0.0074) & 0.2134 (0.0081) & 0.2168 (0.0081) & 0.2122 (0.0077) & 0.2116 (0.0075)\\
Orthogonal & 0.2006 (0.0088) & 0.2059 (0.0100) & 0.2208 (0.0112) & 0.2651 (0.0118) & 0.2924 (0.0119) & 0.3395 (0.0112)\\
BONMI & 0.2369 (0.0125) & 0.2474 (0.0127) & 0.2661 (0.0134) & 0.2917 (0.0129) & 0.3258 (0.0135) & 0.3483 (0.0130)\\
RGCCA & 0.2123 (0.0086) & 0.2293 (0.0090) & 0.2399 (0.0091) & 0.2733 (0.0092) & 0.3069 (0.0095) & 0.3477 (0.0093)\\
Direct & 0.2150 (0.0076) & 0.2243 (0.0089) & 0.2384 (0.0093) & 0.2793 (0.0098) & 0.3139 (0.0095) & 0.3617 (0.0085)\\
Target-only & 0.3156 (0.0069) & 0.3156 (0.0069) & 0.3156 (0.0069) & 0.3156 (0.0069) & 0.3156 (0.0069) & 0.3156 (0.0069)\\
\bottomrule
\end{tabular}%
}
\end{table}

\begin{table}[h]
\centering
\caption{Auxiliary loading rotation. Entries are Monte Carlo means with Monte Carlo standard errors in parentheses from 100 replications at each setting. Higher macro-AUC and lower MSE are preferred; the best point estimate in each column is shown in bold.}
\label{tab:supp_rotation}
\scriptsize
\setlength{\tabcolsep}{2.4pt}
\renewcommand{\arraystretch}{1.05}
\resizebox{\linewidth}{!}{%
\begin{tabular}{@{}lcccccc@{}}
\toprule
Method & $\varepsilon_{\mathrm{rot}}=0$ & $\varepsilon_{\mathrm{rot}}=0.2$ & $\varepsilon_{\mathrm{rot}}=0.4$ & $\varepsilon_{\mathrm{rot}}=0.6$ & $\varepsilon_{\mathrm{rot}}=0.8$ & $\varepsilon_{\mathrm{rot}}=1$\\
\midrule
\multicolumn{7}{@{}l}{\textit{Panel A: Target macro-AUC}}\\
Proposed & 0.9369 (0.0029) & \textbf{0.9245 (0.0037)} & \textbf{0.9299 (0.0027)} & \textbf{0.9272 (0.0031)} & \textbf{0.9311 (0.0026)} & \textbf{0.9297 (0.0032)}\\
Raw-X XGBoost & 0.9058 (0.0021) & 0.7846 (0.0108) & 0.7668 (0.0103) & 0.7543 (0.0094) & 0.7555 (0.0088) & 0.7500 (0.0082)\\
Raw-X LASSO & \textbf{0.9557 (0.0012)} & 0.8000 (0.0131) & 0.7688 (0.0132) & 0.7442 (0.0119) & 0.7398 (0.0117) & 0.7329 (0.0113)\\
Orthogonal & 0.6228 (0.0297) & 0.6498 (0.0271) & 0.6494 (0.0267) & 0.6441 (0.0269) & 0.6411 (0.0270) & 0.6394 (0.0269)\\
BONMI & 0.6819 (0.0227) & 0.6233 (0.0255) & 0.6357 (0.0243) & 0.6387 (0.0237) & 0.6081 (0.0249) & 0.6003 (0.0249)\\
RGCCA & 0.6757 (0.0220) & 0.6048 (0.0231) & 0.6359 (0.0233) & 0.6098 (0.0227) & 0.6116 (0.0231) & 0.5796 (0.0239)\\
Direct & 0.5972 (0.0274) & 0.5471 (0.0268) & 0.5697 (0.0267) & 0.5620 (0.0271) & 0.5442 (0.0277) & 0.5472 (0.0278)\\
Target-only & 0.4996 (0.0124) & 0.5133 (0.0123) & 0.5113 (0.0121) & 0.5155 (0.0130) & 0.5127 (0.0131) & 0.5137 (0.0122)\\
\midrule
\multicolumn{7}{@{}l}{\textit{Panel B: Target MSE}}\\
Proposed & 0.0943 (0.0022) & \textbf{0.1035 (0.0029)} & \textbf{0.0987 (0.0022)} & \textbf{0.1005 (0.0024)} & \textbf{0.0975 (0.0021)} & \textbf{0.0983 (0.0024)}\\
Raw-X XGBoost & 0.1686 (0.0010) & 0.2032 (0.0026) & 0.2085 (0.0023) & 0.2124 (0.0021) & 0.2126 (0.0021) & 0.2145 (0.0019)\\
Raw-X LASSO & \textbf{0.0795 (0.0011)} & 0.1896 (0.0087) & 0.2076 (0.0085) & 0.2225 (0.0074) & 0.2254 (0.0074) & 0.2288 (0.0071)\\
Orthogonal & 0.2678 (0.0122) & 0.2624 (0.0119) & 0.2626 (0.0117) & 0.2650 (0.0118) & 0.2656 (0.0117) & 0.2664 (0.0117)\\
BONMI & 0.2731 (0.0129) & 0.3084 (0.0147) & 0.2981 (0.0138) & 0.2958 (0.0134) & 0.3157 (0.0142) & 0.3188 (0.0141)\\
RGCCA & 0.2485 (0.0096) & 0.2732 (0.0098) & 0.2626 (0.0099) & 0.2732 (0.0097) & 0.2755 (0.0098) & 0.2869 (0.0103)\\
Direct & 0.2701 (0.0101) & 0.2891 (0.0105) & 0.2833 (0.0105) & 0.2867 (0.0103) & 0.2957 (0.0109) & 0.2925 (0.0106)\\
Target-only & 0.3221 (0.0063) & 0.3111 (0.0063) & 0.3146 (0.0066) & 0.3124 (0.0072) & 0.3152 (0.0077) & 0.3138 (0.0073)\\
\bottomrule
\end{tabular}%
}
\end{table}

\begin{table}[h]
\centering
\caption{Missing source gold-standard outcomes. Entries are Monte Carlo means with Monte Carlo standard errors in parentheses from 100 replications at each setting. Higher macro-AUC and lower MSE are preferred; the best point estimate in each column is shown in bold.}
\label{tab:supp_label_missing}
\scriptsize
\setlength{\tabcolsep}{2.4pt}
\renewcommand{\arraystretch}{1.05}
\resizebox{\linewidth}{!}{%
\begin{tabular}{@{}lcccccc@{}}
\toprule
Method & $p_{\mathrm{lab}}=0$ & $p_{\mathrm{lab}}=0.1$ & $p_{\mathrm{lab}}=0.2$ & $p_{\mathrm{lab}}=0.3$ & $p_{\mathrm{lab}}=0.4$ & $p_{\mathrm{lab}}=0.5$\\
\midrule
\multicolumn{7}{@{}l}{\textit{Panel A: Target macro-AUC}}\\
Proposed & \textbf{0.9344 (0.0025)} & \textbf{0.9310 (0.0029)} & \textbf{0.9328 (0.0029)} & \textbf{0.9279 (0.0035)} & \textbf{0.9276 (0.0035)} & \textbf{0.9198 (0.0043)}\\
Raw-X XGBoost & 0.7679 (0.0098) & 0.7684 (0.0096) & 0.7680 (0.0093) & 0.7693 (0.0091) & 0.7652 (0.0095) & 0.7607 (0.0092)\\
Raw-X LASSO & 0.7579 (0.0130) & 0.7570 (0.0128) & 0.7576 (0.0127) & 0.7563 (0.0125) & 0.7531 (0.0126) & 0.7541 (0.0121)\\
Orthogonal & 0.6191 (0.0275) & 0.6263 (0.0275) & 0.6291 (0.0276) & 0.6391 (0.0272) & 0.6457 (0.0269) & 0.6352 (0.0280)\\
BONMI & 0.6456 (0.0228) & 0.6462 (0.0228) & 0.6467 (0.0228) & 0.6476 (0.0229) & 0.6482 (0.0229) & 0.6490 (0.0226)\\
RGCCA & 0.6048 (0.0219) & 0.6044 (0.0220) & 0.6050 (0.0220) & 0.6049 (0.0219) & 0.6038 (0.0220) & 0.6065 (0.0219)\\
Direct & 0.5748 (0.0265) & 0.5764 (0.0263) & 0.5775 (0.0262) & 0.5796 (0.0259) & 0.5772 (0.0255) & 0.5812 (0.0247)\\
Target-only & 0.5126 (0.0129) & 0.5126 (0.0129) & 0.5126 (0.0129) & 0.5126 (0.0129) & 0.5126 (0.0129) & 0.5126 (0.0129)\\
\midrule
\multicolumn{7}{@{}l}{\textit{Panel B: Target MSE}}\\
Proposed & \textbf{0.0950 (0.0020)} & \textbf{0.0970 (0.0020)} & \textbf{0.0961 (0.0022)} & \textbf{0.0989 (0.0026)} & \textbf{0.0993 (0.0024)} & \textbf{0.1054 (0.0028)}\\
Raw-X XGBoost & 0.2084 (0.0022) & 0.2086 (0.0021) & 0.2087 (0.0021) & 0.2084 (0.0021) & 0.2095 (0.0021) & 0.2104 (0.0021)\\
Raw-X LASSO & 0.2156 (0.0085) & 0.2164 (0.0084) & 0.2156 (0.0082) & 0.2167 (0.0082) & 0.2168 (0.0080) & 0.2186 (0.0078)\\
Orthogonal & 0.2764 (0.0124) & 0.2734 (0.0123) & 0.2730 (0.0123) & 0.2662 (0.0119) & 0.2638 (0.0117) & 0.2700 (0.0119)\\
BONMI & 0.2953 (0.0131) & 0.2943 (0.0130) & 0.2934 (0.0130) & 0.2921 (0.0129) & 0.2908 (0.0128) & 0.2882 (0.0126)\\
RGCCA & 0.2750 (0.0094) & 0.2745 (0.0094) & 0.2740 (0.0093) & 0.2733 (0.0092) & 0.2728 (0.0091) & 0.2711 (0.0092)\\
Direct & 0.2839 (0.0106) & 0.2824 (0.0103) & 0.2815 (0.0102) & 0.2793 (0.0098) & 0.2792 (0.0096) & 0.2762 (0.0090)\\
Target-only & 0.3156 (0.0069) & 0.3156 (0.0069) & 0.3156 (0.0069) & 0.3156 (0.0069) & 0.3156 (0.0069) & 0.3156 (0.0069)\\
\bottomrule
\end{tabular}%
}
\end{table}

\begin{table}[h]
\centering
\caption{Source modality-pattern index. Entries are Monte Carlo means with Monte Carlo standard errors in parentheses from 100 replications at each setting. Higher macro-AUC and lower MSE are preferred; the best point estimate in each column is shown in bold.}
\label{tab:supp_modality}
\scriptsize
\setlength{\tabcolsep}{2.4pt}
\renewcommand{\arraystretch}{1.05}
\resizebox{\linewidth}{!}{%
\begin{tabular}{@{}lccccc@{}}
\toprule
Method & $p_{\mathrm{mod}}=0$ & $p_{\mathrm{mod}}=0.1$ & $p_{\mathrm{mod}}=0.2$ & $p_{\mathrm{mod}}=0.3$ & $p_{\mathrm{mod}}=0.4$\\
\midrule
\multicolumn{6}{@{}l}{\textit{Panel A: Target macro-AUC}}\\
Proposed & \textbf{0.9273 (0.0031)} & \textbf{0.9142 (0.0053)} & \textbf{0.9116 (0.0043)} & \textbf{0.9327 (0.0029)} & \textbf{0.9079 (0.0053)}\\
Raw-X XGBoost & 0.7670 (0.0093) & 0.7668 (0.0092) & 0.7653 (0.0091) & 0.7556 (0.0097) & 0.7538 (0.0099)\\
Raw-X LASSO & 0.7549 (0.0126) & 0.7559 (0.0124) & 0.7538 (0.0128) & 0.7586 (0.0121) & 0.7582 (0.0119)\\
Orthogonal & 0.6438 (0.0269) & 0.6789 (0.0272) & 0.6750 (0.0262) & 0.6257 (0.0308) & 0.6349 (0.0290)\\
BONMI & 0.6475 (0.0229) & 0.6537 (0.0233) & 0.6555 (0.0212) & 0.6644 (0.0213) & 0.6396 (0.0255)\\
RGCCA & 0.6041 (0.0219) & 0.6276 (0.0228) & 0.6188 (0.0204) & 0.6424 (0.0227) & 0.6460 (0.0235)\\
Direct & 0.5791 (0.0257) & 0.5860 (0.0265) & 0.6161 (0.0242) & 0.5946 (0.0252) & 0.5915 (0.0268)\\
Target-only & 0.5126 (0.0129) & 0.5049 (0.0130) & 0.4976 (0.0133) & 0.5343 (0.0128) & 0.5066 (0.0136)\\
\midrule
\multicolumn{6}{@{}l}{\textit{Panel B: Target MSE}}\\
Proposed & \textbf{0.1008 (0.0024)} & \textbf{0.1063 (0.0034)} & \textbf{0.1111 (0.0029)} & \textbf{0.0955 (0.0021)} & \textbf{0.1119 (0.0030)}\\
Raw-X XGBoost & 0.2091 (0.0021) & 0.2081 (0.0020) & 0.2090 (0.0019) & 0.2159 (0.0021) & 0.2149 (0.0022)\\
Raw-X LASSO & 0.2168 (0.0081) & 0.2176 (0.0078) & 0.2156 (0.0081) & 0.2151 (0.0075) & 0.2136 (0.0073)\\
Orthogonal & 0.2651 (0.0118) & 0.2517 (0.0117) & 0.2525 (0.0109) & 0.2737 (0.0130) & 0.2826 (0.0131)\\
BONMI & 0.2917 (0.0129) & 0.2886 (0.0133) & 0.2894 (0.0119) & 0.2808 (0.0123) & 0.3037 (0.0149)\\
RGCCA & 0.2733 (0.0092) & 0.2676 (0.0098) & 0.2667 (0.0082) & 0.2585 (0.0095) & 0.2601 (0.0101)\\
Direct & 0.2793 (0.0098) & 0.2754 (0.0101) & 0.2643 (0.0091) & 0.2728 (0.0093) & 0.2810 (0.0109)\\
Target-only & 0.3156 (0.0069) & 0.3181 (0.0070) & 0.3253 (0.0069) & 0.3051 (0.0065) & 0.3179 (0.0069)\\
\bottomrule
\end{tabular}%
}
\end{table}

\subsection{Surrogate-assisted simulation: Design and methods}
\label{subsec:supp_surrogate_design}

The surrogate experiment used a binary restriction of the main data-generating mechanism.  It retained one target and five source domains, the same five raw modality dimensions and source modality patterns, the same rank-deficient auxiliary coordinate sets, reference and auxiliary measurement-error standard deviations 0.26 and 0.41, latent-noise standard deviation 0.70, auxiliary rotation $\varepsilon_{\mathrm{rot}}=0.50$, and independent loading-subspace weight 0.5.  The per-domain sample size varied over
\begin{equation*}
\{25,50,100,150,200,400,800,1600,3200,6400\},
\end{equation*}
with 100 replications at each value.

The target event prevalence was 0.22.  The five source prevalences were
\begin{equation*}
\pi^{(m)}=\operatorname{clip}\{0.22+0.24c_m,0.05,0.95\},
\qquad
(c_1,\ldots,c_5)=(0.6,0.8,1.0,1.2,1.4),
\end{equation*}
which gives $(0.364,0.412,0.460,0.508,0.556)$.  A random unit direction $\mathbf a\in\mathbb R^5$ was shared by the reference and auxiliary latent factors:
\begin{equation*}
\widetilde{\mathbf u}=0.48(2Y-1)\mathbf a+\bm\xi_u,
\qquad
\widetilde{\mathbf v}=1.34(2Y-1)\mathbf a+\bm\xi_v,
\qquad
\bm\xi_u,\bm\xi_v\sim N_5(\mathbf0,0.70^2\mathbf I).
\end{equation*}
A continuous surrogate was observed only in the sources.  With $\Delta=2\Phi^{-1}(0.80)$,
\begin{equation*}
S\mid Y=y\sim N\{(2y-1)\Delta/2,1\}.
\end{equation*}
Neither target $Y$ nor target $S$ was available during fitting.

The surrogate-based variants used pooled rank-five reference PCA, within-domain rank-three auxiliary PCA, a two-component Gaussian mixture for the source surrogate distribution, and XGBoost surrogate regression with 60 trees, depth two, learning rate 0.04, row and column sampling fractions 0.9, ridge penalty one and minimum child weight five.  The methods were defined as follows.
\begin{itemize}[leftmargin=2.0em,itemsep=2pt,topsep=3pt]
\item \textit{Proposed (no $Y$)} used no source gold-standard labels.  A reference-only surrogate bridge initialized the target prevalence.  Within each source, a pre-match-up regression of $S$ on the reference and available auxiliary scores generated working outcome probabilities and expected label-shift weights.  These weights entered the reference-anchored match-up, after which a pooled aligned surrogate bridge produced the target likelihood ratio.
\item \textit{Proposed (semi)} used exactly the same weights and aligned representation as Proposed (no $Y$).  A Bernoulli 20\% subset of source gold-standard outcomes was used only to add a monotone intercept correction to the final working probability.  The calibration was activated only when the pooled gold subset contained at least 150 observations and both outcome classes; its shrinkage parameter was 10.  Consequently, this variant has exactly the same AUC as Proposed (no $Y$), but can improve MSE.
\item \textit{Reference-only} used the same source surrogate information but excluded all auxiliary alignment and transfer terms.  \textit{Direct full-$Y$} used all source gold-standard outcomes with the reference representation but without target calibration or cross-domain alignment.  \textit{Orthogonal full-$Y$} used all source gold-standard outcomes and an unweighted orthogonal alignment, but not the target CCA anchor.  \textit{Proposed full-$Y$} used all source gold-standard outcomes in the complete target-calibrated reference-anchored procedure and serves as the full-label proposed benchmark.
\item The final \textit{Vanilla XGBoost} and \textit{Vanilla LASSO} baselines pooled all five sources by row and concatenated all five raw modality blocks, with an absent source block represented by missing values and no source-site indicator.  Both used the same working label: the true source $Y$ on the Bernoulli 20\% gold mask and $\mathbf1(S>0)$ otherwise.  Vanilla XGBoost used 250 trees, depth three, learning rate 0.05 and row and column sampling fractions 0.9.  Vanilla LASSO used source-only median imputation with missingness indicators, standardization, and $L_1$ logistic cross-validation over 12 $C$ values using negative Brier score.  Neither direct learner used label-shift weights, EM, the proposed alignment or any target-label adaptation \citep{chen2016xgboost,tibshirani1996regression}.
\end{itemize}
Target AUC and target MSE were evaluated against the hidden binary target outcome.

\subsection{Surrogate-assisted simulation: Results}
\label{subsubsec:supp_surrogate_results}

At $n=400$ per domain, Proposed (no $Y$) attained AUC 0.9695 and MSE 0.0880.  Relative to Reference-only, this is a 20.4\% increase in AUC and a 55.3\% reduction in MSE; relative to Direct full-$Y$, the corresponding changes are 18.8\% and 54.6\%.  Proposed (semi) had the same AUC and reduced MSE further to 0.0814, while the Proposed full-$Y$ benchmark attained AUC 0.9794 and MSE 0.0722.

As the sample size increased, the no-gold surrogate bridge approached the full-label proposed benchmark.  At $n=6400$, Proposed (no $Y$) had AUC 0.9822, only 0.29\% below the full-label value 0.9851, and MSE 0.0676 versus 0.0617.  The semi-supervised calibration reduced MSE to 0.0608.  At the two largest sample sizes, its Monte Carlo mean was slightly below the full-label benchmark; this should not be interpreted as a systematic oracle improvement, because the differences are small relative to Monte Carlo uncertainty and the two procedures use different calibrated working models.

The direct learners did not recover the same sample-size trend.  Reference-only and Direct full-$Y$ plateaued near AUC 0.82 and MSE 0.19.  Vanilla XGBoost and Vanilla LASSO improved only modestly, reaching AUC 0.7702 and 0.7965 and MSE 0.1878 and 0.1835, respectively, at $n=6400$.  Orthogonal full-$Y$ eventually improved in ranking but remained less accurate probabilistically, with AUC 0.9335 and MSE 0.1212 at $n=6400$.  At the smallest sample sizes, the proposed full-label and surrogate procedures were themselves variable; the advantage of the reference-anchored pathway became systematic as the representation and bridge estimators stabilized.

\begin{table}[h]
\centering
\caption{Surrogate-assisted simulation at smaller per-domain sample sizes. Entries are Monte Carlo means with Monte Carlo standard errors in parentheses from 100 replications. Higher AUC and lower MSE are preferred; the best point estimate in each column is shown in bold.}
\label{tab:supp_surrogate_small}
\scriptsize
\setlength{\tabcolsep}{2.8pt}
\renewcommand{\arraystretch}{1.05}
\resizebox{\linewidth}{!}{%
\begin{tabular}{@{}lccccc@{}}
\toprule
Method & $n=25$ & $n=50$ & $n=100$ & $n=150$ & $n=200$\\
\midrule
\multicolumn{6}{@{}l}{\textit{Panel A: Target AUC}}\\
Proposed (no $Y$) & 0.7665 (0.0154) & 0.8543 (0.0120) & 0.9171 (0.0066) & 0.9367 (0.0041) & 0.9504 (0.0034)\\
Proposed (semi) & 0.7665 (0.0154) & 0.8543 (0.0120) & 0.9171 (0.0066) & 0.9367 (0.0041) & 0.9504 (0.0034)\\
Vanilla XGBoost & 0.6635 (0.0162) & 0.6844 (0.0136) & 0.7115 (0.0136) & 0.7206 (0.0134) & 0.7299 (0.0139)\\
Vanilla LASSO & 0.6055 (0.0189) & 0.6391 (0.0167) & 0.6725 (0.0171) & 0.6991 (0.0169) & 0.7101 (0.0157)\\
Ref.-only & 0.7308 (0.0129) & 0.7626 (0.0087) & 0.7886 (0.0053) & 0.7873 (0.0050) & 0.8009 (0.0039)\\
Direct full-$Y$ & \textbf{0.8175 (0.0105)} & 0.8162 (0.0076) & 0.8153 (0.0054) & 0.8085 (0.0047) & 0.8154 (0.0035)\\
Orth. full-$Y$ & 0.4692 (0.0291) & 0.4981 (0.0328) & 0.5417 (0.0375) & 0.5477 (0.0352) & 0.5891 (0.0384)\\
Prop. full-$Y$ & 0.8098 (0.0145) & \textbf{0.8782 (0.0124)} & \textbf{0.9345 (0.0055)} & \textbf{0.9546 (0.0032)} & \textbf{0.9676 (0.0027)}\\
\midrule
\multicolumn{6}{@{}l}{\textit{Panel B: Target MSE}}\\
Proposed (no $Y$) & 0.2644 (0.0158) & 0.1797 (0.0078) & 0.1482 (0.0068) & 0.1269 (0.0054) & 0.1078 (0.0046)\\
Proposed (semi) & 0.2644 (0.0158) & 0.1797 (0.0078) & 0.1482 (0.0068) & 0.1197 (0.0051) & 0.0988 (0.0041)\\
Vanilla XGBoost & 0.2361 (0.0076) & 0.2121 (0.0061) & 0.1982 (0.0053) & 0.1995 (0.0053) & 0.1940 (0.0057)\\
Vanilla LASSO & 0.2444 (0.0067) & 0.2382 (0.0063) & 0.2299 (0.0070) & 0.2231 (0.0065) & 0.2196 (0.0067)\\
Ref.-only & 0.2525 (0.0099) & 0.2198 (0.0062) & 0.2121 (0.0043) & 0.2098 (0.0039) & 0.2012 (0.0030)\\
Direct full-$Y$ & \textbf{0.1981 (0.0049)} & 0.1903 (0.0029) & 0.1993 (0.0026) & 0.1988 (0.0019) & 0.1935 (0.0017)\\
Orth. full-$Y$ & 0.2721 (0.0088) & 0.2674 (0.0099) & 0.2643 (0.0123) & 0.2491 (0.0100) & 0.2469 (0.0134)\\
Prop. full-$Y$ & 0.2351 (0.0162) & \textbf{0.1422 (0.0076)} & \textbf{0.1172 (0.0050)} & \textbf{0.1010 (0.0038)} & \textbf{0.0859 (0.0035)}\\
\bottomrule
\end{tabular}%
}
\end{table}

\begin{table}[h]
\centering
\caption{Surrogate-assisted simulation at larger per-domain sample sizes. Entries are Monte Carlo means with Monte Carlo standard errors in parentheses from 100 replications. Higher AUC and lower MSE are preferred; the best point estimate in each column is shown in bold.}
\label{tab:supp_surrogate_large}
\scriptsize
\setlength{\tabcolsep}{2.8pt}
\renewcommand{\arraystretch}{1.05}
\resizebox{\linewidth}{!}{%
\begin{tabular}{@{}lccccc@{}}
\toprule
Method & $n=400$ & $n=800$ & $n=1600$ & $n=3200$ & $n=6400$\\
\midrule
\multicolumn{6}{@{}l}{\textit{Panel A: Target AUC}}\\
Proposed (no $Y$) & 0.9695 (0.0017) & 0.9757 (0.0009) & 0.9786 (0.0009) & 0.9821 (0.0005) & 0.9822 (0.0005)\\
Proposed (semi) & 0.9695 (0.0017) & 0.9757 (0.0009) & 0.9786 (0.0009) & 0.9821 (0.0005) & 0.9822 (0.0005)\\
Vanilla XGBoost & 0.7342 (0.0139) & 0.7578 (0.0137) & 0.7680 (0.0147) & 0.7644 (0.0171) & 0.7702 (0.0163)\\
Vanilla LASSO & 0.7132 (0.0163) & 0.7585 (0.0148) & 0.7677 (0.0164) & 0.7806 (0.0184) & 0.7965 (0.0169)\\
Ref.-only & 0.8055 (0.0024) & 0.8123 (0.0020) & 0.8144 (0.0013) & 0.8172 (0.0009) & 0.8165 (0.0007)\\
Direct full-$Y$ & 0.8159 (0.0024) & 0.8175 (0.0019) & 0.8173 (0.0013) & 0.8188 (0.0009) & 0.8175 (0.0007)\\
Orth. full-$Y$ & 0.6399 (0.0412) & 0.7477 (0.0382) & 0.8645 (0.0318) & 0.8673 (0.0317) & 0.9335 (0.0234)\\
Prop. full-$Y$ & \textbf{0.9794 (0.0010)} & \textbf{0.9822 (0.0006)} & \textbf{0.9832 (0.0004)} & \textbf{0.9848 (0.0004)} & \textbf{0.9851 (0.0003)}\\
\midrule
\multicolumn{6}{@{}l}{\textit{Panel B: Target MSE}}\\
Proposed (no $Y$) & 0.0880 (0.0032) & 0.0749 (0.0014) & 0.0711 (0.0009) & 0.0687 (0.0007) & 0.0676 (0.0005)\\
Proposed (semi) & 0.0814 (0.0031) & 0.0682 (0.0012) & 0.0646 (0.0010) & \textbf{0.0620 (0.0005)} & \textbf{0.0608 (0.0004)}\\
Vanilla XGBoost & 0.1931 (0.0046) & 0.1879 (0.0043) & 0.1854 (0.0043) & 0.1873 (0.0049) & 0.1878 (0.0045)\\
Vanilla LASSO & 0.2232 (0.0070) & 0.1997 (0.0057) & 0.1959 (0.0064) & 0.1910 (0.0070) & 0.1835 (0.0065)\\
Ref.-only & 0.1968 (0.0016) & 0.1953 (0.0012) & 0.1948 (0.0009) & 0.1946 (0.0006) & 0.1953 (0.0004)\\
Direct full-$Y$ & 0.1941 (0.0012) & 0.1933 (0.0009) & 0.1939 (0.0006) & 0.1937 (0.0004) & 0.1939 (0.0003)\\
Orth. full-$Y$ & 0.2309 (0.0139) & 0.1895 (0.0124) & 0.1477 (0.0125) & 0.1539 (0.0141) & 0.1212 (0.0098)\\
Prop. full-$Y$ & \textbf{0.0722 (0.0016)} & \textbf{0.0657 (0.0010)} & \textbf{0.0640 (0.0007)} & 0.0632 (0.0005) & 0.0617 (0.0003)\\
\bottomrule
\end{tabular}%
}
\end{table}

\subsection{Crossed ablation of the target-anchor CCA and target update}
\label{subsec:supp_ablation}

The crossed ablation was conducted at $\varepsilon_{\mathrm{rot}}=1$, with $\delta=0.24$, $p_{\mathrm{lab}}=0.35$, the base source modality patterns, and 400 subjects in every domain.  The same 100 Monte Carlo seeds were used in all four cells.  The full method used pooled reference PCA, within-domain auxiliary PCA, raw reference scores for the initial target calibration, outcome-reweighted block alignment, the rank-five single-joint target-anchor CCA regression, source-overlap augmentation and the stabilized target update.

The CCA switch disabled only the target-anchor CCA regression step.  The preceding outcome-reweighted block alignment, initial target calibration, density-ratio fitting and all other components remained active.  Thus this ablation isolates the extra target-defined CCA/ridge match-up from the reweighted preliminary alignment; it does not remove every alignment operation.  The target-update switch bypassed the stabilized update while retaining the initial reference-based target distribution estimate.  Crossing the two switches gave the full method, target-anchor CCA ablated, target update ablated, and both components ablated.  Whenever the target update was active, the same convergence and marginal-shift safeguard described above was applied.  It was activated in 95 of 100 full-method replications and in all 100 target-anchor-CCA-ablated replications.

Removing the target-anchor CCA component caused the larger deterioration.  Macro-AUC fell from 0.9297 to 0.8635 and MSE increased from 0.0983 to 0.1361.  The paired full-minus-ablation AUC difference was 0.0662 (95\% Monte Carlo interval 0.0591 to 0.0733), while the paired MSE reduction from retaining the component was 0.0379 (0.0332 to 0.0425).  Hamming loss increased from 0.1306 to 0.1948 and exact-match accuracy decreased from 0.6786 to 0.5724.

The target update had a smaller, calibration-specific effect.  Relative to the no-update variant, the full method reduced MSE by 0.0069 (0.0044 to 0.0094), Hamming loss from 0.1395 to 0.1306, and increased exact-match accuracy from 0.6635 to 0.6786.  Macro-AUC was 0.0030 lower with the update, with the paired interval $-0.0047$ to $-0.0012$ for full minus no update.  The state-prevalence $L_1$ reduction was 0.0213, with interval $-0.0022$ to 0.0448.  Removing both components produced the largest MSE, state-prevalence error and Hamming loss.  Together, these comparisons show that the target-anchor CCA regression supplies the main discrimination gain under loading rotation, while the stabilized target update primarily refines the posterior probability scale.

\begin{table}[h]
\centering
\caption{Crossed ablation at $\varepsilon_{\mathrm{rot}}=1$. Performance entries are Monte Carlo means with Monte Carlo standard errors in parentheses from 100 paired replications. Paired contrasts are normal-approximation 95\% Monte Carlo confidence intervals and are oriented so that positive values favor the full method: full minus ablation for macro-AUC and ablation minus full for the error metrics.}
\label{tab:supp_ablation}
\scriptsize
\setlength{\tabcolsep}{2.5pt}
\renewcommand{\arraystretch}{1.08}
\begin{tabular*}{\linewidth}{@{\extracolsep{\fill}}lccccc@{}}
\toprule
Variant & \shortstack{Target-anchor\\CCA} & \shortstack{Target\\update} & \shortstack{Target\\macro-AUC} & \shortstack{Target\\MSE} & \shortstack{State\\$L_1$ error}\\
\midrule
Full proposed & On & On & 0.9297 (0.0032) & \textbf{0.0983 (0.0024)} & \textbf{0.3423 (0.0133)}\\
Target update ablated & On & Off & \textbf{0.9326 (0.0032)} & 0.1052 (0.0024) & 0.3636 (0.0083)\\
Target-anchor CCA ablated & Off & On & 0.8635 (0.0033) & 0.1361 (0.0019) & 0.3745 (0.0113)\\
Both components ablated & Off & Off & 0.8627 (0.0033) & 0.1417 (0.0020) & 0.3886 (0.0100)\\
\bottomrule
\end{tabular*}
\vspace{0.6em}
\begin{tabular*}{\linewidth}{@{\extracolsep{\fill}}lcc@{}}
\toprule
Variant & Hamming loss & Exact match accuracy\\
\midrule
Full proposed & \textbf{0.1306 (0.0032)} & \textbf{0.6786 (0.0063)}\\
Target update ablated & 0.1395 (0.0032) & 0.6635 (0.0062)\\
Target-anchor CCA ablated & 0.1948 (0.0029) & 0.5724 (0.0047)\\
Both components ablated & 0.2024 (0.0031) & 0.5620 (0.0049)\\
\bottomrule
\end{tabular*}
\vspace{0.8em}
\resizebox{\linewidth}{!}{%
\begin{tabular}{@{}lccc@{}}
\toprule
Full method versus & $\Delta$macro-AUC (95\% CI) & $\Delta$MSE (95\% CI) & $\Delta$state $L_1$ (95\% CI)\\
\midrule
Target-anchor CCA ablated & 0.0662 (0.0591, 0.0733) & 0.0379 (0.0332, 0.0425) & 0.0322 (0.0087, 0.0558)\\
Target update ablated & -0.0030 (-0.0047, -0.0012) & 0.0069 (0.0044, 0.0094) & 0.0213 (-0.0022, 0.0448)\\
Both components ablated & 0.0670 (0.0599, 0.0740) & 0.0435 (0.0387, 0.0483) & 0.0464 (0.0252, 0.0675)\\
\bottomrule
\end{tabular}%
}
\end{table}

\clearpage
\section{Real-data analysis protocol and supporting results}
\label{sec:supp_real_data}

\subsection{Data assets, cohort construction and 12-month endpoint}
\label{subsec:supp_real_cohort}

The analysis reused patient-level assets developed in the two RCC studies cited in the main
manuscript.  The EHR study supplied the phenotyped RCC cohort, NLP-derived stage, grade and
histology variables, and longitudinal \textsc{latte} recurrence outputs.  The companion imaging
study supplied linked preoperative CT scans, report-derived concepts, conventional radiomic
features and residual radiomic features.  The current analysis imposed a new endpoint, temporal
split and modality-domain protocol on those source files; it did not reuse the previous papers'
prediction targets or train--test splits.

After demographic, nephrectomy and stage-based exclusions, 8235 patients remained in the
working cohort.  A patient was eligible for the 12-month analysis when recurrence occurred by 12
months or at least 12 months of post-nephrectomy follow-up was available.  Requiring a nonmissing
\texttt{year\_first\_ICD} proxy left 7713 eligible patients.  For chart-reviewed patients, the
analysis endpoint was the physician-adjudicated indicator \texttt{event12}.  For all other patients,
the endpoint was the \textsc{latte}-derived indicator $\mathbf 1\{S_{\rm rec}=1,\ S_T\leq 12\}.$ The resulting mixed endpoint contained 240 physician-adjudicated labels and 7473 surrogate
labels.  Variables directly encoding recurrence status, recurrence time or death within 12 months
were removed before preprocessing and model fitting.

\subsection{Temporal split and modality-domain assignment}
\label{subsec:supp_real_split}

The temporal split was pre-specified before any model was fitted.  Patients with
\texttt{year\_first\_ICD}$\leq 2016$ formed the source/training set, and patients with
\texttt{year\_first\_ICD}$\geq 2017$ formed the late target/test set.  The five modality blocks were:
(1) demographics and NLP-derived clinicopathologic variables; (2) longitudinal codified EHR and
healthcare-utilization summaries; (3) radiology-report concepts; (4) conventional CT radiomics;
and (5) residual radiomic features.  The three source domains were $1+2$, $1+2+3$ and $1+4+5$.  The primary late-target domain contained all five modalities, $1+2+3+4+5$.  The smaller late-target configurations $1+2$ and $1+2+3$, together with their pooled analysis, were retained as supplementary configurations.
Rare partial-extra patterns were assigned downward to the largest fully observed compatible
domain: $1+2+4$ to $1+2$, and $1+2+3+4$ or $1+2+3+5$ to $1+2+3$.  No patient was assigned
upward by imputing an entirely missing modality block.

\begin{table}[H]
\centering
\caption{Cohort and label composition under the pre-specified temporal split.  The pooled source and complete late-target rows form the primary analysis ($N=6053$); the remaining late-target rows define supplementary configurations.  Events refer to the mixed 12-month endpoint: physician-adjudicated \texttt{event12} for chart-reviewed patients and \textsc{latte}-derived $S12$ otherwise.}
\label{tab:supp_rcc_counts}
\scriptsize
\setlength{\tabcolsep}{3.1pt}
\begin{tabular}{llrrrrrr}
\toprule
Set & Domain & $n$ & Events & Gold $n$ & Gold events & Surrogate $n$ & Surrogate events \\
\midrule
Source ($\leq 2016$) & $1+2$ & 265 & 56 & 5 & 0 & 260 & 56 \\
Source ($\leq 2016$) & $1+2+3$ & 4248 & 805 & 130 & 17 & 4118 & 788 \\
Source ($\leq 2016$) & $1+4+5$ & 1161 & 172 & 43 & 4 & 1118 & 168 \\
Source ($\leq 2016$) & Pooled & 5674 & 1033 & 178 & 21 & 5496 & 1012 \\
\midrule
Target ($\geq 2017$) & $1+2$ & 136 & 29 & 3 & 0 & 133 & 29 \\
Target ($\geq 2017$) & $1+2+3$ & 1524 & 326 & 41 & 8 & 1483 & 318 \\
Target ($\geq 2017$) & $1+2+3+4+5$ & 379 & 100 & 18 & 2 & 361 & 98 \\
Target ($\geq 2017$) & Pooled & 2039 & 455 & 62 & 10 & 1977 & 445 \\
\bottomrule
\end{tabular}
\end{table}

All target outcomes were withheld from centering, feature processing, component construction,
CCA alignment, density-ratio estimation, target-EM, model fitting and hyperparameter selection.
Unlabelled target covariates entered only the prespecified unsupervised adaptation steps.  Target
outcomes were merged back after all predictions had been fixed.

\subsection{Primary analysis and supplementary target configurations}
\label{subsec:supp_real_configurations}

The primary test reported in Section 6 of the main paper used all 5674 source patients in their observed heterogeneous modality domains and the 379 patients in the complete late-target domain, for a total analysis cohort of 6053 patients.  Thus, no source domain shown in the main paper was discarded: each contributed through its available blocks, while the target supplied modalities 1--5.  The late-target $1+2$ stratum ($n=136$, 29 events), the late-target $1+2+3$ stratum ($n=1524$, 326 events), and the pooled late-target cohort ($n=2039$, 455 events) are secondary configurations documented only in this supplement.  They are useful for checking robustness to target-side modality availability but do not replace the complete-target primary analysis.

\subsection{Model implementations, metrics and uncertainty}
\label{subsec:supp_real_models}

The proposed row used deterministic within-site centering without PCA in the initial representation
step, modality 1 as the reference block, a rank-10 CCA anchor, target-EM enabled and
source-overlap augmentation disabled.  In the primary complete-target experiment, all 5674 source rows were used in their observed $1+2$, $1+2+3$ or $1+4+5$ domains, while the 379 complete-target rows retained modalities 1--5.
The comparator set comprised direct-transfer XGBoost, category-specific logistic LASSO, RGCCA
and BONMI.  Each comparator used only allowable source rows and received no source reweighting,
label-shift correction, target-EM or target-label calibration.

The two primary metrics were the Brier skill score
\begin{equation*}
  \operatorname{BSS}=1-
  \frac{n^{-1}\sum_{i=1}^n(y_i-\widehat p_i)^2}
       {n^{-1}\sum_{i=1}^n(y_i-\bar y)^2},
\end{equation*}
where $\bar y$ is the event rate in the evaluated target sample, and the absolute calibration gap
$|\bar{\widehat p}-\bar y|$.  AUC summarized ranking performance, and average precision (AP)
summarized precision across recall levels.  The real-data tables report these metrics in the order
BSS, calibration gap, AUC and AP, except that the ablation table focuses on the two core rubrics,
BSS and calibration gap.  Higher BSS, AUC and AP and lower calibration gap are preferred.
Each table gives point estimates with standard errors in parentheses from 1,000 stratified
patient-level bootstrap resamples of the fixed predictions without model refitting.  Resampling was
stratified by event status within a single target configuration and jointly by event status and target
modality configuration for pooled analyses.  The ablation used 1,000 common event-stratified
resamples so that both cell-specific standard errors and paired-difference standard errors were
computed from the same resamples.

\subsection{Surrogate-row-only endpoint sensitivity}
\label{subsec:supp_real_surrogate_only}

Table~\ref{tab:supp_rcc_surrogate_only} restricts evaluation to target rows with a
\textsc{latte}-derived surrogate endpoint while retaining the predictions from the primary fits.
In the complete target domain, the proposed method retained the highest BSS, 0.4389 (SE 0.0476),
the smallest calibration gap, 0.0080 (SE 0.0084), and the highest AUC, 0.9082 (SE 0.0151).
XGBoost retained the highest AP, 0.7662 (SE 0.0361), compared with 0.7238 (SE 0.0454) for the
proposed method.  Results in the smaller and pooled target configurations varied by metric, so this
sensitivity analysis supports the primary complete-target calibration finding without implying
uniform superiority in every supplementary configuration.

\begin{table}[H]
\centering
\caption{Surrogate-row-only endpoint sensitivity using the fixed predictions from the primary fits.
Parentheses give standard errors from 1,000 stratified patient-level bootstrap resamples without
model refitting.  Higher BSS, AUC and AP and lower calibration gap are preferred.  The best point
estimate within each target domain is shown in \textbf{bold}.}
\label{tab:supp_rcc_surrogate_only}
\scriptsize
\setlength{\tabcolsep}{4.2pt}
\begin{tabular}{llcccc}
\toprule
Target domain & Method & BSS & Cal. gap & AUC & AP \\
\midrule
$1+2$ & Proposed & 0.2335 (0.0914) & 0.0397 (0.0199) & 0.8488 (0.0344) & 0.6429 (0.0678) \\
& XGBoost & 0.3404 (0.0789) & 0.0408 (0.0156) & 0.8773 (0.0317) & 0.6932 (0.0737) \\
& Category-specific LASSO & \textbf{0.3818 (0.0610)} & 0.0220 (0.0121) & \textbf{0.9082 (0.0250)} & \textbf{0.7444 (0.0648)} \\
& RGCCA & 0.0169 (0.0964) & \textbf{0.0150 (0.0146)} & 0.7225 (0.0501) & 0.4199 (0.0708) \\
& BONMI & -0.0199 (0.0200) & 0.0394 (0.0037) & 0.5305 (0.0649) & 0.2492 (0.0437) \\
\addlinespace
$1+2+3$ & Proposed & 0.3007 (0.0281) & 0.0238 (0.0057) & 0.8653 (0.0108) & 0.6163 (0.0256) \\
& XGBoost & \textbf{0.3912 (0.0246)} & 0.0355 (0.0050) & \textbf{0.9003 (0.0087)} & \textbf{0.7229 (0.0215)} \\
& Category-specific LASSO & 0.3243 (0.0298) & \textbf{0.0116 (0.0058)} & 0.8796 (0.0098) & 0.6180 (0.0267) \\
& RGCCA & 0.2377 (0.0257) & 0.0189 (0.0049) & 0.8341 (0.0126) & 0.5628 (0.0258) \\
& BONMI & 0.0420 (0.0085) & 0.0300 (0.0018) & 0.6619 (0.0180) & 0.3762 (0.0239) \\
\addlinespace
$1+2+3+4+5$ & Proposed & \textbf{0.4389 (0.0476)} & \textbf{0.0080 (0.0084)} & \textbf{0.9082 (0.0151)} & 0.7238 (0.0454) \\
& XGBoost & 0.3461 (0.0463) & 0.0907 (0.0103) & 0.9044 (0.0160) & \textbf{0.7662 (0.0361)} \\
& Category-specific LASSO & 0.3018 (0.0505) & 0.0822 (0.0112) & 0.8771 (0.0209) & 0.6989 (0.0450) \\
& RGCCA & 0.1404 (0.0359) & 0.1131 (0.0072) & 0.8680 (0.0181) & 0.6242 (0.0448) \\
& BONMI & -0.0430 (0.0102) & 0.0941 (0.0023) & 0.5273 (0.0345) & 0.3056 (0.0302) \\
\addlinespace
Pooled & Proposed & 0.3268 (0.0223) & \textbf{0.0191 (0.0051)} & 0.8739 (0.0083) & 0.6405 (0.0205) \\
& XGBoost & \textbf{0.3802 (0.0210)} & 0.0459 (0.0040) & \textbf{0.8987 (0.0076)} & \textbf{0.7262 (0.0181)} \\
& Category-specific LASSO & 0.3253 (0.0257) & 0.0252 (0.0050) & 0.8771 (0.0094) & 0.6324 (0.0235) \\
& RGCCA & 0.2051 (0.0201) & 0.0358 (0.0042) & 0.8261 (0.0102) & 0.5445 (0.0206) \\
& BONMI & 0.0230 (0.0065) & 0.0423 (0.0014) & 0.6228 (0.0149) & 0.3389 (0.0178) \\
\bottomrule
\end{tabular}
\end{table}

\subsection{Complete-domain matched-modality sensitivity}
\label{subsec:supp_real_sensitivity}

Table~\ref{tab:supp_rcc_table4} reports the sensitivity analysis in which the $1+4+5$ source domain and complete-target rows were both
restricted to modalities 1, 4 and 5.  All five methods were refitted under this matched-modality
specification, with target outcomes still withheld.  The proposed method had the best point estimate
for each reported metric: BSS 0.3487 (SE 0.0438), calibration gap 0.0379 (SE 0.0103), AUC 0.8858
(SE 0.0167) and AP 0.6666 (SE 0.0456).  This analysis separates the value of reference-anchored
integration from the additional target-side availability of modalities 2 and 3 in the primary
complete-target analysis.

\begin{table}[H]
\centering
\caption{Complete-domain matched-modality sensitivity retaining modalities 1, 4 and 5 in both
source and target ($n=379$, 100 events).  Parentheses give event-stratified bootstrap standard
errors from 1,000 resamples.  Higher BSS, AUC and AP and lower calibration gap are preferred.  The
best point estimate for each metric is shown in \textbf{bold}.}
\label{tab:supp_rcc_table4}
\small
\setlength{\tabcolsep}{5.1pt}
\begin{tabular}{lcccc}
\toprule
Method & BSS & Cal. gap & AUC & AP \\
\midrule
Proposed & \textbf{0.3487 (0.0438)} & \textbf{0.0379 (0.0103)} & \textbf{0.8858 (0.0167)} & \textbf{0.6666 (0.0456)} \\
XGBoost & 0.2288 (0.0538) & 0.0862 (0.0106) & 0.8520 (0.0209) & 0.6109 (0.0476) \\
Category-specific LASSO & 0.2529 (0.0451) & 0.0720 (0.0095) & 0.8481 (0.0213) & 0.5941 (0.0448) \\
RGCCA & 0.1103 (0.0371) & 0.1160 (0.0078) & 0.8253 (0.0215) & 0.5581 (0.0432) \\
BONMI & 0.1197 (0.0395) & 0.1131 (0.0088) & 0.8215 (0.0234) & 0.5523 (0.0422) \\
\bottomrule
\end{tabular}
\end{table}

\subsection{Complete-domain sensitivity using modalities 1, 2 and 3}
\label{subsec:supp_real_sensitivity_123}

Table~\ref{tab:supp_rcc_table5} reports the complementary matched-modality analysis retaining modalities 1, 2 and 3.
The proposed method had the highest BSS, 0.3938 (SE 0.0471), and the smallest calibration gap,
0.0113 (SE 0.0092).  RGCCA had the highest AUC, 0.9068 (SE 0.0162), and AP, 0.7952
(SE 0.0348), compared with 0.8946 (SE 0.0159) and 0.7103 (SE 0.0448), respectively, for the
proposed method.  Thus, this configuration again favors the proposed method for probability
accuracy and calibration, but not for every ranking metric.

\begin{table}[H]
\centering
\caption{Complete-domain matched-modality sensitivity retaining modalities 1, 2 and 3 in both
source and target ($n=379$, 100 events).  Parentheses give event-stratified bootstrap standard
errors from 1,000 resamples.  Higher BSS, AUC and AP and lower calibration gap are preferred.  The
best point estimate for each metric is shown in \textbf{bold}.}
\label{tab:supp_rcc_table5}
\small
\setlength{\tabcolsep}{5.1pt}
\begin{tabular}{lcccc}
\toprule
Method & BSS & Cal. gap & AUC & AP \\
\midrule
Proposed & \textbf{0.3938 (0.0471)} & \textbf{0.0113 (0.0092)} & 0.8946 (0.0159) & 0.7103 (0.0448) \\
XGBoost & 0.3430 (0.0512) & 0.0756 (0.0110) & 0.8896 (0.0168) & 0.7358 (0.0404) \\
Category-specific LASSO & 0.3306 (0.0503) & 0.0738 (0.0104) & 0.8836 (0.0203) & 0.7118 (0.0458) \\
RGCCA & 0.3782 (0.0477) & 0.0942 (0.0100) & \textbf{0.9068 (0.0162)} & \textbf{0.7952 (0.0348)} \\
BONMI & 0.3093 (0.0457) & 0.0950 (0.0093) & 0.8742 (0.0214) & 0.7357 (0.0408) \\
\bottomrule
\end{tabular}
\end{table}

\subsection{Crossed CCA and target-EM ablation}
\label{subsec:supp_real_ablation}

The ablation study held the primary complete-target cohort, endpoint, modality rules, preprocessing and
all non-ablated settings fixed.  Two binary components were crossed: the target-anchor CCA/ridge match-up and the target outcome-distribution update implemented by target-EM.  The full-model cell used the same complete-domain predictions as the reference
configuration; only the other three cells were refitted.  As shown in Table~\ref{tab:supp_rcc_table6}, removal of the target-anchor CCA/ridge match-up
reduced BSS by 0.0710 (paired SE 0.0260) and increased the calibration gap by 0.0628 (paired
SE 0.0116).  Removing target-EM reduced BSS by 0.0195 (paired SE 0.0080) and increased the
calibration gap by 0.0366 (paired SE 0.0118).  Removing both components reduced BSS by 0.0702
(paired SE 0.0257) and increased the calibration gap by 0.0621 (paired SE 0.0110).  The target-EM
update improved probability accuracy and calibration when the target-anchor CCA/ridge match-up was retained;
without this match-up, its effect was negligible.  This crossed pattern indicates that the target update relies
on a sufficiently aligned representation, with the target-anchor CCA/ridge match-up producing the larger contribution.

\begin{table}[H]
\centering
\caption{Crossed ablation of the target-anchor CCA/ridge match-up and target-EM in the complete target domain
($n=379$, 100 events).  Parentheses give standard errors from 1,000 common event-stratified
bootstrap resamples.  The second panel reports full-model-minus-ablation point differences with
paired bootstrap standard errors.  The table reports the two core ablation rubrics: higher BSS and
lower calibration gap are preferred.  For calibration-gap differences, a negative value favors the
full model.}
\label{tab:supp_rcc_table6}
\small
\setlength{\tabcolsep}{7pt}
\begin{tabular}{lcccc}
\toprule
Variant & CCA & Target-EM & BSS & Cal. gap \\
\midrule
Full model & On & On & \textbf{0.4194 (0.0485)} & \textbf{0.0015 (0.0070)} \\
CCA ablated & Off & On & 0.3484 (0.0447) & 0.0642 (0.0097) \\
Target-EM ablated & On & Off & 0.3999 (0.0444) & 0.0381 (0.0105) \\
CCA + target-EM ablated & Off & Off & 0.3492 (0.0460) & 0.0636 (0.0101) \\
\bottomrule
\end{tabular}

\vspace{0.7em}
\setlength{\tabcolsep}{9pt}
\begin{tabular}{lcc}
\toprule
Full model minus & $\Delta$BSS & $\Delta$Cal. gap \\
\midrule
CCA ablated & 0.0710 (0.0260) & $-0.0628$ (0.0116) \\
Target-EM ablated & 0.0195 (0.0080) & $-0.0366$ (0.0118) \\
Both components ablated & 0.0702 (0.0257) & $-0.0621$ (0.0110) \\
\bottomrule
\end{tabular}
\end{table}

\subsection{Missingness and interpretation}
\label{subsec:supp_real_missingness}

Whole missing modality blocks were handled through the canonical domain assignment rather than
upward imputation.  Within an observed block, the pre-specified method-specific preprocessing was used.
XGBoost used its native missing-value mechanism.  Category-specific LASSO used source-only
median imputation, missingness indicators and source-only standardization.  All transformation
parameters estimated from labelled data were learned from source rows only.

Among the 379 patients in the primary complete-target evaluation, 18 have labels adjudicated by physicians and 361 have \textsc{latte}-derived surrogate labels.  The pooled secondary late-target evaluation contains 62 adjudicated labels and 1977 surrogate labels.  Accordingly, the real-data analysis quantifies transport performance for the prespecified mixed endpoint.  It should not be interpreted as a fully adjudicated external validation of 12-month RCC recurrence.


\bibliographystyle{plainnat}
\bibliography{paper-ref}

\end{document}